\documentclass[onecolumn,aps,prd,superscriptaddress,nofootinbib,amsmath,amssymb,floats,floatfix,showkeys,notitlepage,longbibliography]{revtex4-1}

\usepackage{amssymb}
\usepackage{amsmath}
\usepackage{amsfonts}
\usepackage{graphicx}
\usepackage{epsfig}
\usepackage[utf8]{inputenc}
\DeclareUnicodeCharacter{FF0B}{+}
\usepackage{float}
\usepackage{palatino}
\usepackage{mwe,tikz}
\usepackage{subcaption}
\usepackage{color}
\usepackage{enumerate}
\usepackage{changes}
\usepackage{cancel}
\usepackage[colorlinks=true,
            linkcolor=blue,
            urlcolor=blue,
            citecolor=blue]{hyperref}
\usepackage{nicefrac}
\newcommand{\beq}{\begin{equation}}
\newcommand{\eeq}{\end{equation}}
\newcommand{\bea}{\begin{eqnarray}}
\newcommand{\eea}{\end{eqnarray}}

\begin{document}
\title{ Shadow and deflection angle of asymptotic, magnetically-charged, non-singular black hole }
\author{Yashmitha Kumaran }
\email{y.kumaran13@gmail.com}
\affiliation{Physics Department, Eastern Mediterranean
University, Famagusta, 99628 North Cyprus via Mersin 10, Turkey.}

\author{Ali \"{O}vg\"{u}n}
\email{ali.ovgun@emu.edu.tr}
\homepage{https://aovgun.weebly.com} 
\affiliation{Physics Department, Eastern Mediterranean
University, Famagusta, 99628 North Cyprus via Mersin 10, Turkey.}

\begin{abstract}
In this paper, we present a detailed analysis of an asymptotic, magnetically-charged, non-singular (AMCNS) black hole. By utilizing the Gauss-bonnet theorems, we aim to unravel the intricate astrophysics associated with this unique black hole. The study explored various aspects including the black hole's gravitational field, intrinsic properties, light bending, the shadow and greybody bounding of the black hole. Through rigorous calculations and simulations, we derive the weak deflection angle of the optical metric of AMCNS black hole. Additionally, we investigate the impact of the dark matter medium on the deflection angle, examined the distinctive features of the black hole's shadow, and bound its greybody factors. Our findings not only deepen our understanding of gravitational lensing but also pave the way for future improvements in black hole theories by minimizing restrictive assumptions and incorporating a more realistic representation of these cosmic phenomena.

\end{abstract}
\date{\today}

\keywords{ Deflection of light; gravitational lensing; black hole; Gauss-bonnet theorem; Shadow.}
\pacs{04.40.-b, 95.30.Sf, 98.62.Sb}

\maketitle

%\tableofcontents
\section{INTRODUCTION}
The evolution of black hole physics can be traced all the way back to the year 1784 when the English philosopher John Michell \cite{michell}. He theorized that all light emitted by an invisible central body having a radius more than five hundred times that of the sun but with a density equal to that of the sun would return to itself due to its own gravity. Essentially, when an object falls from infinity into the central body, it acquires a velocity greater than that of light at the surface of the central body owing to the latter's force of attraction.

Laplace followed this by proposing an invisible star that could potentially be the largest and highly gravitating luminous object in the universe in the year 1796 \cite{de1799exposition}. His speculations were connected to the observations on an astronomical body with a density equal to that of the Earth but with a diameter more than 250 times that of the sun, not allowing its light to reach us, thanks to its strong attraction.

Einstein's ground-breaking theory formulated in 1915 suggests that the apparent gravitational force originates from the curvature of the fabric of spacetime \cite{Einstein:1916vd}.
Schwarzschild gave the first solution for the simplest black holes in the year 1916 depicting the gravitational field from a spherical mass with no angular momentum, charge, or universal cosmological constant \cite{Schwarzschild:1916uq}.
Between then and the year 1918, the combined works of four European physicists discovered the Reissner-Nordstr\"om metric solved using the Einstein-Maxwell equations representing an electrically charged black holes \cite{1918KNAB...20.1238N}.

Centuries of research have enabled humanity to understand the concept of a star collapsing due to its own intense gravity into a black hole to an extent that nothing can escape its gravitational pull beyond its horizon.
Being a consequence of General Relativity, black holes are enigmatic objects of the universe that started off as a bunch of mathematical equations which were written down only to take form in the sky hundred years later, only to be confirmed with recent observations by Event Horizon Telescopes (EHT) \cite{EventHorizonTelescope:2019dse,EventHorizonTelescope:2022xnr,EventHorizonTelescope:2021dqv,Vagnozzi:2022moj}.

Now that their existence is proven, the notion of singularity inside a black hole becomes the next question of interest \cite{Penrose:1964wq}. Quantum Field Theory is sought for answers in this regard with the novelty idea of Hawking radiation for a stationary black hole suggesting that the black holes shrink until the point of singularity is reached at its center. This stems from a semi-classical approximation that radiation has a negative flux of energy contradicting the singularity assumptions \cite{Hawking:1975vcx}.

When there is no existing singularity, a "regular" black hole is a good prospect to proceed with, as pioneered by Bardeen. These non-singular black holes are deemed to possess regular centers with a core of collapsed charged matter instead of singularity \cite{bardeen1968proceedings,Carleo:2022ukm}.
%*ref* J. M. Bardeen, “Non-singular general-relativistic gravitational collapse,” in Proceedings of International Conference GR5 (Tbilisi, USSR, 1968) p. 174.
The ensuing Einstein tensor is not only reasonable for a physical domain but also satisfies the necessary conditions that eventually give rise to a static, spherically symmetric, and asymptotically flat metric. A non-singular spacetime is defined locally for a black hole forming initially from a vacuum region and evaporating subsequently to a vacuum region, with its quiescence described as a static region \cite{Hayward:2005gi}.

The invisibility of a black hole has not stopped its distant observers from seeing it, thanks to the existence of a shadow \cite{Luminet:1979nyg}. The accretion disk that is formed due to the immense gravity of a black hole attracting everything in its path such as dust, photons, radiation, etc., constitutes the shadow shape and size, hence, indirectly making it visible and observable \cite{Falcke:1999pj,Bronzwaer:2021lzo}. The shadow is imagined to be a circular ring of light around its host; however, in reality, this is not the case. It is a region that is geometrically thick and optically thin, especially for regular black holes, that is not so circular. In relativistic models, the size and shape of a shadow is found to be dependant on the geometry of the spacetime, rather than the characteristics of the accretion \cite{Narayan:2019imo}.

The extreme gravitational pull of the black hole compels the light rays to be deflected towards the singularity, causing those that skim the photon sphere to start looping around it. When a photon ends up precisely on the photon sphere, it will continue to encircle the black hole forever. This phenomenon occurring for the light rays passing in the neighborhood of the unstable photon region abets the intensity of the original source through the extended path length of the light rays, thus, increasing its brightness around the shadow's edge; consecutively, the cloud's brightness just outside the shadow appears to be enhanced as well. The captured light rays spiraling into a black hole and the scattered light rays veering away from it are separated by a shadow, giving it the facade of a bright surrounding illuminating a dark disk. Therefore, it is also known as the critical curve that manifests as an isotropic and homogenous emission ring. For a given black hole, its intrinsic parameters play a major role in determining the size of the shadow, whereas, the light rays and the instabilities of their orbits in the photosphere affect its contour. Numerous researchers have explored the distinct imprints that alternative gravitational theories leave behind by studying the phenomenon of shadows \cite{Symmergent-bh3,Symmergent-bh,Symmergent-bh2,Ghosh:2020spb,Allahyari:2019jqz,Bambi:2019tjh,Vagnozzi:2022moj,Prashant2021,Ovgun:2018tua,Ovgun:2020gjz,Ovgun:2019jdo,Kuang:2022xjp,Kumaran:2022soh,Mustafa:2022xod,Okyay:2021nnh,Atamurotov:2022knb,Abdikamalov:2019ztb,Abdujabbarov:2016efm,Atamurotov:2015nra,Papnoi:2014aaa,Abdujabbarov:2012bn,Atamurotov:2013sca,Cunha:2018acu,Gralla:2019xty,Belhaj:2020okh,Belhaj:2020rdb,Konoplya2019,Wei2019,Ling:2021vgk,Kumar:2020hgm,Kumar2017EPJC,Cunha:2016wzk,Cunha:2016bpi,Cunha:2016bjh,Zakharov:2014lqa,Tsukamoto:2017fxq,Chakhchi:2022fls,Li2020,EventHorizonTelescope:2021dqv,Vagnozzi:2022moj,Pantig:2022ely,Pantig:2022gih,Lobos:2022jsz,Uniyal:2022vdu,Ovgun:2023ego,Uniyal:2023inx,Panotopoulos:2021tkk,Panotopoulos:2022bky,Khodadi:2022pqh,Khodadi:2021gbc,Meng:2023unt,Pantig:2022whj,Pantig:2022sjb, Pantig:2023yer, Wang:2019skw, Roy:2020dyy,Konoplya:2021ube,Anjum:2023axh,Hou:2018bar,Lambiase:2023fbd,
Shaikh:2022ivr,Shaikh:2021cvl,Shaikh:2018lcc,Shaikh:2019fpu,Shaikh:2019hbm,Shaikh:2021yux,Rahaman:2021kge}.

Apart from accretion, the other factor that enables the visibility of the shadow is a phenomenon that occurs due to the deflection of light rays by the gravity fields of a massive object in their path to a distant observer. Called gravitational lensing, the light rays from the source in the background are distorted owing to the massive object acting as a lens \cite{Virbhadra:1999nm,Virbhadra:2002ju,Adler:2022qtb,Bozza:2001xd,Bozza:2002zj,Perlick:2003vg,He:2020eah}. When the lensing is weak, the magnifications are too minute to be detected, yet, enable the observer to study the structural aspects in its periphery. The effects of lensing were seen to influence the shadow dramatically through the radiation from the accretion \cite{Falcke:1999pj,Bronzwaer:2021lzo}. In the realm of astrophysics, the determination of object distances plays a pivotal role in unraveling their inherent qualities and quantities. Remarkably, Virbhadra demonstrated that solely by observing relativistic images, devoid of any knowledge regarding masses and distances, it is feasible to accurately establish an upper limit on the compactness of massive dark entities \cite{Virbhadra:2022ybp}. Furthermore, Virbhadra unveiled a distortion parameter, leading to the cancellation of the algebraic sum of all singular gravitational lensing images. This groundbreaking discovery has undergone rigorous testing using Schwarzschild lensing within both weak and strong gravity fields \cite{Virbhadra:2022iiy}.
Weak gravitational lensing exhibits the property of differential deflection and can be utilized to distinguish between mass distributions. The extent of the light deflection and lensing is denoted by the deflection angle or the bending angle. In order to evaluate the weak deflection angle, the Gauss-Bonnet theorem is used to derive the optical geometry of the corresponding spacetime by Gibbons and Werner in 2008 \cite{Gibbons:2008rj}. Subsequently, this methodology has been extensively employed to investigate a wide array of phenomena \cite{Ovgun:2022opq,Kumaran:2021rgj,Javed:2020pyz,Kumaran:2019qqp,Ovgun:2018fnk,Ovgun:2019wej,Ovgun:2018oxk,Javed:2019ynm,Werner2012,Ishihara:2016vdc,Ono:2017pie,Li:2020dln,Li:2020wvn,Belhaj:2022vte, Pantig:2022toh, Javed:2023iih, Javed:2023IJGMMP, Javed:2022fsn, Javed:2022gtz}. The specialty of this formula is the relationship it establishes between the surface curvature and the underlying topology for a given differential geometry as  \cite{Gibbons:2008rj}:
\begin{equation}
\iint_{D_{R}}\mathcal{K} \mathrm{~d}S+\oint_{\partial D_{R}}\hat{k} \mathrm{~d}\sigma+
\sum_{j}\tilde{\theta}_{j}=2\pi ~\mathcal{X}(D_{R}),
\label{gbtwb}
\end{equation}
where $\mathcal{K}$ is the Gaussian optical curvature, $\hat{k}$ is the geodesic curvature, $\tilde{\theta}_j$ is an exterior angle at the vertex $j$, and $D_R$ is the boundary along which the line element$\sigma$ pertains. For a two-dimensional manifold, a regular domain $D_{R}$ aligned by the two-dimensional surface $S$ with the Riemannian metric $\hat{g}_{ij}$ is considered, along with its piece-wise smooth boundary $\partial D_{R}=\gamma_{g}\cup C_{R}$. Also, for a regular domain, it is known that the Euler characteristic is unity, $\mathcal{X}_{D_{R}}=1$.
\\ \\
\textbf{The Gibbons and Werner method:}
If $\gamma$ is taken to be a smooth curve in the said domain, then $\dot{\gamma}$ becomes the unit-speed vector. %*ref* \cite{Gibbons:2008rj}. 
The geodesic curvature $\kappa$ is given by  \cite{Gibbons:2008rj,Kumaran:2021rgj}:
\begin{equation}
\kappa=g^{opt}(\nabla_{\dot{\gamma}}\dot{\gamma}, \ddot{\gamma}),
\end{equation}
having the unit-speed condition $g^{opt}(\dot{\gamma}, \dot{\gamma})=1$. Here, $\ddot{\gamma}$ is the unit-acceleration vector that is perpendicular to $\dot{\gamma}$.
When $R\rightarrow\infty$, the relevant jump angles are considered to be $\pi/2$; that is to say, the angles respective to the source and the observer sum up as $\tilde{\theta}_{S}+\tilde{\theta}_{O}\rightarrow \pi$.
Putting all this together, Eq.~(\ref{gbtwb}) now becomes:
\begin{equation}
\iint\limits_{\mathcal{D}_{R}}\mathcal{K}\,\mathrm{~d}S+\oint\limits_{C_{R}}\kappa \,%
\mathrm{~d}t\overset{{R\rightarrow \infty }}{=}\iint\limits_{\mathcal{D}
_{\infty }}\mathcal{K}\,\mathrm{~d}S+\int\limits_{0}^{\pi +\hat{\alpha}}\mathrm{~d}\phi
=\pi,
\end{equation}
where, $\phi$ is the angular coordinate centered at the lens and $\hat{\alpha}$ is the small, positive, non-trivial, asymptotic angle of deflection. Knowing that $\gamma _{\tilde{g}}$ is a geodesic and the geodesic curvature offers zero contribution i.e., $\kappa(\gamma_{\tilde{g}})=0$, the curve $C_{R}$ is pursued which contributes in such a manner that:
\begin{equation}
\kappa(C_{R})=\mid \nabla_{\dot{C}_{R}}\dot{C}_{R}\mid.
\end{equation}
Considering $C_{R}:=r(\phi)=R= \text{constant}$, $R$ is the distance
from the origin of the selected coordinate system. The radial component of $\kappa$ becomes:
\begin{equation}
(\nabla_{\dot{C}_{R}}\dot{C}_{R})^{r}=\dot{C}_{R}^{\phi}(\partial_{\phi}
\dot{C}_{R}^{\phi})+\Gamma^{\, r}_{\phi\phi}(\dot{C}_{R}^{\phi})^{2}.
\end{equation}

Evidently, the first term disappears from the presumptions and the second term can be obtained using the unit-speed condition. Then:
$$\lim_{R \to \infty} \kappa (C_{R}) =\lim_{R\to \infty}\left\vert \nabla_{\dot{C}_{R}}\dot{C}_{R}\right\vert\to\frac{1}{R} \quad \text{and at higher limits of radial distance,} \,\, \lim_{R \to \infty } \mathrm{~d}t\to  R \mathrm{~d}\phi.$$

When gravitational lensing is approached geometrically, this proves effective in finding the deflection caused by any curved surface. In the weak lensing regime, it becomes possible to relate the geometry and the topology with the Gauss-Bonnet theorem to determine the bending angle through the optical metric because:
\begin{itemize}
    \item the metric evolves from the surface curvature relating it to its geometry
    \item the geodesics of the metric are spatial light rays regarding the focused light rays as a topological effect.
\end{itemize}
%*ref* https://arxiv.org/pdf/0807.0854.pdf
Gibbons and Werner obtained the bending angle as a consequence of weak lensing using the Gauss-Bonnet theorem with the Gaussian curvature of the optical metric directed outwards from the light ray \cite{Gibbons:2008rj}. For a simply connected and an asymptotically flat domain $D_R$ in which the lens is not an element of the regime but the source and the observer are, with the differential domain consisting of a bounding geodesic at the source, the weak deflection angle is evaluated as  \cite{Gibbons:2008rj,Kumaran:2021rgj}:
\begin{equation}
    \hat{\alpha} = -\iint_{D_R} \mathcal{K} \mathrm{~d}S.
\end{equation} 

These are the foundations of this paper with the target of finding the weak deflection angle for an asymptotic magnetically charged non-singular (AMCNS) black hole. The paper is organized as follows: \S ~\ref{Samcns-sol}, the solution of the AMCNS black hole is analyzed, followed by its application in \S ~\ref{Swdagbt} using the Gauss-Bonnet theorem and the Gibbons and Werner method. \S ~\ref{Swdapm} deals with the influence of dark matter on the bending angle. We then carry on to explore the shadow of the AMCNS black hole in \S ~\ref{Sshadow}. Finally, the greybody factor is investigated in \S ~\ref{Sgreybodyfactor} before concluding in \S ~\ref{Sconc}.

\section{ASYMPTOTIC, MAGNETICALLY-CHARGED, NON-SINGULAR (AMCNS) BLACK HOLE}
\label{Samcns-sol}

Extensive research has been conducted on the solutions that depict black holes within the framework of non-linear electrodynamics in the context of general relativity
\cite{Breton:2003tk,Hendi:2013dwa,Kruglov:2015yua,Kruglov:2016ymq,Kruglov:2016ezw,Kruglov:2017fck,Ovgun:2021ttv}.
To describe the solution for a magnetically charged black hole, consider the non-linear electrodynamics \cite{Kruglov:2017fck}
where the Lagrangian density is written in the exponential form as: 
\begin{equation}
\pounds =-P\exp (-\beta P),  \label{A1}
\end{equation}%
where, $P \equiv \nicefrac{1}{4} \left(F_{\mu \nu }\, F^{\mu \nu }\right)=\nicefrac{1}{2}\left( \mathbf{B}^{2}-\mathbf{E}^{2}\right),~\text{and}~F^{\mu \nu }=\partial ^{\mu}A^{\nu }-\partial
^{\nu }A^{\mu }.$ $F^{\mu \nu }$ is the electromagnetic field tensor, $\mathbf{B}$ is the magnetic field, $\mathbf{E}$ is the electric field, $A^{\mu }$ is the four-potential, and $\beta $ is a parameter with the dimensions of [Length]$^{4}$ having an upper bound of $\left( \beta \leq 1\times 10^{-23}\text{T}^{-2}\right) $ %*ref* PVLAS experiment.
The Euler-Lagrange equation is formulated as:
\begin{equation}
\partial _{\mu }\left( \frac{\partial \pounds }{\partial \left( \partial
_{\mu }A_{\nu}\right) }\right) -\frac{\partial \pounds }{\partial A_{\nu}}=0,
\end{equation}
where $\mu,~\nu \in [0,3]$. Thus, the field equations come to be: 
\begin{equation}
\partial _{\mu }\left[ \left( \beta P-1\right) \exp \left( -\beta P\right)
F^{\mu \nu }\right] =0, \label{A2}
\end{equation}

The energy-momentum tensor takes the form of: 
\begin{equation}
\tau ^{\mu \upsilon }=H^{\mu \lambda }F_{\lambda }^{\upsilon }-g^{\mu
\upsilon }\pounds,  \label{A3}
\end{equation}
where, $g^{\mu \upsilon }={1}/{g_{\mu \upsilon}}$ and:
\begin{equation}
H^{\mu \lambda }\equiv\frac{\partial \pounds }{\partial F_{\mu \lambda }}=-\left(1-\beta P\right) \exp \left( -\beta P\right) F^{\mu \lambda }.  \label{A4}
\end{equation}%
Thus, the energy-momentum tensor derived from the Lagrangian density is:
\begin{equation}
\tau ^{\mu \upsilon }=\exp \left( -\beta P\right) \left[ \left( \beta
P-1\right) F^{\mu \lambda }F_{\lambda }^{\upsilon }+g^{\mu \upsilon }P\right],  \label{A5}
\end{equation}
whose trace is:
\begin{equation}
\tau =4\beta P^{2}\exp \left( -\beta P\right).  \label{A6}
\end{equation} 
The instance of $\beta \rightarrow 0$ not only signifies weak limits, but also reverts to classical electrodynamics as $\pounds \rightarrow -P$ and $\tau=0$ in Eq.~\eqref{A6}.

Typically, when $\beta$ is not zero and the energy-momentum tensor has a non-zero trace, it entails that the scale invariance is violated. That's why, if any forms of nonlinear electrodynamics incorporate a dimensional parameter, they would also break the scale invariance, prompting the dilation current to be non-trivial: $\partial_{\upsilon}\, x^{\mu }\,\tau _{\mu }^{\upsilon} = \tau$ 

This can be overruled by the general principles of causality together with unitarity
according to which the group velocity of the background perturbations are less than $c$. The sustaining of the causality principle necessitates that %*ref*\cite{11, 12, 13} 
$\pounds _{P}=\partial \pounds /\partial P \leq 0 \Rightarrow \beta P\leq 1$. For a purely magnetic field, $B\leq \sqrt{\nicefrac{2}{\beta}}$ ought to be fulfilled. Besides, the constraint $\pounds _{P}+2P\pounds _{PP}\leq 0$ allows the unitarity principle to hold provided $\pounds _{PP}\geq 0$. While Eq.~\eqref{A1} construes $\beta P\leq 0.219$ agreeing with %*ref*\cite{13}. 
, it is clear that the limitations for both the causality and unitarity principles incur for $\beta P\leq 0.219$ giving: 
\begin{equation}
\mathbf{B}\leq \sqrt{\frac{5-\sqrt{17}}{2\beta }}\simeq \frac{0.66}{\sqrt{\beta }}.
\end{equation}

Deriving the metric of a static AMCNS black hole in a spherically symmetric spacetime is our next goal. For a pure magnetic field, 
the governing equations are as follows: \newline
The invariant $P$ is deliberated in terms of an electric charge $q$ to be:
\begin{equation}
P=\frac{q^{2}}{2r^{4}}.  \label{A7}
\end{equation}
The generic line element of a spherically symmetric black hole with a time coordinate $t$, radial coordinate $r$, co-latitude $\theta$ and longitude $\phi$ both outlined for a point on the two-sphere, and mass $M$ is given by:
\begin{equation}
    \label{le} %line element
    \mathrm{~d}s^2 = - f(r) \mathrm{~d}t^2 + f(r)^{-1} \mathrm{~d}r^2 + r^2 (\mathrm{~d}\theta^2 + \sin^2 \theta \mathrm{~d}\phi^2).
\end{equation}
Computing the function $f(r)$ specific to the AMCNS black hole is necessary to go on. To pursue this, putting: 
\begin{equation}
f=1-\frac{2mr^{2}}{r^{3}+2ml^{2}},  \label{A9}
\end{equation}%
engenders the Hayward metric \cite{Hayward:2005gi} 
of a non-singular black hole that is static, has no charge, and emerges as the solution of a certain modified gravity theory. The quantity $l$ is an approximate parameter in the length-scale under which the corollaries of the cosmological constant prevail. \newline
If $M$ is assumed to vary with $r$, then:
\begin{equation}
M(r)=\int_{0}^{r}\rho (r)r^{2} \mathrm{~d}r=m-\int_{r}^{\infty }\rho
(r)r^{2}\mathrm{~d}r, \label{A10}
\end{equation} 
where, $m=\int_{0}^{\infty }\rho (r)r^{2}\mathrm{~d}r$ describes the magnetic mass of the black hole that is responsible for screening the magnetic interactions so that the perturbative divergences of magnetostatics can be eliminated. \newline
The energy density, in the absence of an electric field, ensues from Eq.~\eqref{A5} to be:
\begin{equation}
\rho =\frac{q^{2}}{2r^{4}}\exp \left( \frac{-\beta q^{2}}{2r^{4}}\right).
\label{A11}
\end{equation}

This transforms the mass function to:
\begin{equation}
M(r)=\frac{q^{2}}{2}\int_{0}^{r}\exp \left( \frac{-\beta q^{2}}{%
2r^{4}}\right) \frac{\mathrm{d}r}{r^{2}}=\frac{{q^{3/2}}\, \Gamma \left(\frac{1}{4},\frac{\beta q^{2}}{2r^{4}}\right) }{2^{11/4}\, \beta ^{1/4}},  \label{A12}
\end{equation}
where, the incomplete gamma function is characterized by:
\begin{equation}
\Gamma (s,x)=\int_{x}^{\infty }t^{s-1}e^{-t}\mathrm{~d}t. \label{A13}
\end{equation}
Therefore, rewriting the magnetic mass as:
\begin{equation}
m=M(\infty )=\frac{q^{\frac{3}{2}}\Gamma \left( \frac{1}{4}\right) }{2^{\frac{11}{4}}\beta ^{\frac{1}{4}}}\simeq \frac{0.54q^{\frac{3}{2}}}{\beta ^{\frac{1}{4}}}. \label{A15}
\end{equation}

Consolidating all of the above, the metric function is discovered to be \cite{Ali:2019bcn}:
\begin{equation}
f\left( r\right) =1-\frac{r^{2}q^{\frac{3}{2}}\Gamma \left( \frac{1}{4},\frac{\beta q^{2}}{2r^{4}}\right) 2^{\frac{-7}{4}}\beta ^{\frac{-1}{4}}}{r^{3}+l^{2}q^{\frac{3}{2}}2^{\frac{-7}{4}}\beta ^{\frac{-1}{4}}\Gamma \left(\frac{1}{4},\frac{\beta q^{2}}{2r^{4}}\right) }. \label{H1}
\end{equation}
The metric function in the circumstance of $l=0$ reduces to Reissner-Nordström solution  \cite{Kruglov:2017fck}. In the vicinity of radial infinity, its asymptotic value is attained using the expansion of the series:
\begin{equation}
\Gamma (s,z)=\Gamma (s)-z^{s}\left[ \frac{1}{s}-\frac{z}{s+1}+\frac{z^{2}}{2\left( s+2\right) }+O\left( z^{3}\right) \right] ,\text{ \ }z\rightarrow 0.
\label{A16}
\end{equation}

\begin{figure}[htp!]
   \centering
   \includegraphics[scale=1]{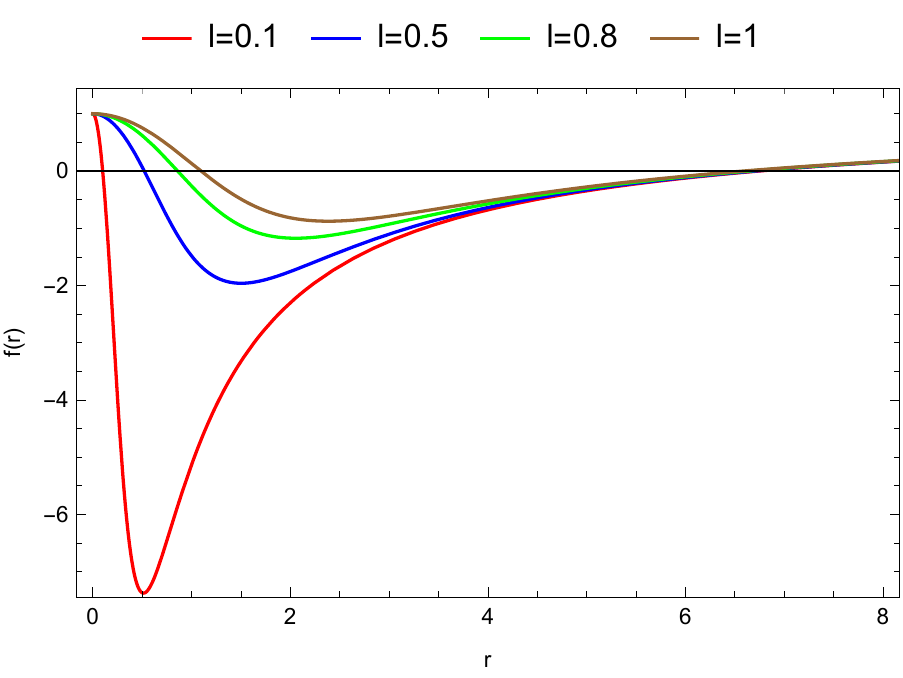}
    \caption{Figure shows the lapse function $f(r)$ as a function of $r$ for  $q=0.5$, $\beta=0.00001$ and for the different values of $l$.}
    \label{fig:hr1}
\end{figure}

Ultimately, the metric function $f(r)$ at $r\rightarrow \infty $ is structured as \cite{Ali:2019bcn}:
\begin{equation}
f\left( r\right) =1-\frac{r^{2}\left[ 2m-\frac{q^{2}}{r}+\frac{\beta q^{4}}{%
20r^{5}}+O(r^{-9})\right] }{r^{3}+l^{2}\left[ 2m-\frac{q^{2}}{r}+\frac{\beta
q^{4}}{20r^{5}}+O(r^{-9})\right] }. \label{A17}
\end{equation}
This will be used for the subsequent computations of the AMCNS black hole. The black hole horizon radius $r_H$, the distance measured from the center of the black hole to its event horizon, can be evaluated by finding the larger root of $f(r)=0$ from this equation which locates the outer horizon. As shown in Fig. \eqref{fig:hr1}, the number of horizons are dependent on the parameters of $l$.

\subsection{Calculation of Hawking Radiation in Jacobi Metric formalism}

To  calculate a black hole's Hawking temperature, we use the Jacobi metric corresponding to the four-dimensional covariant metrics to calculate Hawking temperature by obtaining the tunneling probability of the particle through the horizon \cite{Bera:2019oxg}. In this semi-classical method, we use  the wave function of the particle as $\psi=e^{(i/\hbar)S}$. The corresponding Jacobi metric will be \cite{Gibbons:2015qja,Chanda:2016aph,Das:2016opi}, 
	\begin{equation}\label{10}
	\mathrm{d}s^2 = j_{ij} \mathrm{~d}x^{i}\mathrm{~d}x^{j}= \Big(E^2-m^2f(r)\Big) \left(\frac{\mathrm{d}r^2}{f^2(r)} + \frac{r^2}{f(r)}(\mathrm{d}\theta^2 + \sin^2\theta \mathrm{~d}\phi^2)\right)~.
	\end{equation}

Now, the action for a moving particle in this scenario where Hawking radiation occurs at the near-horizon via radial tunneling is found to be:
 \begin{equation}\label{11}
	S = -\int \sqrt{j_{ij}\frac{\mathrm{d}x^i}{\mathrm{d}s}\frac{\mathrm{d}x^j}{\mathrm{d}s}} \mathrm{d}s~,
	\end{equation}
where the integrand is
	\begin{equation}\label{12}
	\sqrt{j_{ij}\frac{\mathrm{d}x^i}{\mathrm{d}s}\frac{\mathrm{d}x^j}{\mathrm{d}s}} = \pm \left(E^2-m^2f(r)\right)^{\frac{1}{2}} \left(f^{-1}(r)\right) \left(\frac{\mathrm{d}r}{\mathrm{d}s}\right)~,
	\end{equation}

The radial momentum of the particle using the (\ref{12}) in (\ref{11}) is calculated as \cite{Bera:2019oxg}
\begin{equation}\label{13}
p_r=\partial_r S= \mp(E^2-m^2f(r))^{\frac{1}{2}} (f^{-1}(r))~.
\end{equation}

As in our tunneling model, the particle is located near the horizon and so the condition $f(r)<0$ must be satisfied. Therefore, since $p_r>0$ is negative in our equation, it corresponds to the particle going outwards (identical to the conventional tunneling approaches in \cite{Srinivasan:1998ty}). Since tunneling occurs near the horizon, the metric is effectively simplified to $(1+1)$-dimensions near the horizon, where only the radial movement is significant \cite{Robinson:2005pd,Iso:2006wa,Majhi:2010onr} so one can use the Taylor series expansion and apply it to $f(r)$ for $r=r_H$ to expand it around the horizon as: $f(r) = f(r_H) + f'(r_H)(r-r_H) + \mathcal{O}(r-r_H)^2 = 2\tilde\kappa(r-r_H) + \mathcal{O}(r-r_H)^2~,$ where $\tilde\kappa = h^\prime(r_H)/2$ is the surface gravity of the black hole. Afterward, we substitute this in (\ref{12}), and obtain near horizon action for radial motion is  \cite{Bera:2019oxg}	

\begin{equation}\label{17}
    S = \mp \frac{E}{2\tilde\kappa} \int_{r_H-\epsilon}^{r_H+\epsilon} \frac{1}{(r-r_H)} \mathrm{d}r \pm \frac{m^2}{2E}\int_{r_H-\epsilon}^{r_H+\epsilon} \mathrm{d}r \mp \int_{r_H-\epsilon}^{r_H+\epsilon} \mathcal{O}(r-r_H) \mathrm{d}r~,
\end{equation}
with note that $r_H-\epsilon$ is close to horizon and  $r_H+\epsilon$ is across the horizon, with $\epsilon(>0)$. Using the change of variable  $r-r_H = \epsilon e^{i\theta}$ in Eq. (\ref{17}) and $
	\int_{r_H-\epsilon}^{r_H+\epsilon} \frac{1}{(r-r_H)} \mathrm{d}r = - i\pi~,$ for first integral, $\sim \epsilon \approx 0$ for second integral, we obtain the action as
		\begin{equation}\label{19}
	S = \pm \frac{i\pi E}{2\tilde\kappa} + \textrm{real part}~,
	\end{equation}
where the positive $+$ is for the outgoing trajectory and negative $-$ is for the incoming trajectory of tunneling particles. We then write the WKB wave function for outgoing and incoming particles $
	\psi_{out} =A e^{\frac{i}{\hbar}S_{out}}, ~~~
	\psi_{in} = Ae^{\frac{i}{\hbar}S_{in}}~,$ with $A$ being a  normalization constant and corresponding probabilities of the outgoing and incoming particles, become
		\begin{equation} \label{22}
	P_{out} = |\psi_{out}|^2 = \mid A\mid ^2|e^{\frac{i}{\hbar}S_{out}}|^2 
	%=\mid A\mid ^2||e^{\frac{i}{\hbar}(\frac{i\pi E}{2k} + real part)}|^2 = 
	=\mid A\mid ^2 e^{-\frac{\pi E}{\hbar \tilde\kappa}}~,
	\end{equation}

 and 

 \begin{equation} \label{23}
	P_{in} = |\psi_{in}|^2 = \mid A\mid ^2|e^{\frac{i}{\hbar}S_{in}}|^2 		
	%=\mid A\mid ^2|e^{\frac{i}{\hbar}(-\frac{i\pi E}{2k} + real part)}|^2 
	= \mid A\mid ^2 e^{\frac{\pi E}{\hbar \tilde\kappa}}.
	\end{equation}
Note that real part has no contribution to probability. Using the outgoing and ingoing particles probabilities, one can calculate the tunneling rate as
		
		\begin{equation}\label{24}
	\Gamma = \frac{P_{out}}{P_{in}} = e^{-\frac{2\pi E}{\hbar \tilde\kappa}} \equiv  e^{-\frac{E}{T_{H}}},
	\end{equation}
which is similar to the Boltzmann factor with $T_H$ recognized as the Hawking temperature,
\begin{equation}\label{25}
	 T = \frac{\tilde\kappa}{2\pi}.
\end{equation}

\begin{figure}[ht!]
   \centering
    \includegraphics[scale=0.45]{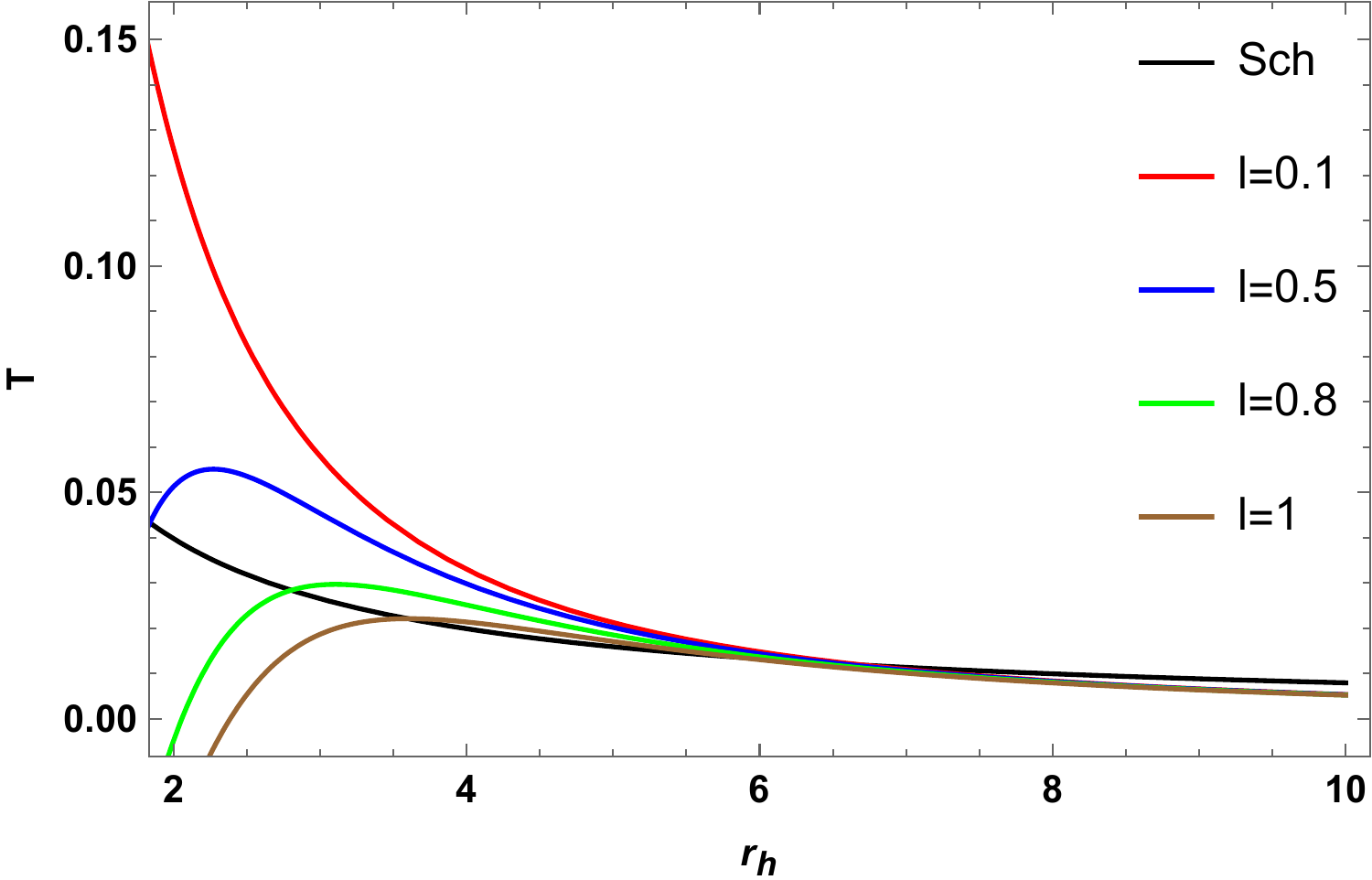}
    \includegraphics[scale=0.45]{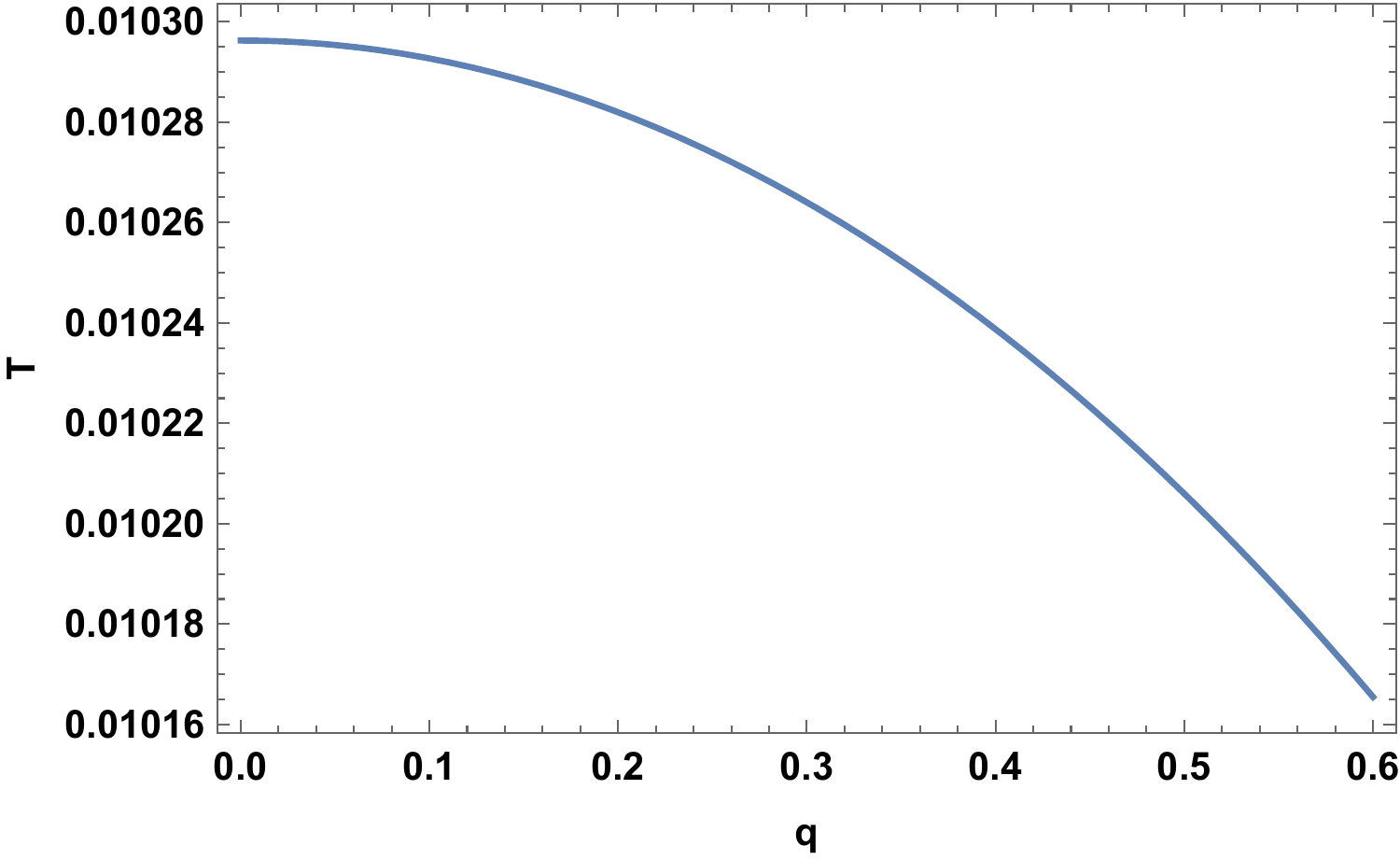}
        \caption{ Left: Hawking temperature $T$ versus $r$ for $q=0.5$, $\beta=0.00001$ and for the various values of $\l$.  Right: Hawking temperature $T$ versus $q$ for $q=0.5$, and $\beta=0.00001$.}
    \label{fig:temp}
\end{figure}

\begin{figure}[ht!]
   \centering
    \includegraphics[scale=0.45]{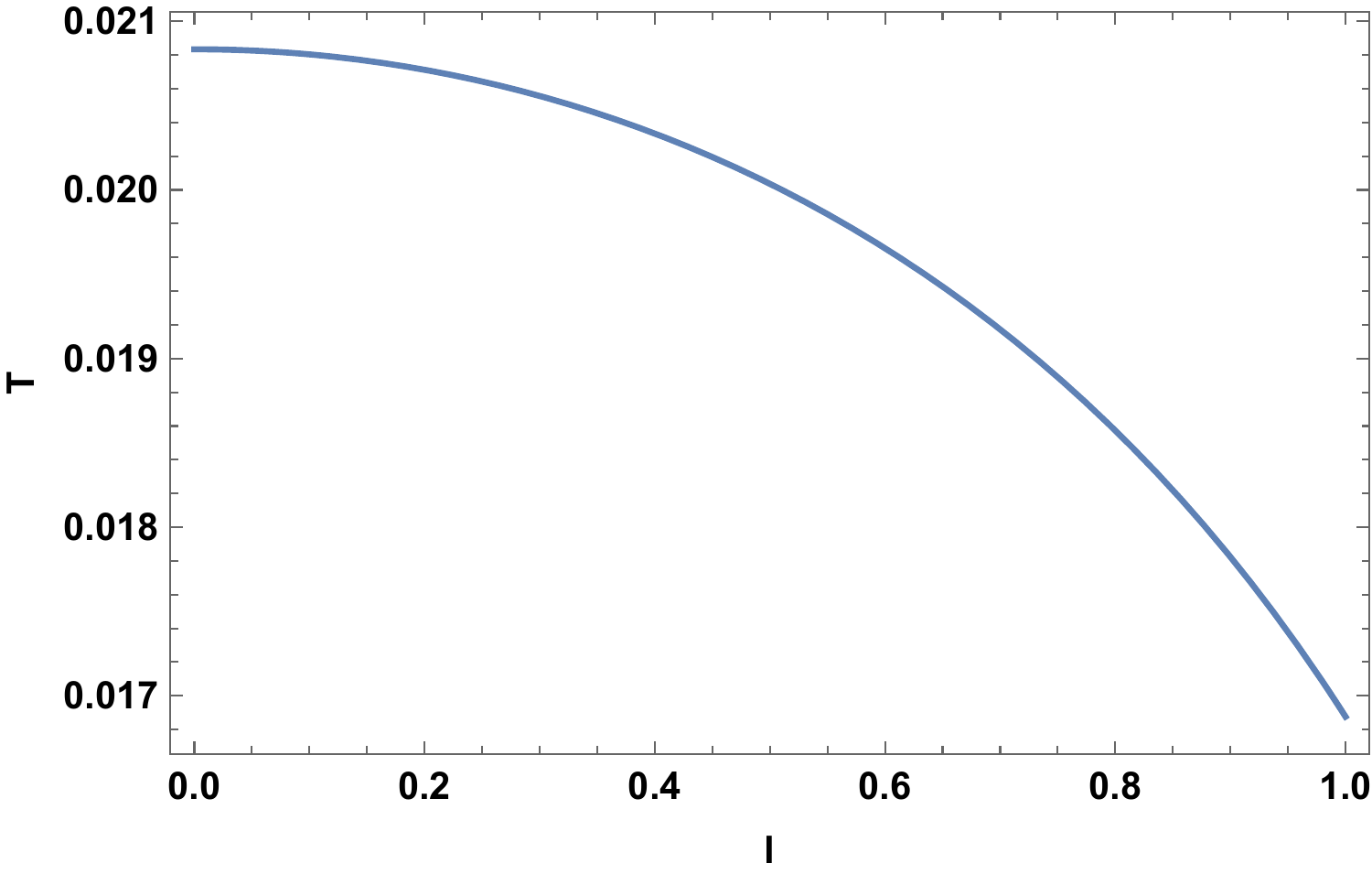}
        \includegraphics[scale=0.45]{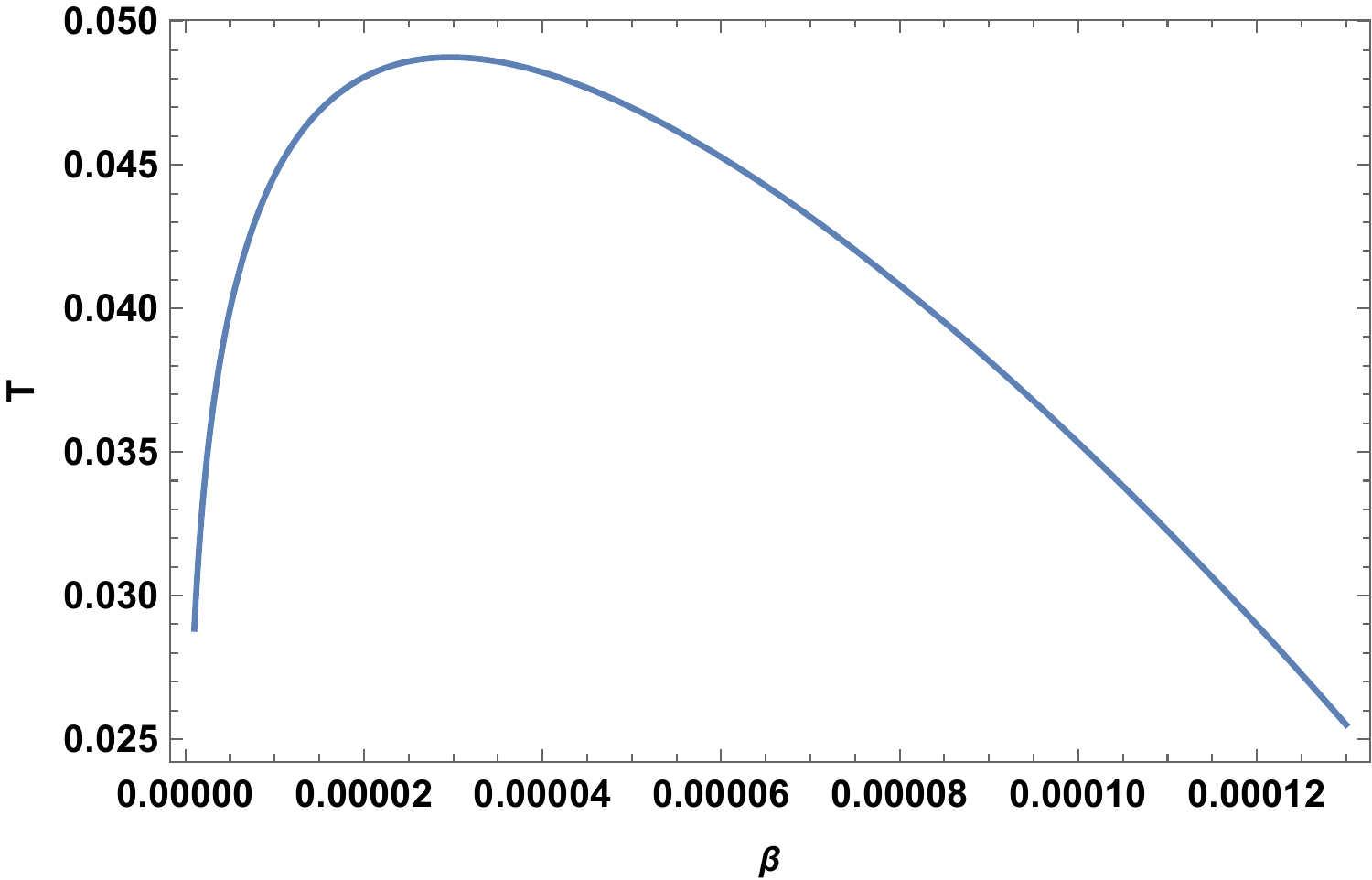}
    \caption{Left: Hawking temperature $T$ versus $l$ for $q=0.5$, and $\beta=0.0001$. Right: Hawking temperature $T$ versus $\beta$ for $q=0.5$, and $l=0.5$.}
    \label{fig:temp2}
\end{figure}

With the surface gravity expressed as:
\begin{equation} 
    \tilde\kappa=\left.\sqrt{-\frac{1}{2} \nabla_{\mu} \chi_{\nu} \nabla^{\mu} \chi^{\nu}} \equiv \frac{1}{2} \frac{\partial f(r)}{\partial r}\right|_{r=r_{H}},
\end{equation}
the Hawking temperature related to $r_{H}$ is determined to be:
\begin{eqnarray}
T=\frac{-\frac{r_H^2 \left(\frac{q^2}{r_H^2}-\frac{\beta  q^4}{4 r_H^6}\right)}{l^2 \left(\frac{\beta  q^4}{20 r_H^5}-\frac{q^2}{r_H}+\frac{1.08 q^{3/2}}{\sqrt[4]{\beta }}\right)+r_HH^3}+\frac{r_H^2 \left(\frac{\beta  q^4}{20 r_H^5}-\frac{q^2}{r_H}+\frac{1.08 q^{3/2}}{\sqrt[4]{\beta }}\right) \left(l^2 \left(\frac{q^2}{r_H^2}-\frac{\beta  q^4}{4 r_H^6}\right)+3 r_H^2\right)}{\left(l^2 \left(\frac{\beta  q^4}{20 r_H^5}-\frac{q^2}{r_H}+\frac{1.08 q^{3/2}}{\sqrt[4]{\beta }}\right)+r_H^3\right){}^2}-\frac{2 r_H \left(\frac{\beta  q^4}{20 r_H^5}-\frac{q^2}{r_H}+\frac{1.08 q^{3/2}}{\sqrt[4]{\beta }}\right)}{l^2 \left(\frac{\beta  q^4}{20 r_H^5}-\frac{q^2}{r_H}+\frac{1.08 q^{3/2}}{\sqrt[4]{\beta }}\right)+r_H^3}}{4 \pi },
\end{eqnarray}
and is plotted in Figs. \eqref{fig:temp} and \ref{fig:temp2} for different values of $l$, $q$ and $\beta$.  Moreover, the horizon area is plainly calculated as $A=\int_{0}^{2 \pi} \mathrm{d} \phi \int_{0}^{\pi} \sqrt{-g} \mathrm{~d} \theta=4 \pi r_{H}^2,$
which gives the black hole entropy to be $S_{B H}=\frac{A}{4}=\pi r_{H}^2$.

\section{WEAK DEFLECTION ANGLE OF AMCNS BLACK HOLE USING THE GAUSS-BONNET THEOREM}
\label{Swdagbt}
 The deflection angle of a photon approaching the said black hole with the distance of closest approach $r_0$ corresponds to \cite{Virbhadra:1998dy,Weinberg:1972kfs,Lu:2019ush}:
\begin{equation}
    \hat{\alpha}(r_0) = -\pi + 2 \int_{r_0}^\infty \mathrm{~d}r \frac{\sqrt{f(r)^{-1}}}{r \sqrt{\frac{r^2}{r_0^2}\frac{f(r_0)}{f(r)}-1}}.
    %%%= -\pi + 2 \int_{r_0}^\infty dr \textbf{I}\,(r)
\end{equation}
The condition $M/r_0 \ll 1$ is satisfied for weak lensing since the deflection angle will be too small \cite{keeton}. As $r_0$ verges on the photon sphere, $\hat{\alpha}$ increases until it diverges, hence, giving rise to strong lensing. Considering the equatorial plane in which $\theta=\pi/2$ equation Eq.~\eqref{le} reduces to the optical metric for null geodesics:
\begin{equation}
    \label{om} %optical metric
    \mathrm{~d}t^2 = \frac{\mathrm{~d}r^2}{f(r)^2} + \frac{r^2}{f(r)} \mathrm{~d}\phi^2.
\end{equation}

\begin{figure}
\centering
\includegraphics[scale=0.5]{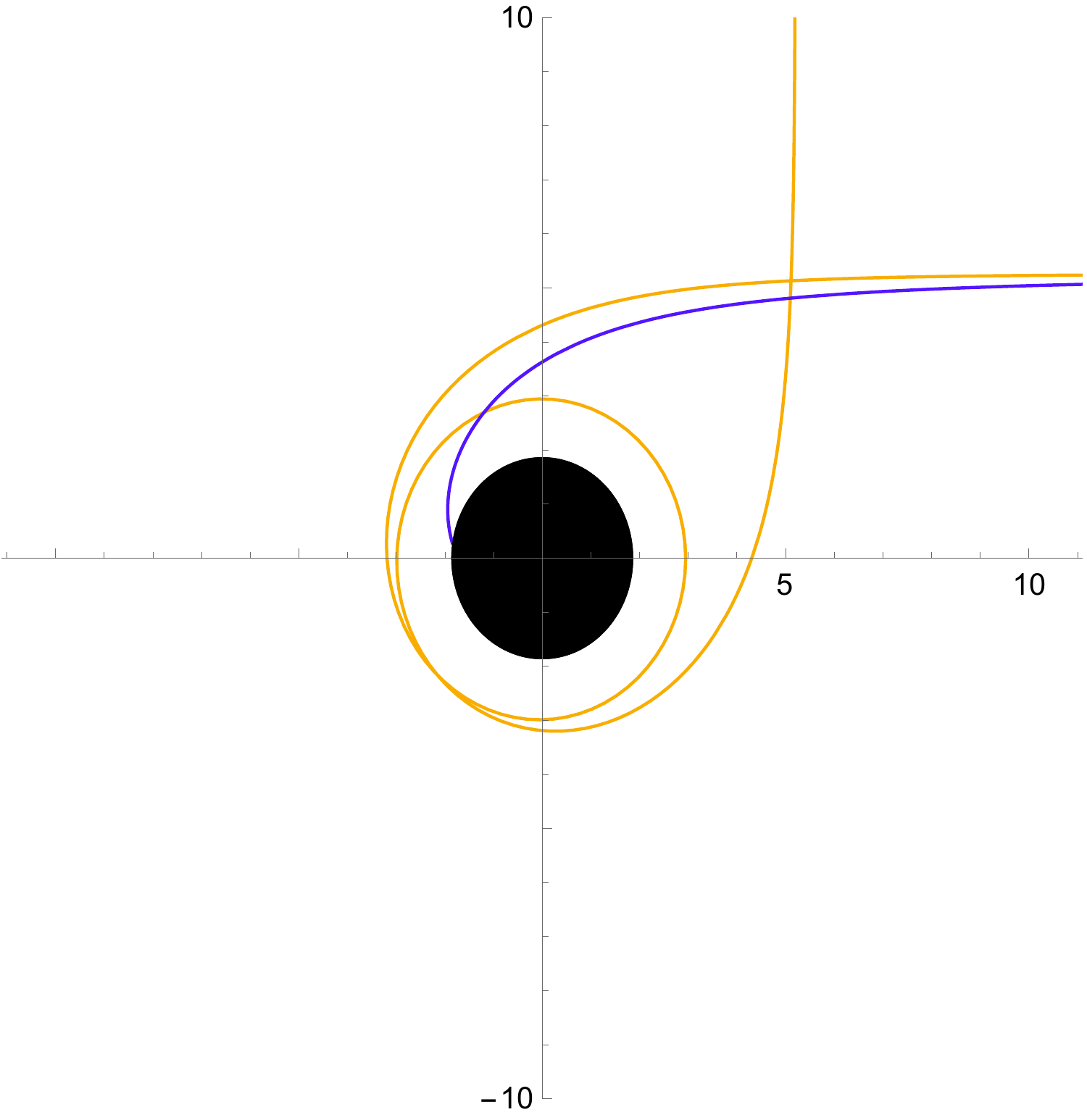}
\includegraphics[scale=0.5]{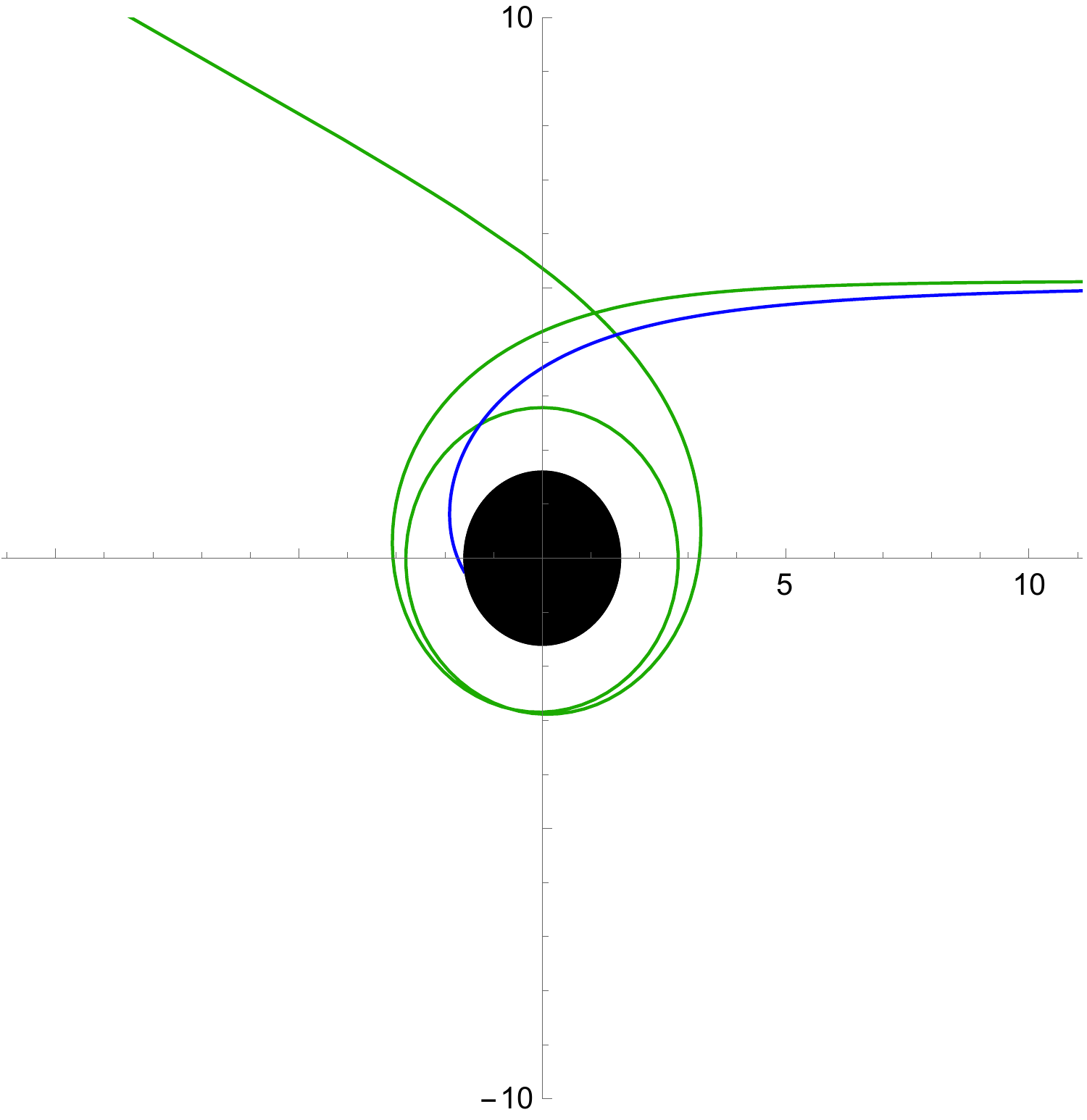}
\caption{\label{geod} Geodesics of AMCNS black hole for $q=0.5$, $\beta=0.001$, and varying $l$; $l=0.5$ (left), and $l=0.7$ (right).  }
\end{figure}

\begin{figure}
\centering
\includegraphics[scale=0.5]{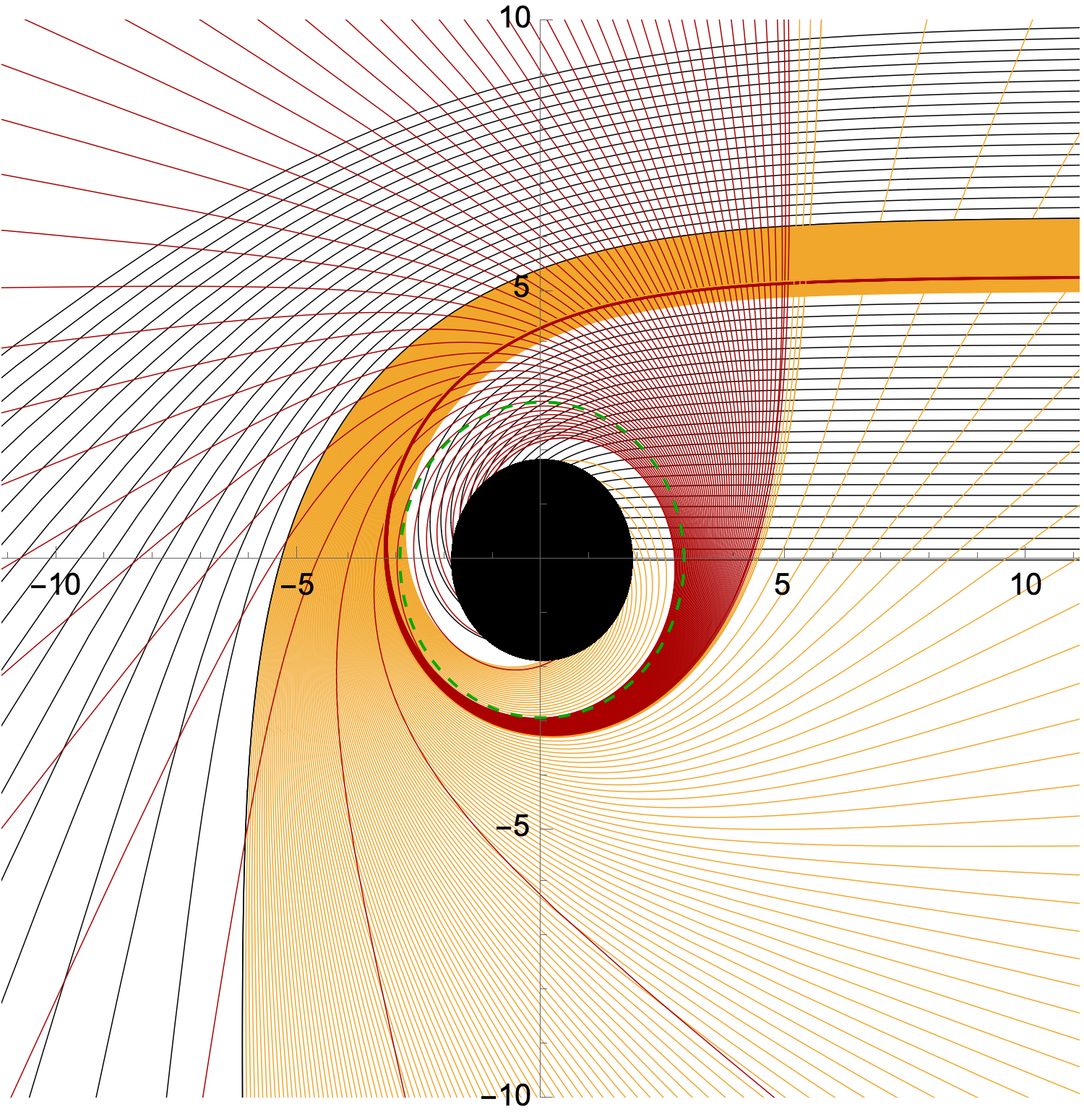}
\includegraphics[scale=0.5]{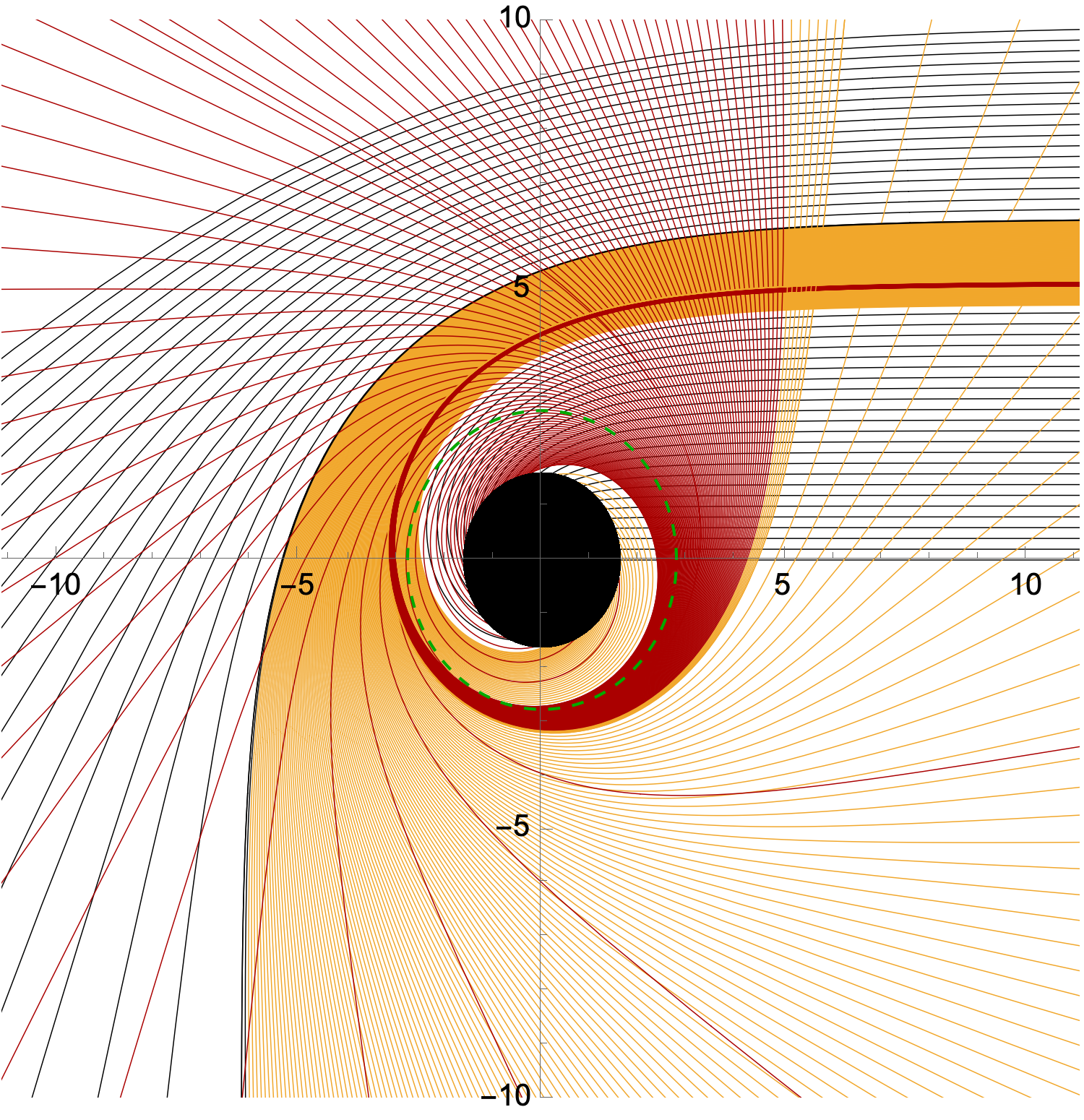}
\caption{\label{rt} Raytracing of AMCNS black hole for $q=0.5$, $\beta=0.001$, and varying $l$; $l=0.5$ (left), and $l=0.7$ (right). The lines colored in black, gold, and red represent the direct, lensed, and photon rings correspondingly. On the panel located to the right, a chosen assortment of related trajectories in Euclidean polar coordinates, denoted as ($r$, $\phi$), is depicted. The black hole is symbolized by a black disk, and the circular orbit of light is a dashed yellow circular.}
\end{figure}

As mentioned above, the optical metric takes advantage of its geodesics for the topological effect that relates it to the Gauss-Bonnet theorem. The photon behaviors within the AMCHS black hole for a range of l values are illustrated in Figure \ref{geod}. Subsequently, Figure \ref{rt} presents the paths followed by light rays as they orbit around the AMCHS black hole using the method defined in \cite{Gralla:2019xty}.

In that context, we delve deeper into the geodesics of the optical metric that governs the AMCNS black hole in Fig.~\ref{geod}. Notice the slightest difference in the light ray trajectory approaching from infinity towards the black hole that determines the fate of the light ray, an expected repercussion of the electric charge to starkly either pull the ray in or divert it around. Besides, an increase in $l$ is observed to retract the light rays more swiftly towards the black hole center. Extrapolating this further, Fig.~\ref{rt} encapsulates the raytracing of the AMCNS metric. The abrupt coiling is more obvious in the myriad of light rays in this figure.

The Gaussian curvature is computed from the non-zero Christoffel symbols due to its proportionality$ \mathcal{K} = {R}/{2}$ where $R$ is the Ricci scalar and is found to be:
\begin{equation}
    \begin{split}
        \label{gc}
   \mathcal{K} =\frac{q^4 \left(-21 \beta +40 r^2+1050\right)}{20 r^8}+\frac{3 q^2}{r^4} +m \left(\frac{q^4 \left(225 \beta +19 \beta  r^2-1830 r^2-22950\right)}{10 r^{11}}-\frac{6 q^2 \left(r^2+25\right)}{r^7}-\frac{2}{r^3}\right) \\+ m^2 \left(-\frac{3 q^4 \left(1326 \beta -130 r^4+124 \beta  r^2-18410 r^2-202800\right)}{10 r^{14}}+\frac{6 q^2 \left(50 r^2+612\right)}{r^{10}}+\frac{3 r^2+100}{r^6}\right) \\ +l^2 \left(m^2 \left(\frac{30 q^2}{r^8}-\frac{3 q^4 \left(23 \beta +65 r^2-3640\right)}{10 r^{12}}\right)-\frac{21 m q^4}{r^9}\right).
    \end{split}
\end{equation}

\begin{figure}
\centering
\includegraphics[scale=0.8]{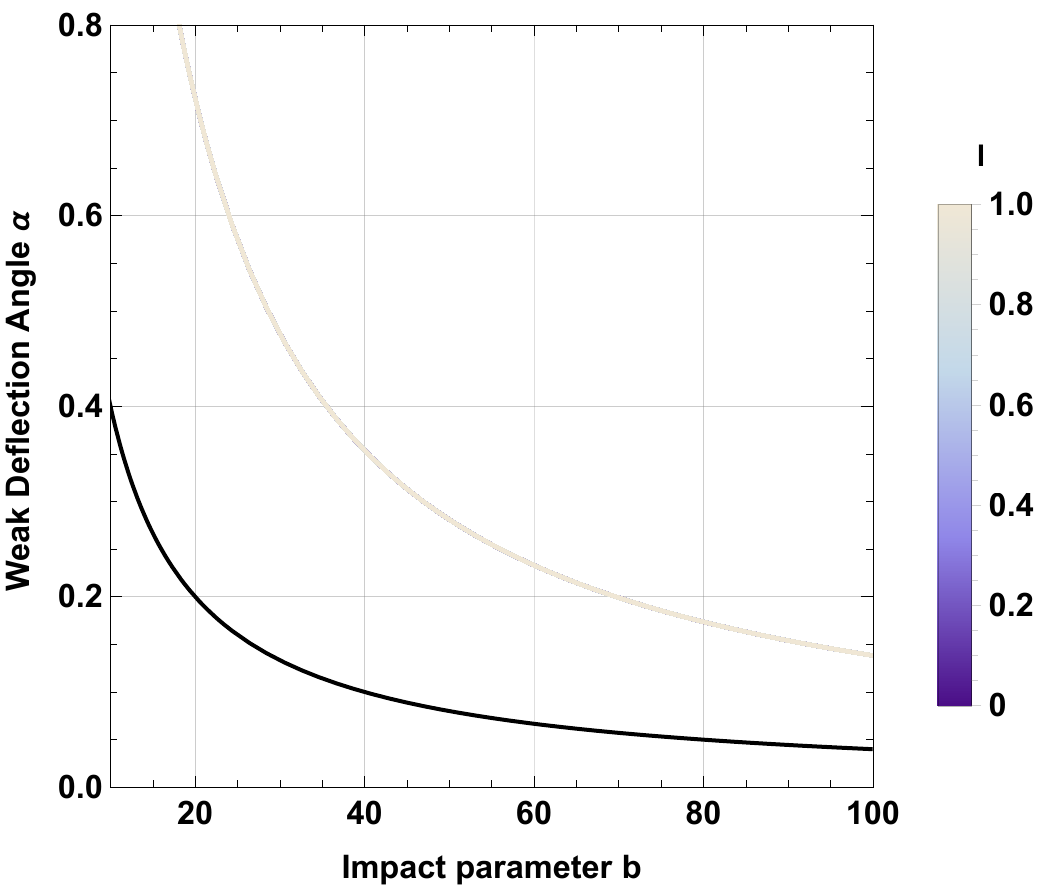}
\includegraphics[scale=0.8]{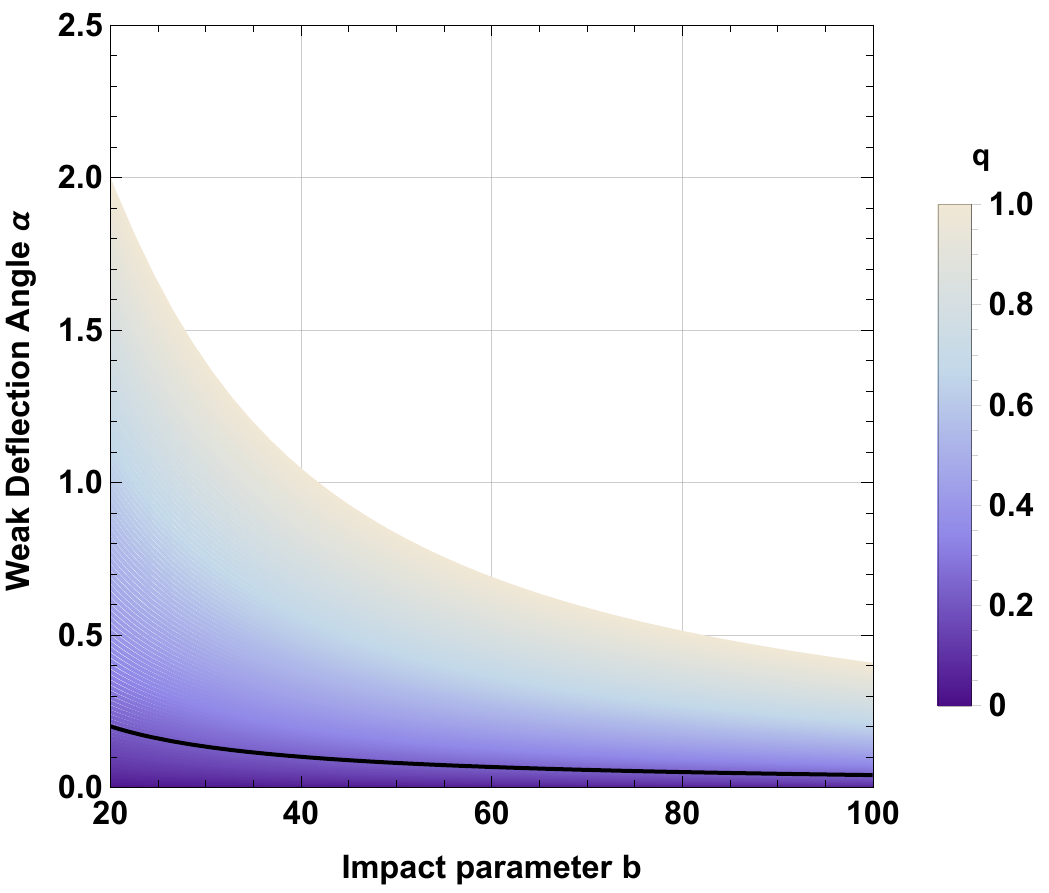}
\caption{\label{wdabv} Weak deflection angle plotted against the impact parameter. The solid black line denotes the Schwarzschild case and the gradient of grey shows the case of the  asymptotic, magnetically-charged, non-singular black hole, Left:  with $\beta=0.00001$, $q=0.5$, and $0 < l< 1$. Right: Figure is plotted for  $\beta=0.00001$, $l=0.5$, and $0 < q< 1$.  }
\end{figure}

 Employing the straight-line approximation $r = b/ \sin \phi$ for the impact parameter $b$, the Gauss-Bonnet theorem suggests that \cite{Gibbons:2008rj,Kumaran:2021rgj}:
\begin{equation}
     \label{allim}
     \hat{\alpha} = - \int_0^\pi \int_{{b}/{\sin \phi}}^\infty \mathcal{K} \mathrm{~d}S,
\end{equation}
where $\mathrm{~d}S\approx (r+3m) \mathrm{~d}r \mathrm{~d}\phi$.
Owing to the complexity of this calculation, the Gaussian curvature was reduced to $\mathcal{O}\left(m^2\right)$ and the integral was simplified by ignoring the higher-order terms $\mathcal{O}\left(q^5\right)$. Thus, the deflection angle due to weak lensing for an asymptotic, magnetically charged, non-singular black hole is estimated to be:
\begin{equation}
\begin{split}
    \hat{\alpha} = -\frac{1.43139 l^2 q^5}{b^6 \sqrt{\beta }}+\frac{7.15694 q^5}{b^6 \sqrt{\beta }}+\frac{7 \pi  \beta  q^4}{128 b^6}-\frac{175 \pi  q^4}{64 b^6}-\frac{10.0777 q^{9/2}}{b^5 \beta ^{3/4}}+\frac{17.28 q^{7/2}}{b^5 \sqrt[4]{\beta }}-\frac{0.6912 q^{11/2}}{b^5 \sqrt[4]{\beta }}+\frac{1.5459 q^5}{b^4 \sqrt{\beta }}\\
    -\frac{3 \pi  q^4}{16 b^4}-\frac{8.58833 q^3}{b^4 \sqrt{\beta }}-\frac{0.629856 q^{9/2}}{b^3 \beta ^{3/4}}-\frac{0.72 q^{7/2}}{b^3 \sqrt[4]{\beta }}+\frac{0.687066 q^3}{b^2 \sqrt{\beta }}-\frac{3 \pi  q^2}{4 b^2}+\frac{2.16 q^{3/2}}{b \sqrt[4]{\beta }}, \label{wdav}
\end{split}
\end{equation}
in the weak-field limit. Note that the mass function defined by Eq.~\eqref{A15} is employed here in the place of $m$. From this result, the deflection angle is expected to change significantly because of the parameters that govern the black hole, reducing it to the case of a Schwarzschild black hole when $\beta=0$, $q=0$, and $l=0$. This relation is depicted in Fig.~\ref{wdabv}. 

Taking a look at this equation graphically, the deflection angle is seen to be vividly affected by $l$ and more intensely for $q$. The ranges of values for $l$ and $q$ are chosen according to the acquirable potentialities of a charged particle in the black hole locale. The deviating curve shows that the deflection angle is more than the Schwarzschild case when there is an electric charge, and it keeps increasing as $q$ increases. The deceiving large contribution from $l$ is seemingly increasing the deflection angle.

Therefore, the presence of the charge $q$ appears to be enhancing the deflection angle. This suggests that the distortions will be stronger, more evident, and could contain more information about the characteristics of the black hole's structure and of the light source.

\subsection{Weak Deflection Angle in dark matter medium}
\label{Swdapm}
The presence of a medium significantly affects the dispersive properties of light rays compared to those in a vacuum. One such medium that is known to exist almost always is the one surrounded by dark matter. The observations from the Event Horizon Telescope of enormous dark matter halos around the black holes in the galactic centers. %*ref*K. Akiyama and et al. (Event Horizon Telescope Collaboration), Astrophys. J. 875, L1 (2019), arXiv:1906.11238 [astro-ph.GA]
Its ability to engulf galaxies and to invade the interstellar and intergalactic media makes it a matter of inquisitivity. When surrounding a black hole, the extreme gravity experienced by the dark matter added to its already enduring gravitational impacts constitutes an important component of astrophysics that would greatly aid the gravitational lensing analyses.

To evaluate the consequences of dark matter on the deflection angle, the refractive index computed from the scatterers is defined as outlined in references \cite{Latimer:2013rja, Ovgun:2018oxk, Ovgun:2020yuv}:
\begin{equation}
n = 1 + B u + v w^2.
\label{dm}
\end{equation}

In the given equation, $w$ represents the frequency of light. The term $B \equiv { \rho_0 }/{ 4 m ^ { 2 }w ^ { 2 } } $ introduces a quantity denoted as $B$, where $\rho_0$ denotes the mass density of the scattered particles of dark matter. The parameter $u= - 2 \varepsilon ^ { 2 } e ^ { 2 }$ is defined, where $\varepsilon$ represents the charge of the scatterer in units of $e$, and $v$ is a non-negative value. The terms of order $\mathcal{O} \left(w^2\right)$ and higher account for the polarizability of the dark matter particle. This expression presents the refractive index expected for an optically inactive medium. The term of order $w^{-2}$ corresponds to a charged dark matter candidate, while $w^{2}$ corresponds to a neutral dark matter candidate. Additionally, a linear term in $w$ may be present if there exist parity and charge-parity asymmetries.

The line element in equation (\ref{le}) can be reformulated using the optical metric as follows \cite{Crisnejo:2018uyn,Ovgun:2019wej,Javed:2019ynm}: 
\begin{widetext} 
\begin{equation} 
\mathrm{~d} \sigma^2 = g _ { i j }^{ \mathrm { opt } } \mathrm{~d} x^i \mathrm{~d} x^j = \frac {n^2} {f (r)} \left[ \frac{\mathrm{~d} r^2}{f(r)} + r^2 \mathrm{~d} \phi ^ { 2 }\right]. 
\end{equation} 
\end{widetext}

As in the previous section, the weak deflection angle of an asymptotic, magnetically charged, non-singular black hole enveloped by dark matter can be expressed as:
\begin{equation}
    \begin{split}
         \hat{\alpha} = -\frac{1.43139 l^2 q^5}{b^6 \sqrt{\beta } \left(B u+v w^2+1\right)^2}-\frac{14.3139 q^5}{b^6 \sqrt{\beta } \left(B u+v w^2+1\right)^2}+\frac{7 \pi  \beta  q^4}{128 b^6 \left(B u+v w^2+1\right)^2}-\frac{175 \pi  q^4}{64 b^6 \left(B u+v w^2+1\right)^2}\\+\frac{17.28 q^{7/2}}{b^5 \sqrt[4]{\beta } \left(B u+v w^2+1\right)^2}-\frac{3 \pi  q^4}{16 b^4 \left(B u+v w^2+1\right)^2}-\frac{8.58833 q^3}{b^4 \sqrt{\beta } \left(B u+v w^2+1\right)^2}\\+\frac{1.44 q^{7/2}}{b^3 \sqrt[4]{\beta } \left(B u+v w^2+1\right)^2}-\frac{0.687066 q^3}{b^2 \sqrt{\beta } \left(B u+v w^2+1\right)^2}-\frac{3 \pi  q^2}{4 b^2 \left(B u+v w^2+1\right)^2}+\frac{2.16 q^{3/2}}{b \sqrt[4]{\beta } \left(B u+v w^2+1\right)^2}.
    \end{split}
\end{equation}

\begin{figure}
\centering
\includegraphics[width=10cm]{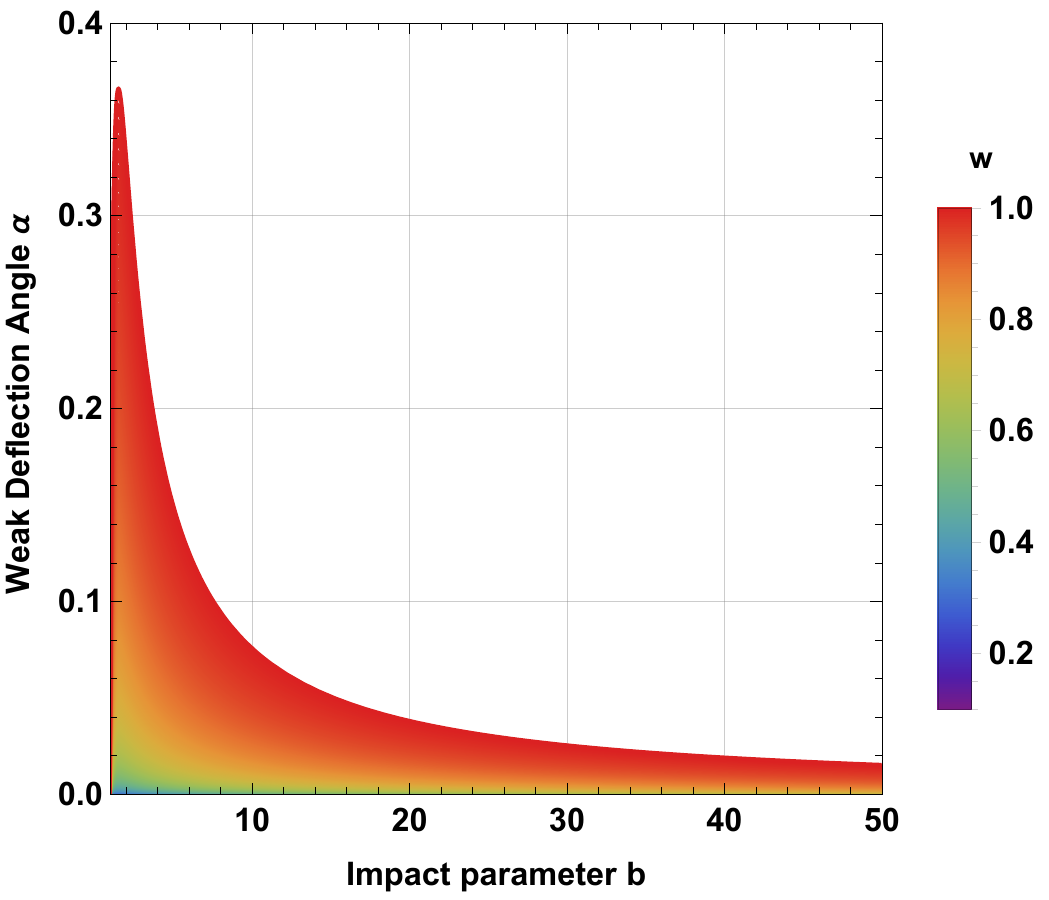}
\caption{\label{wdabp} Weak deflection angle plotted against the impact parameter in the presence of dark matter medium. The rainbow gradient represents the AMCNS black hole when the light frequency $w=1$ at the red end to the black hole in vaccum at the blue end, with  $\beta=0.00001$, $l=0.5$, and $q=0.5$. }
\end{figure}

In Figure \ref{wdabp}, it is demonstrated that the deflection angle exhibits an upward trend as the parameter $w$ increases within the weak field limits, implying more dark matter activity engenders more distortions in the lensing profile. While it is not possible to predict how dark matter behaves around a black hole as opposed to the well-understood outer-galactic distribution, its fundamentality of augmenting the distortions, and by extension, the deflection angle, is unchanging. %*ref*https://arxiv.org/pdf/2303.13554.pdf

\section{ACCRETION DISK AND SHADOW CAST OF AMCNS BLACK HOLE}
\label{Sshadow}
In this section, the shadow that is formed by a black hole that has a thin-accretion disk is examined. An accretion disk refers to the structure that arises from the gravitational pull of a central gravitating object on the adjacent matter like gas or dust \cite{Bambi:2012tg,Bambi:2013nla,Jaroszynski:1997bw,Gralla:2019xty}. As this matter moves toward the central object, it gains kinetic energy and generates a rapidly rotating disk around it. The temperature and density of the disk can cause it to emit radiation in different forms, such as X-rays, visible light, or radio waves. Accretion disks serve as an essential component in the development and progression of gravitating objects since they facilitate the transfer of mass and angular momentum.

When the accretion disk concurs with the lensing effect, produces the appearance of a shadow of a black hole. In essence, the phenomenon of gravitational lensing is the key factor that distinguishes the black hole shadow in the Newtonian case in contrast to that of the general-relativistic case, thus, exaggerating the shadow with the bending of light. The critical curve segregating the capture orbits and the scattering orbits forms the shadow with a geometrically thick, optically thin region filled with emitters (that either spiral into the black hole or swerve away from it) and is associated with a distant, homogeneous, isotropic emission ring. The former property is important while talking about an emission region since the intensity takes a dip that is coincident with the shadow, making it observable through this strong visual signature. The size of the shadow is mainly determined by the inherent parameters of the black hole, and its contour is shape-shifting owing to the instability in the orbits of the light rays from the photon sphere. For a distant observer, the shadow appears as a dark, two-dimensional disk, which is illuminated by its bright and uniform surroundings.

For the line element given by Eq.~\eqref{le}, introducing the function $h(r)$ such that:
\begin{equation}
h^2(r) = \frac{r^2}{A(r)},\end{equation}
which is equivalent to the effective potential for photon motion \cite{Perlick:2021aok}. Not that according to line element in \ref{le}, $A(r)=\frac{1}{B(r)}=f(r)$.
Eq.~\eqref{rphi} can be rearranged and rewritten as:
\begin{equation}
\left(\frac{\mathrm{d} r}{\mathrm{d} \phi}\right)^2=\frac{r^4}{B(r)}\left(\frac{1}{b^{2} A(r)}-\frac{1}{r^{2}}\right)= \frac{r^2}{B(r)}\left(\frac{r^2}{b^{2} A(r)}-1\right)
\label{rphi},
\end{equation}
where, the impact parameter $b \equiv L / E$ yet again. This equation is analogous to the traditional energy-conservation law described for one-dimensional motion in classical mechanics:
\begin{equation} \left(\frac{\mathrm{d}r}{\mathrm{d}\phi}\right)^2 + V(r) =0, \end{equation}
with $\phi$ taking the place of the temporal variable and the effective potential $V(r)$ displaying its dependence on $r$, and by extension, $b$. This is also known as the orbit equation. 

To determine the circular trajectories, the equations $V = 0$ and $\nicefrac{\mathrm{d}V}{\mathrm{d}r} = 0$ can be solved. For a light ray approaching the center, if there exists a point at which it turns back around to exit the orbit after the ray reaches the minimum radius $R$, then the condition $\nicefrac{\mathrm{d}r}{\mathrm{d}\phi}|_R = 0$ needs to be complied with, yielding the orbit equation to be:
\begin{equation}\frac{1}{b^2} = \frac{A(R)}{R^2}.\end{equation}

This relation between the constant of motion and $R$ instigates the following relationship:
\begin{equation}
    b = h(R),
    \label{beqh}
\end{equation}
which in turn yields:
\begin{equation}
    \left(\frac{\mathrm{d} r}{\mathrm{d} \phi}\right)^2=\frac{r^2}{B(r)}\left(\frac{h^2(r)}{h^2(R)}-1\right),
    \label{ritoh}
\end{equation}

\begin{figure}
\centering
\includegraphics[scale=0.4]{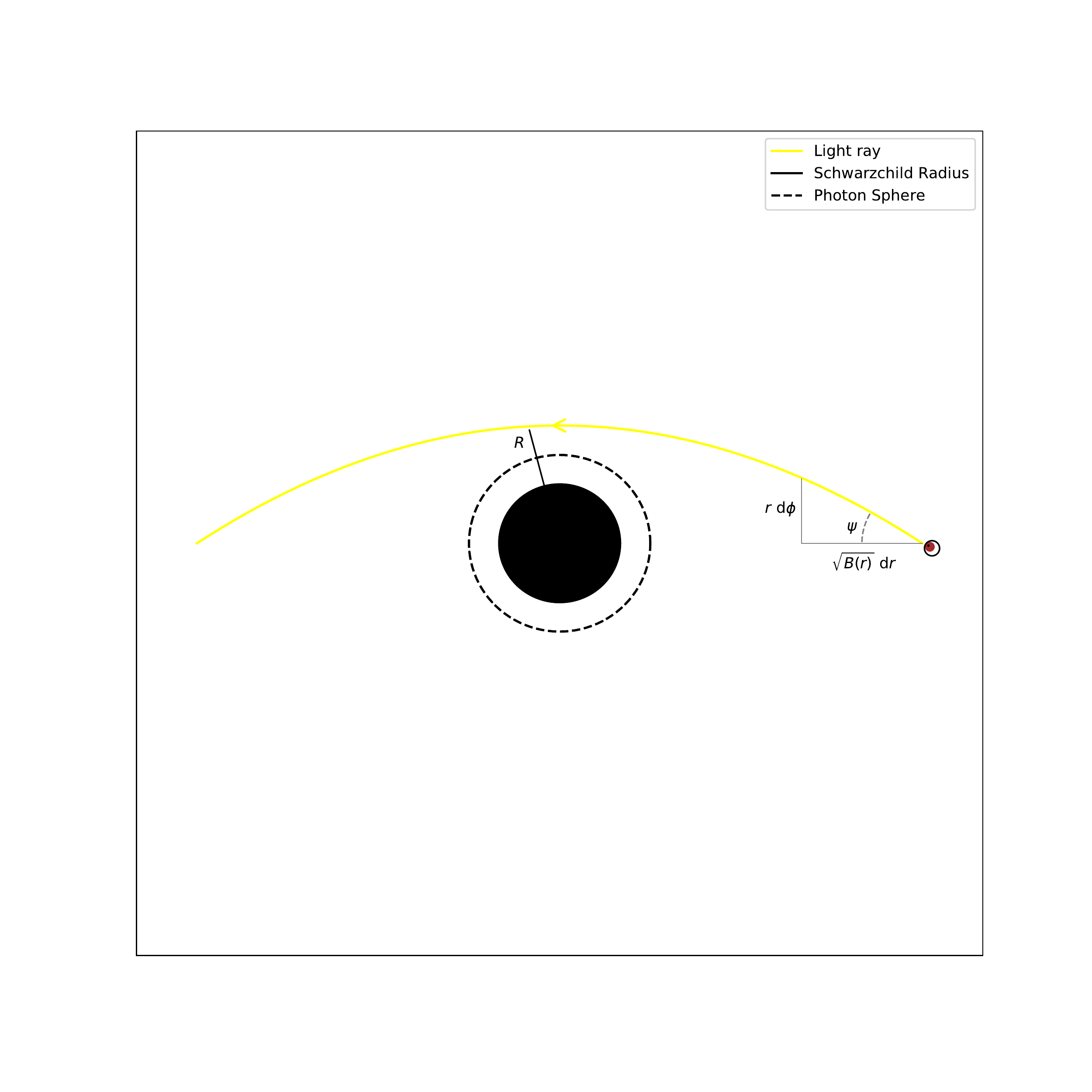}
\caption{\label{sh-diag} Description of the turning point with the minimum radius $R$ and the shadow's angular radius $\psi$ in the Schwarzschild spacetime when a light ray is sent to the past by an observer located at $r_o$ in the present.}
\end{figure}

Say, a static observer located at a radial coordinate $r_o$ shoots one light ray to the past. The dimensionless quantity $\psi$ is identified in terms of the cotangent to be:
\begin{equation} \cot \psi=\left.\sqrt{\frac{g_{r r}}{g_{\phi \phi}}}\, \frac{\mathrm{d} r}{\mathrm{d} \phi}\right|_{r=r_O}=\left.\frac{\sqrt{B(r)}}{r} \,\frac{\mathrm{d} r}{\mathrm{d} \phi}\right|_{r=r_O},\end{equation}
as represented by Fig.~\ref{sh-diag} to be the measurable angle between the light ray and the radial coordinate which is directed towards the point of closest approach. It is termed as the angular radius of the shadow.

Squaring this equation and substituting Eqs.~\eqref{beqh} and \eqref{ritoh} in it gives:
\begin{equation}\cot ^2 \psi=\frac{h^2 r_o)}{h^2(R)}-1 \quad \quad\text{i.e.,}\quad\quad \sin ^2 \psi=\frac{h^2(R)}{h^2\left(r_o\right)}.\end{equation}

Here, the light ray approaching the circular (with radius $r_{ph}$), unstable photon orbits asymptotically is oriented to the past, and so is the shadow's boundary curve. In the limit where where $R \rightarrow r_{ph}$, the above expression can be rewritten as:
\begin{equation}
    \sin ^2 \psi_{sh}=\frac{h^2\left(r_{ph}\right)}{h^2\left(r_o\right)}\quad \quad \text{and} \quad \quad b_{cr}=h\left(r_{ph}\right).
    \label{heqb}
\end{equation}
This implies:
\begin{equation}\sin ^2 \psi_{sh}=\frac{b_{cr}^2}{h^2(r_o)} =\frac{b_{\mathrm{cr}}^2}{r_o^2/ A(r_o)}.\end{equation}

In the pursuit of calculating $r_{ph}$ for the metric in Eq.~\eqref{le} next, it is essential to realize that both $\mathrm{d}r/\mathrm{d}\phi$ and $\mathrm{d}^2 r/ \mathrm{d}\phi ^2$ should simultaneously be zero. Differentiating Eq.~\eqref{rphi}, all terms vanish following the conditions except the third term acquired from the parentheses which are equated to zero resulting in:
\begin{equation}0=\frac{\mathrm{d}}{\mathrm{d} r} h^2(r).\end{equation}
The range of values that $r$ can take is extensive inferring that there could be multiple photon orbits -- stable and unstable, existing together, with the light rays oscillating and spiraling respectively -- impinging on the construction of the black hole shadow. For  $r=r_{ph}$, the shadow can be determined for any distance, small or large, in a static, spherically-symmetric, asymptotically-flat spacetime with $h(r)$ from Eq.~\eqref{heqb} by:
\begin{equation}b_{cr}=\frac{r_{ph}}{\sqrt{A\left(r_{ph}\right)}} \quad\quad \text{and} \quad \quad R_{sh}=\psi_{sh} \approx  \frac{b_{cr}}{r_o}\sqrt{A(r_o)}.\end{equation}

\begin{figure}[htp]
   \centering
          \includegraphics[scale=0.8]{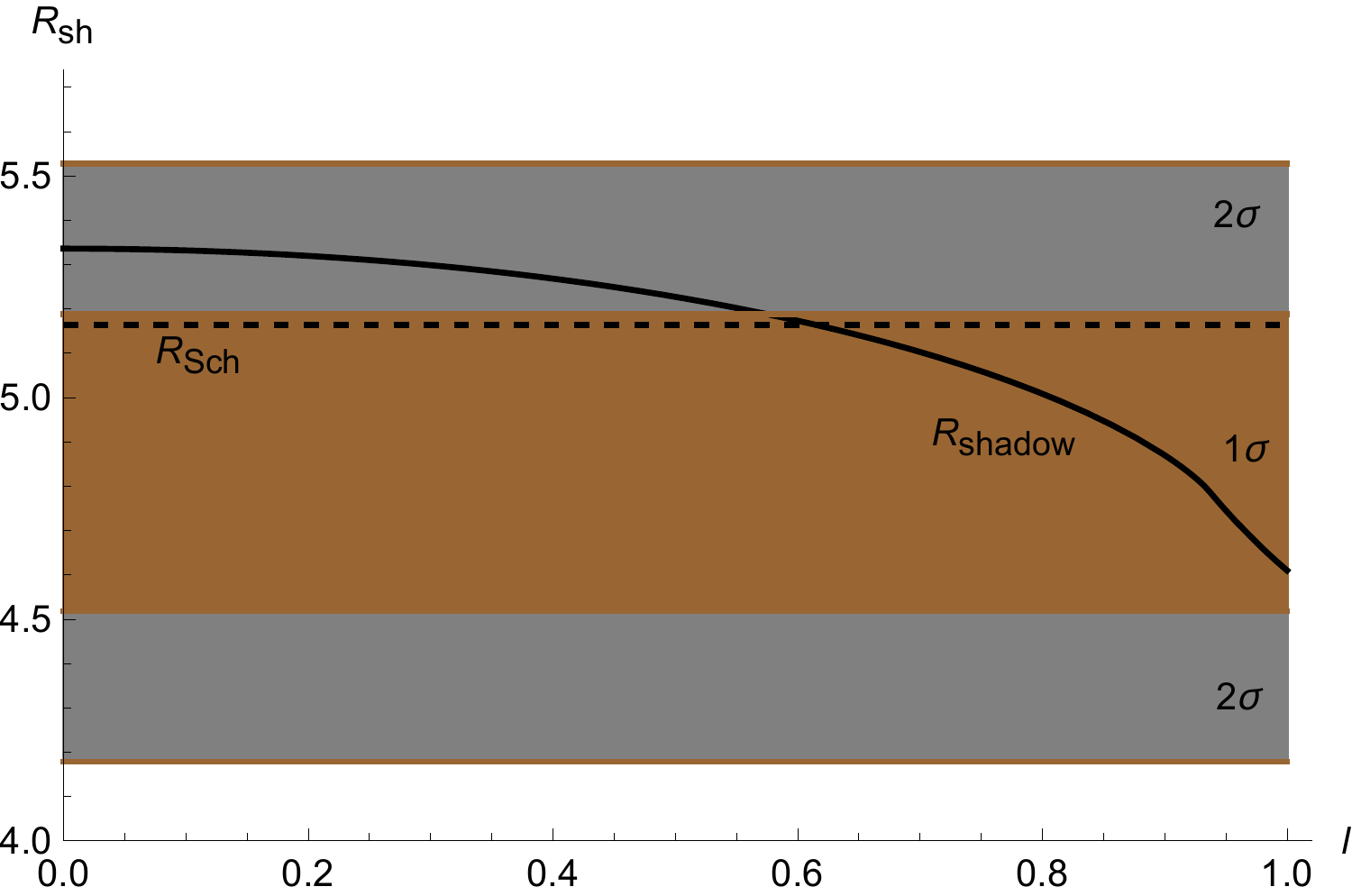}
    \caption{Constraints from the Event Horizon Telescope horizon-scale image of Sagıttarius A* at $1\sigma$\cite{Vagnozzi:2022moj}, after averaging the Keck and VLTI mass-to-distance ratio priors for the same with $q=0.5$, $\beta=0.001$, and varying $l$.}
    \label{fig:shadow.eht}
\end{figure}

In the numerical plot depicted in Fig. \ref{fig:shadow.eht}, the radius of the black hole's shadow (\(R_{sh}\)) is calculated for different values of \(l\) for asymptotically flat case $A(r_o=1)$, and $(r_0=1$). It is intriguing to observe the exponential and inverse relationship exhibited by \(l\), irrespective of the range within which the parameters vary. In Fig. \ref{fig:shadow.eht}, the upper limits of \(l\) are presented based on EHT observations. According to the \(68\%\) confidence level (C.L.) \cite{Vagnozzi:2022moj}, the upper limit for \(l\) is \(0.6\).

\subsection{Spherically infalling accretion} \label{sec4}
%AO 

In this section, we explore the model of spherically free-falling accretion around AMCNS black hole from an infinite distance. The approach described in references \cite{Jaroszynski:1997bw, Bambi:2012tg} is employed for this investigation. By utilizing this method, we can generate a realistic representation of the shadow produced by the accretion disk. However, in reality, the actual image of the black hole cannot be observed as a discernible boundary in the universe. Additionally, employing a static accretion-disk model is not feasible due to the presence of a dynamic accretion disk surrounding the black hole. This dynamic disk also generates synchrotron emission as part of the accretion process. To accomplish this, our initial focus is on studying the specific intensity observed at the frequency of the photon $\nu_\text{obs}$ by solving the following integral along the path of light:
\begin{equation}
    I(\nu_\text{obs},b_\gamma) = \int_\gamma g^3 j(\nu_e) \mathrm{d} l_\text{prop}.
    \label{eq:bambiI}
\end{equation}

We should mention that $b_{\gamma}$ represents the impact parameter, $j(\nu_e)$ corresponds to the emissivity per unit volume, $\mathrm{d} l_\text{prop}$ represents the infinitesimal proper length, and $\nu_e$ denotes the frequency of the emitted photon. In this context, we introduce the redshift factor for the accretion in free-fall using the following definition:

\begin{equation}
    g = \frac{k_\mu u^\mu_o}{k_\mu u^\mu_e},
\end{equation}

The equation provided states that the 4-velocity of the photon is represented by $k^\mu=\dot{x}_\mu$, while the 4-velocity of the distant observer is denoted as $u^\mu_o=(1,0,0,0)$. Furthermore, $u^\mu_e$ represents the 4-velocity of the infalling accretion:
\begin{equation}
u_{\mathrm{e}}^{t}=\frac{1}{A(r)}, \quad u_{\mathrm{e}}^{r}=-\sqrt{\frac{1-A(r)}{A(r) B(r)}}, \quad u_{\mathrm{e}}^{\theta}=u_{\mathrm{e}}^{\phi}=0.
\end{equation}

By utilizing $k_{\alpha} k^{\alpha}=0$, one can derive the constants of motion $k_{r}$ and $k_{t}$ for the photons.
\begin{equation}
k_{r}=\pm k_{t} \sqrt{B(r)\left(\frac{1}{A(r)}-\frac{b^{2}}{r^{2}}\right)}.
\end{equation}
It is worth noting that the sign of $\pm$ indicates whether the photon approaches or deviates away from the black hole. Based on this, the redshift factor $g$ and proper distance $\mathrm{d} l_\gamma$ can be expressed as follows:
   \begin{equation}
   g = \Big( u_e^t + \frac{k_r}{k_t}u_e^r \Big)^{-1},
  \end{equation}
  and
 \begin{equation}
  \mathrm{d} l_\gamma = k_\mu u^\mu_e \mathrm{d} \lambda = \frac{k^t}{g |k_r|}\mathrm{d} r.
\end{equation}
Next, we restrict our analysis to monochromatic emission, where the specific emissivity is considered with a rest-frame frequency $\nu_*$.
        \begin{equation}
            j(\nu_e) \propto \frac{\delta(\nu_e - \nu_*)}{r^2}.
        \end{equation}
        
Subsequently, the intensity equation, given in equation \eqref{eq:bambiI}, can be expressed as follows:
        
        \begin{equation}
            F(b_\gamma) \propto \int_\gamma \frac{g^3}{r^2} \frac{k_e^t}{k_e^r} \mathrm{d} r.
        \end{equation}
        
\begin{figure}[htp]
   \centering
          \includegraphics[scale=1.2]{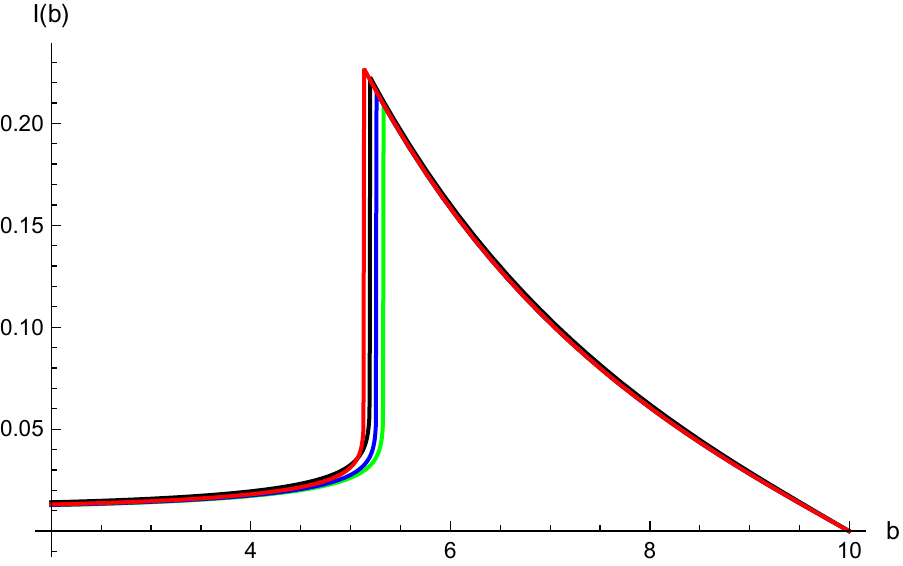}
    \caption{  $q=0.5$, $\beta=0.001$, and varying $l$; $l=0.3$ (blue), $l=0.5$ (green), and $l=0.7$ (red).}
    \label{fig:acc1}
\end{figure}

\begin{figure}[htp]
   \centering
          \includegraphics[scale=0.4]{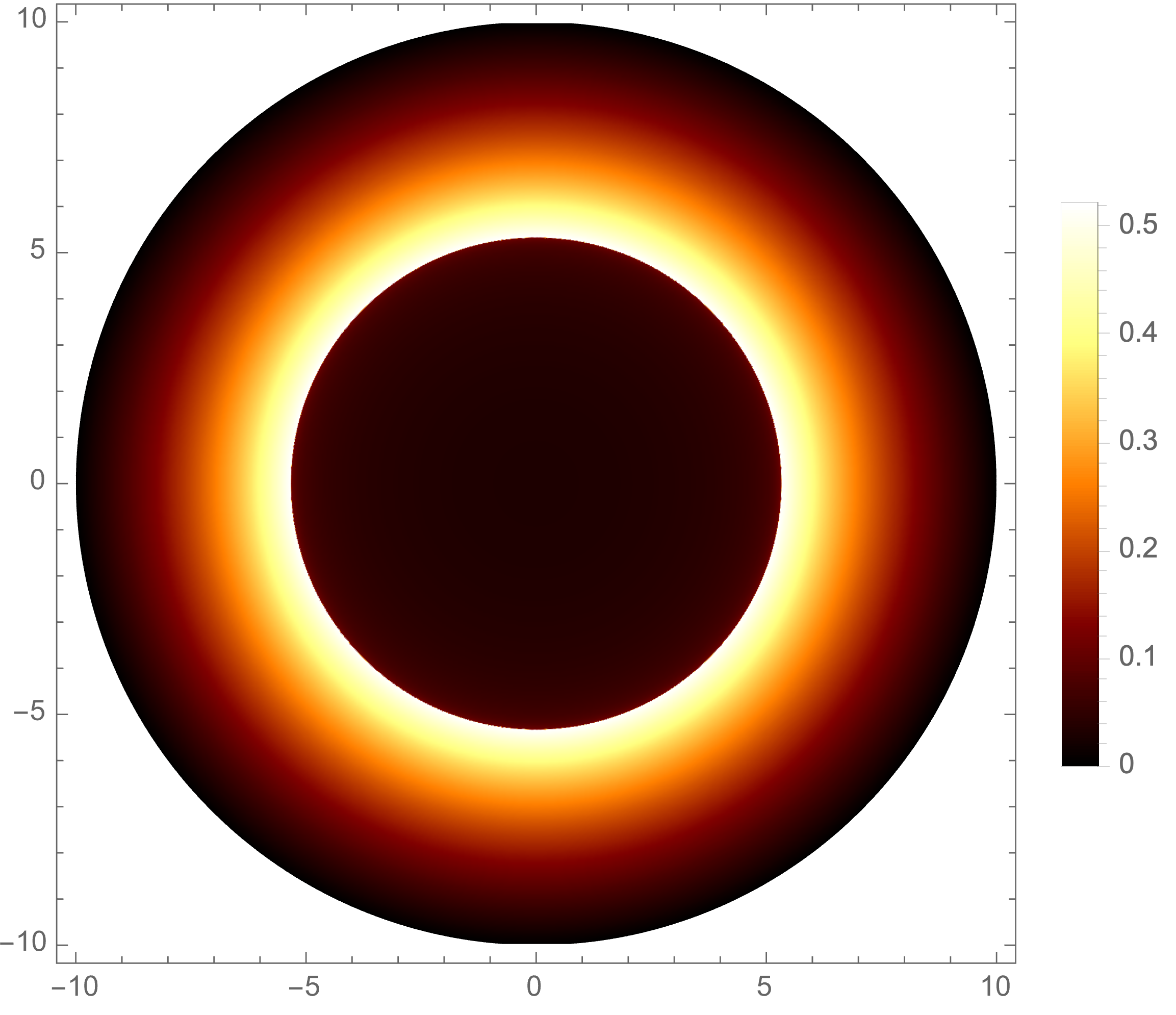}
          \includegraphics[scale=0.4]{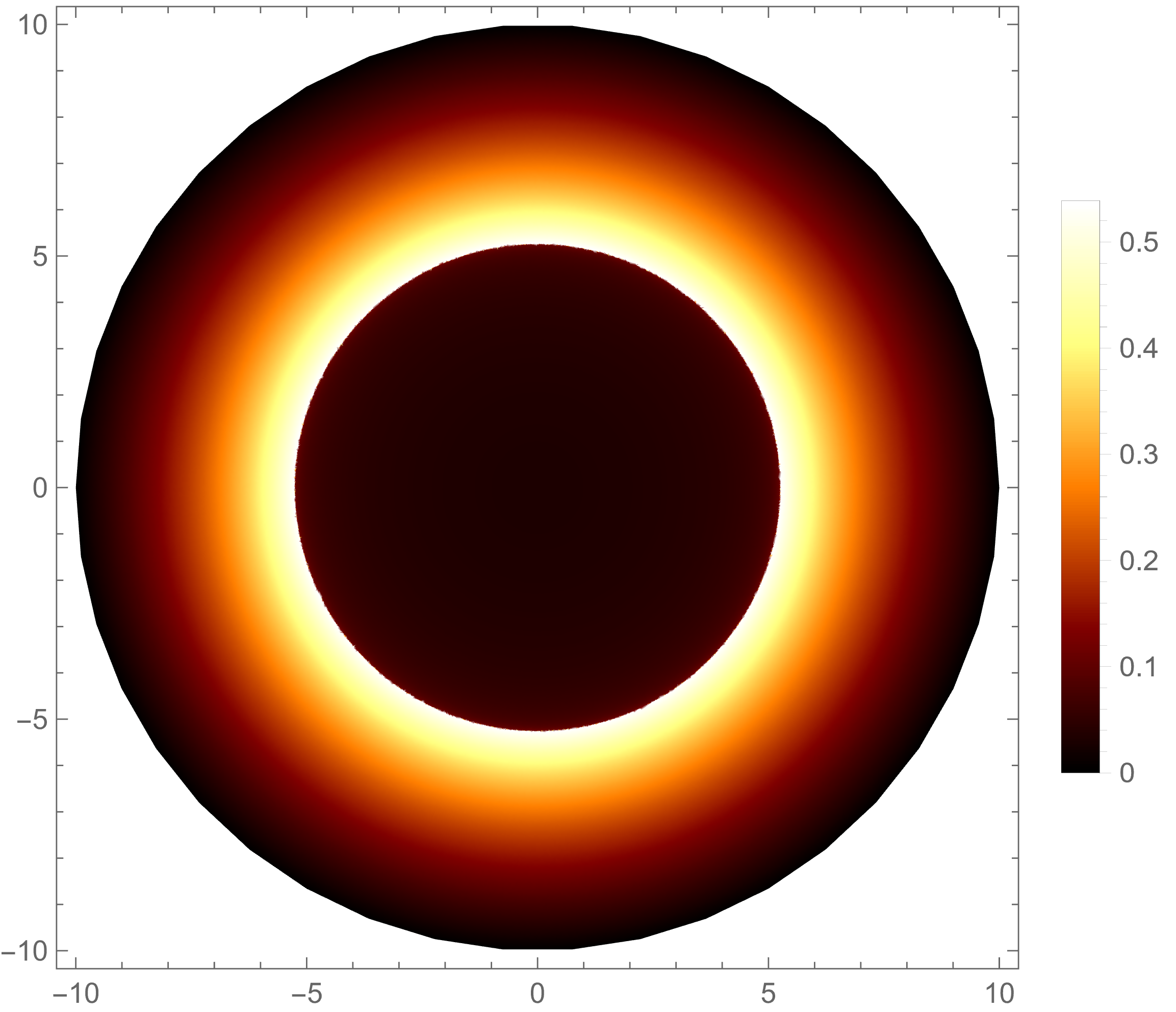}
         \includegraphics[scale=0.4]{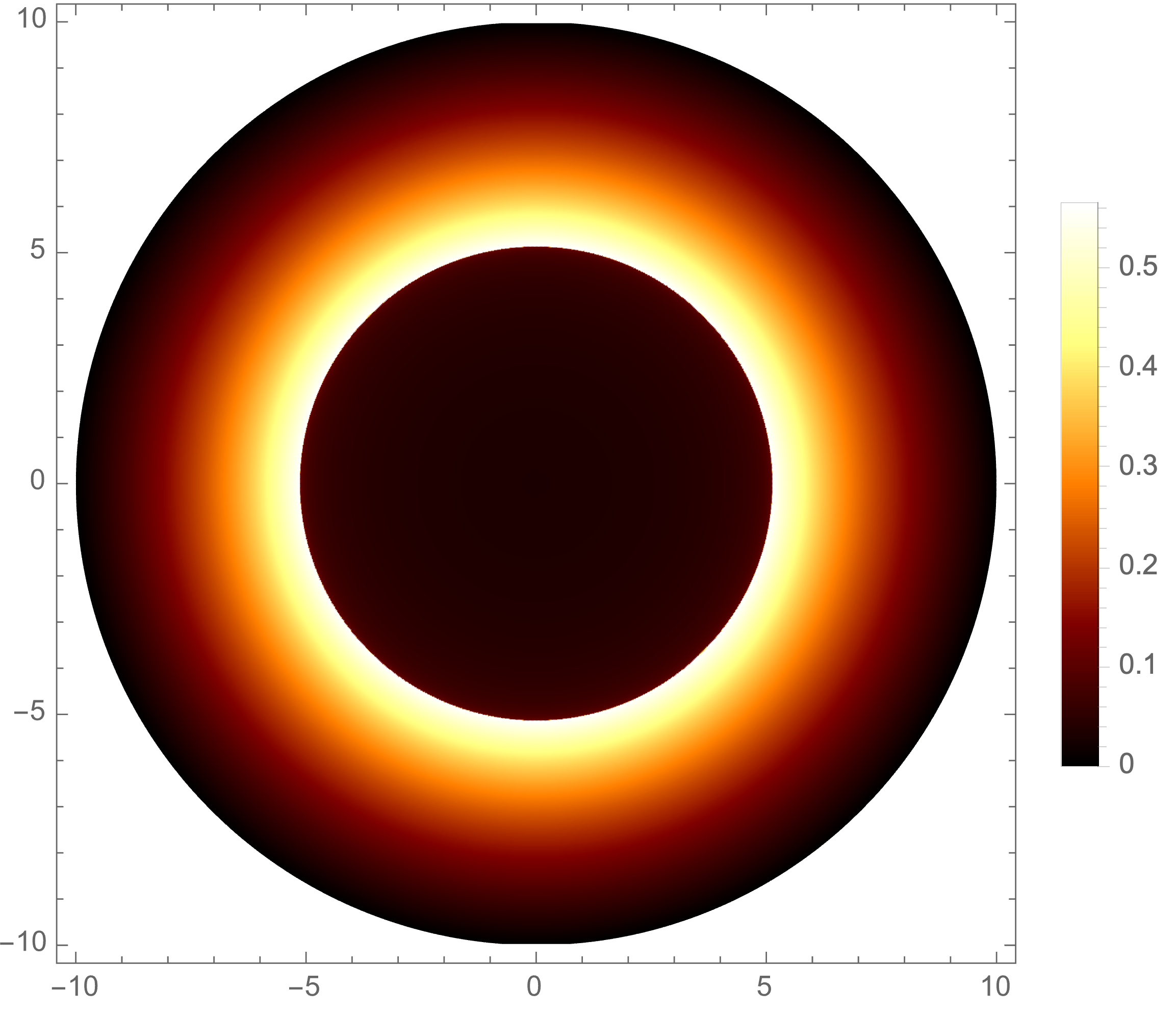}
    \caption{  $q=0.5$, $\beta=0.001$, and $l=0.3$ (left), $l=0.5$ (middle), and $l=0.7$ (right).}
    \label{fig:acc12}
\end{figure}

\begin{figure}[htp]
   \centering
          \includegraphics[scale=0.3]{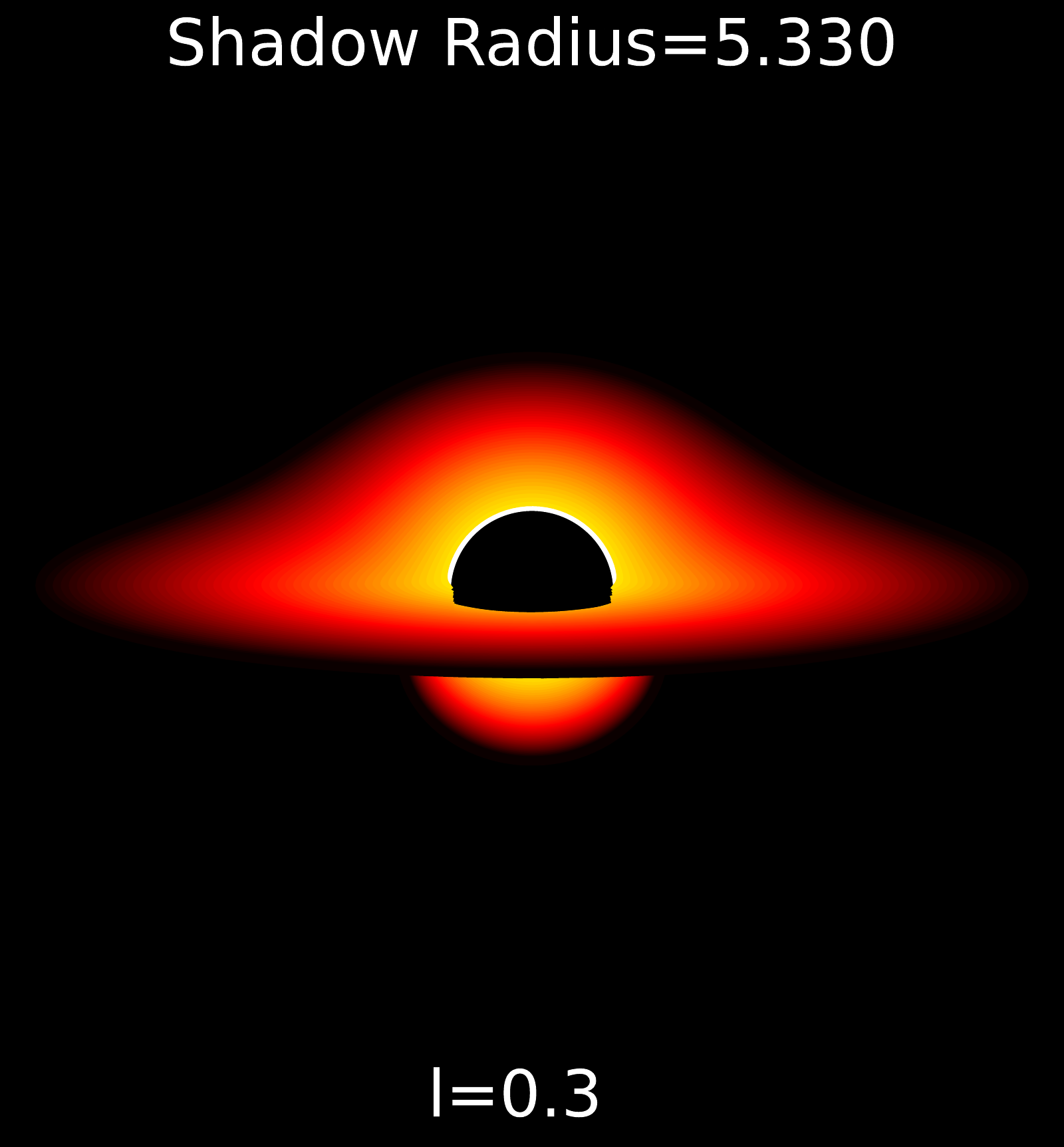}
          \includegraphics[scale=0.3]{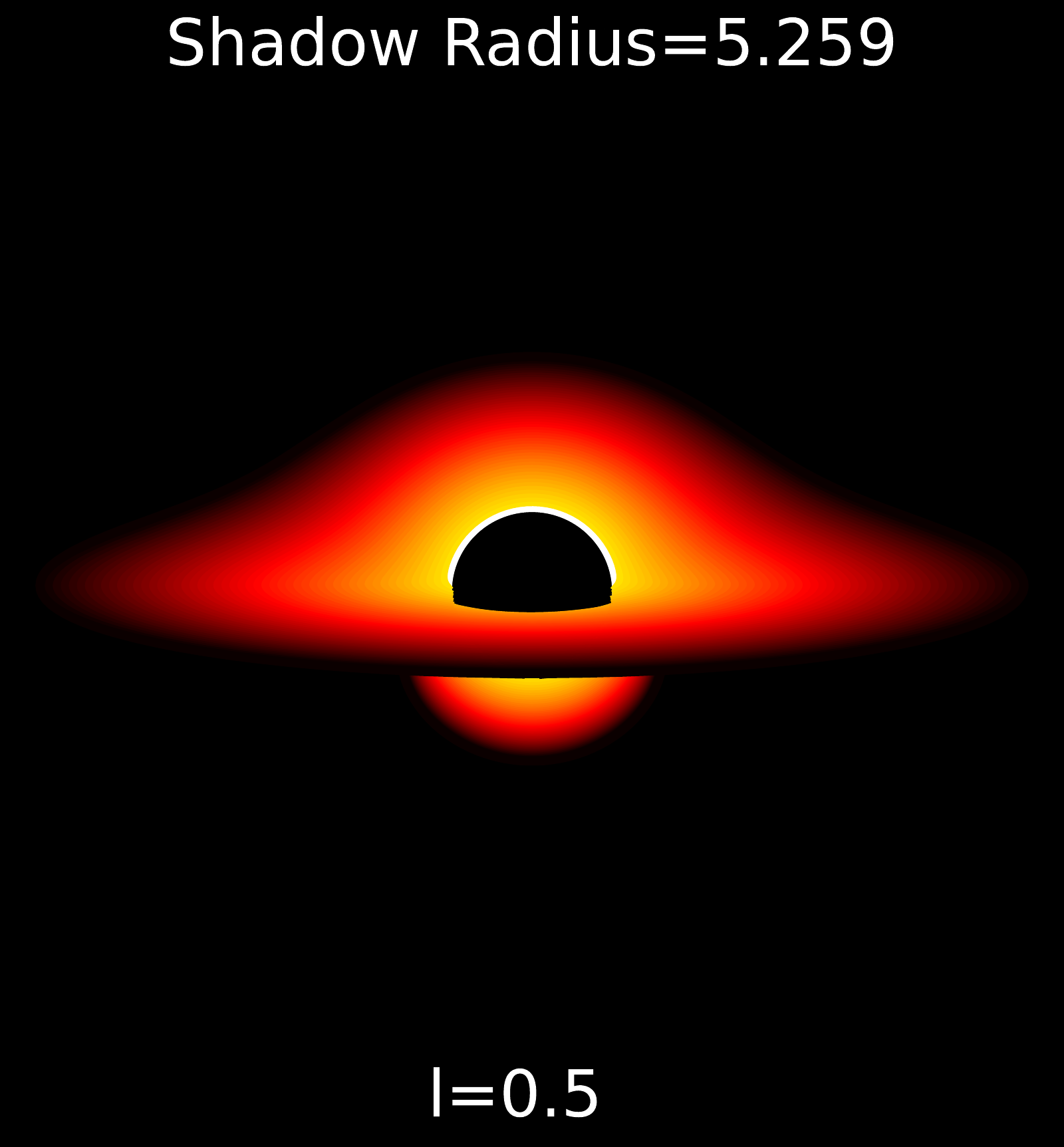}
         \includegraphics[scale=0.3]{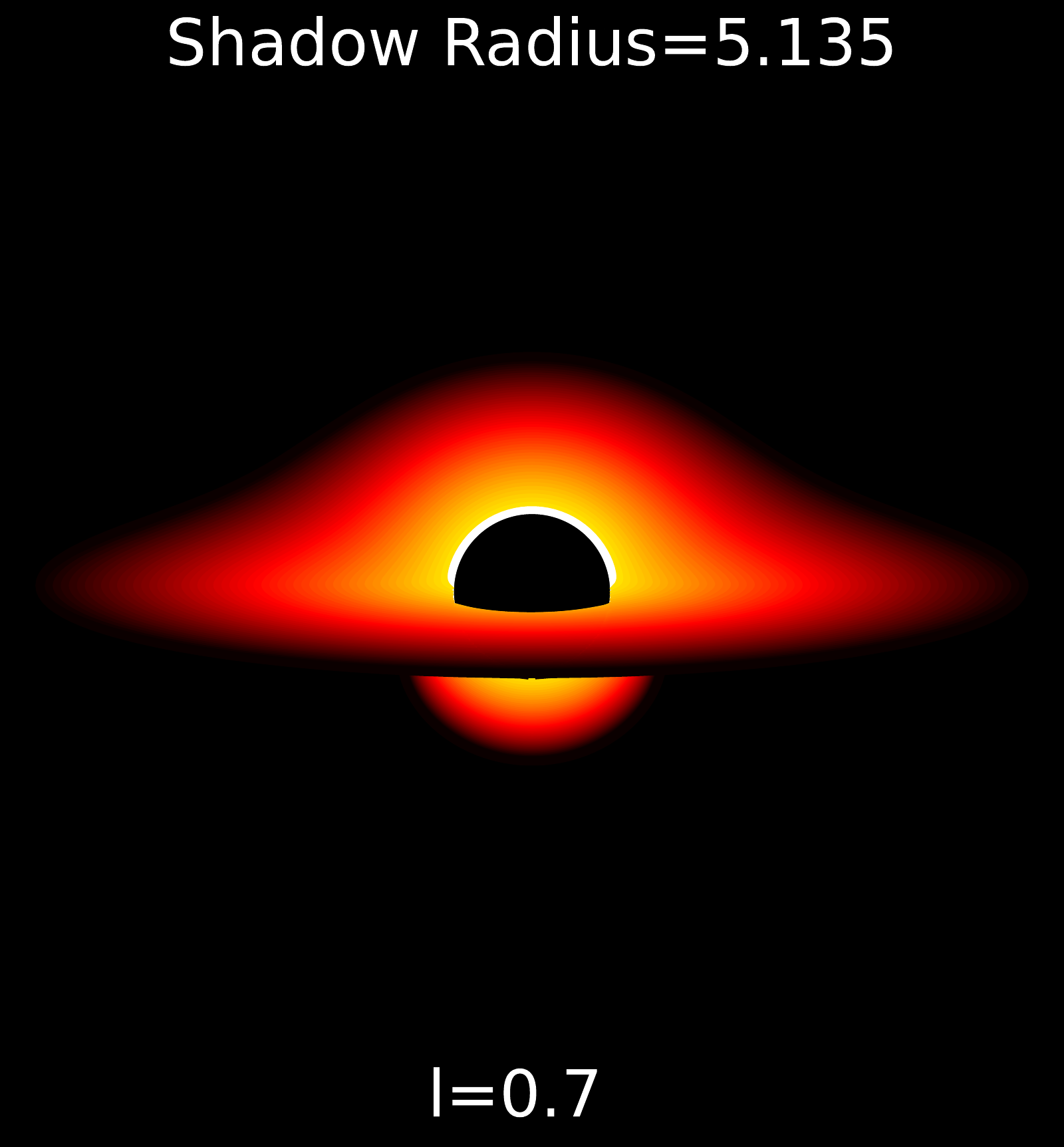}
    \caption{  $q=0.5$, $\beta=0.001$, and $l=0.3$ (left), $l=0.5$ (middle), and $l=0.7$ (right).}
    \label{fig:acc13}
\end{figure}

To analyze the shadow created by the thin accretion disk around the black hole, we begin by numerically solving the aforementioned equation. This numerical calculation is performed using the Einsteinpy library \cite{shreyas_bapat_2021_4739508} and the Mathematica notebook package \cite{Okyay:2021nnh}, which have also been utilized in previous studies \cite{Chakhchi:2022fls, Kuang:2022xjp, Uniyal:2022vdu, Pantig:2022ely}. The integration of the flux reveals the impact of the parameter $l$ on the specific intensity observed by a distant observer for an in-falling accretion. The corresponding results are presented in Figures \ref{fig:acc1}, \ref{fig:acc12}, and \ref{fig:acc13}.

The figures in Fig.~\ref{fig:acc12}, and \ref{fig:acc13} show the intensity plots and stereographic projections of the AMCNS black hole described following the method of \cite{Bambi:2013nla,Jaroszynski:1997bw,Gralla:2019xty}. The intensity plot shows the smooth transition between the layers in the journey directed away from the event horizon. The sharpness in every photon band in Fig.~\ref{fig:acc13} indicates the effect of the parameters $q$, $l$, and $\beta$ appearing to be brighter.

The difference in the appearance of the shadow sizes between the three images in Fig. \ref{fig:acc12} is hard to miss. The emissivity from the innermost photon orbit is seen to get brighter as $l$ increases. Consequently, the size of dark shadow region appears to be decreasing. Numerically speaking, Fig. \ref{fig:acc1} depicts how small yet non-trivial this difference is. The shadow outline seen in the images of Fig. \ref{fig:acc13} clearly exemplifies that the manifestation of additional inner-orbits of photons proportionally increasing with $l$ instigates the peak intensity to grow.

When compared with the shadow of a Schwarzschild black hole, the crucial factor that brings about a striking distinction between them is singularity. The Schwarzschild black hole is singular at its center meaning that it is a point of infinite density and zero volume. Among all the properties of a black hole that are disturbed by this (such as curvature, geometry, density profile, etc.), the one that is sizably impacted is the event horizon which is a mathematical construct that assumes the black hole is a point singularity. For a non-singular black hole, this assumption does not hold; in that case, the event horizon becomes an apparent boundary beyond which light cannot escape to infinity, but can still escape to some finite distance. This leads the non-singular black hole to have a smaller event horizon than the Schwarzschild black hole, which in turn explains the former's smaller shadow size.

In summary, any entity that skirts around a black hole rendering a time-dependent appearance of the shadow will inexorably give rise to distortions, big or small, contingent on its characteristics. Also, the singularity at the center of a black hole can cause its space-time geometry to curve more bringing about a wider event horizon and a bigger shadow size. On the other hand, a non-singular black hole would have a smoother geometry and a smaller horizon that results in a smaller shadow size. Furthermore, dark matter intensifies the extent of distortions because of dispersion.

\section{BOUNDING THE GREYBODY FACTORS OF AMCNS BLACK HOLE}
\label{Sgreybodyfactor}
The equivalence of black holes to black bodies inspires curiosity about their relationship. This section is devoted to inspecting that correlation. A black body is an idealized object that absorbs all radiation that strikes it and emits radiation without incurring any loss at all wavelengths and at the maximum frequency; for a given spectrum, although the wavelength is dependent on temperature, its efficiency is 100\%. Whereas, black holes absorb all radiation and matter inside the event horizon and emit Hawking radiation that is dependent on the mass of the black hole preserving the quantum information of the engulfed particles \cite{Hawking:1975vcx,Akhmedov:2006pg}. While the gravity field and the matter encompassing a black hole deteriorates its energy eventually impairing the Hawking radiation emitted by the black hole from being fully efficient, its resemblance to a pure black body at its horizon, when isolated (resulting from quantum effects), is impeccable.

The deteriorating aspects of a radiating black hole are given by the greybody factor: the nomenclature underlines the departure from the black body behavior. It is a function of frequency and angular momentum that typifies the degree of deviation in the emission spectrum of a black hole from a perfect black body spectrum. The main culprit for this digression is the scattering of the radiation by the geometry of the black hole itself relating to the captured/radiated particles that fall back into the extreme force. This causes the spectrum emitted by the black hole to deviate and can be observed as the greybody factor \cite{Maldacena:1996ix,Harmark:2007jy,Rincon:2018ktz,Rincon:2020cos,Panotopoulos:2018pvu,Panotopoulos:2017yoe,Klebanov:1997cx,Liu:2022ygf,Fernando:2004ay,Konoplya:2019hlu,Javed:2022rrs,Javed:2022kzf,Yang:2022ifo,Javed:2021ymu,Mangut:2023oou,Kanzi:2023itu,Al-Badawi:2021wdm,Al-Badawi:2022uwh,Al-Badawi:2022aby}.

By the scattering of a wave packet off a black hole, the greybody factor can be determined with the help of the classical Klein-Gordon equation -- the Schr\"{o}dinger-like equation that governs the modes for a wave function $\varphi(r)$ relating the position of an electron to its wave amplitude is written as:
\begin{equation}\left(\frac{\mathrm{d}^2}{\mathrm{d} r_{*}^2}+\omega^2-{V}(r)\right) \varphi(r)=0, \end{equation}
where, $r_*$ is commonly known as the tortoise coordinate. Its purpose is to evolve to infinity in a manner that counters the singularity of the metric under scrutiny. The time of an object looming towards the event horizon in the constructed system of coordinates grows to infinity... so does $r_*$ but at a rate that is appropriate. Mathematically:
\begin{equation}\mathrm{d} r_*=\frac{1}{f(r)} \mathrm{~d} r, \end{equation}
and the potential ${V}(r)$ for the azimuthal quantum number $\ell$ specifying the mode can accordingly be formulated as:
\begin{equation}{V}(r)=\frac{f(r) f^{\prime}(r)}{r}+\ell(\ell+1) \frac{f(r)}{r^2}.\end{equation}
This embodies the potential energy of the particle when there is an external force/field. The information about the wavefunction, the bound and scattering states, etc. can be begotten from this quantity.

Assuming that the energy of the particles emitted by the black hole falls in the range $\hbar\omega$ to $\hbar (\omega + \mathrm{d}\omega)$ for a given wave frequency $\omega$, let $\Upsilon$ be a positive function that obeys $\Upsilon(-\infty) = \sqrt{\omega^2 -V_{-\infty}}$ and $\Upsilon(\infty) = \sqrt{\omega^2 -V_\infty}$ when $V_{\pm\infty} = V\pm \infty$. To find the lower bound, the condition $V_{\pm\infty} =0$ must be satisfied, implying that $\Upsilon=\omega$ and the transmission probability is deduced as follows \cite{Visser:1998ke,Boonserm:2008zg,Boonserm:2017qcq}:

\begin{equation}
\begin{aligned}
T_b & \geq \mathrm{sech}^2\left(\int_{-\infty}^{\infty} {V}(r) \mathrm{~d} r_*\right) \\
& =\mathrm{sech}^2\left[\frac{1}{2 \omega} \int_{r_H}^{\infty}\left(\frac{f^{\prime}(r)}{r}+\frac{\ell(\ell+1)}{r^2}\right) \mathrm{~d} r\right] 
%& =\mathrm{sech}^2\left[\frac{1}{2 \omega}\left\{\int_{r_H}^{\infty} \left(\frac{r^2 \left[3 r^2+l^2 \left(\frac{q^2}{r^2}-\frac{\beta  q^4}{4 r^6}\right)\right] \zeta}{\left(r^3+l^2\zeta\right)^2} -\frac{2 r \zeta}{r^3 + l^2 \zeta}-\frac{r^2 \left(\frac{q^2}{r^2}-\frac{\beta  q^4}{4 r^6}\right)}{r^3 + l^2 \zeta} \right)\frac{\mathrm{d}r}{r}\right\} + \frac{\ell(\ell+1)}{r_H}\right].
\end{aligned}
\end{equation}
This expression for $T$ offers insight into the probability of a particle with a specific energy and angular momentum escaping the black hole and being detected at infinity. Yet again, $r_H$ is the radius of the event horizon of the black hole: it directly affects the details of the emitted radiation spectrum. Because of the intricacy of solving for the polynomial, we procure $r_H$ numerically. 

\begin{figure}
\centering  
\begin{subfigure}{0.45\textwidth}
  \centering
  \includegraphics[width=1\linewidth]{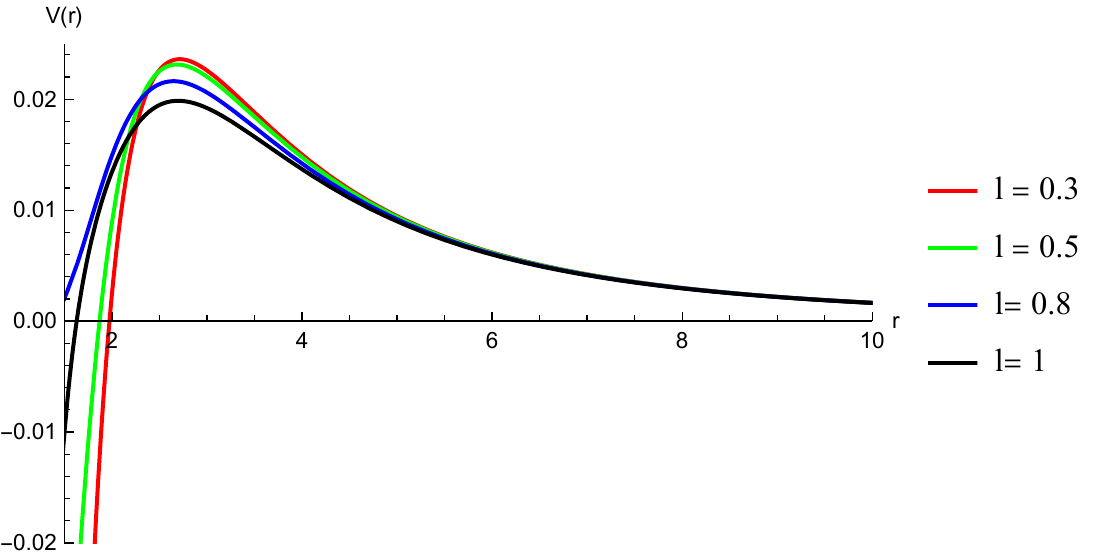}
    \caption{AMCNS black hole for $q=0.5$, $\beta=0.001$, and $\ell=0$ .}
  \label{pot1}
\end{subfigure}
\begin{subfigure}{0.45\textwidth}
  \centering
  \includegraphics[width=1\linewidth]{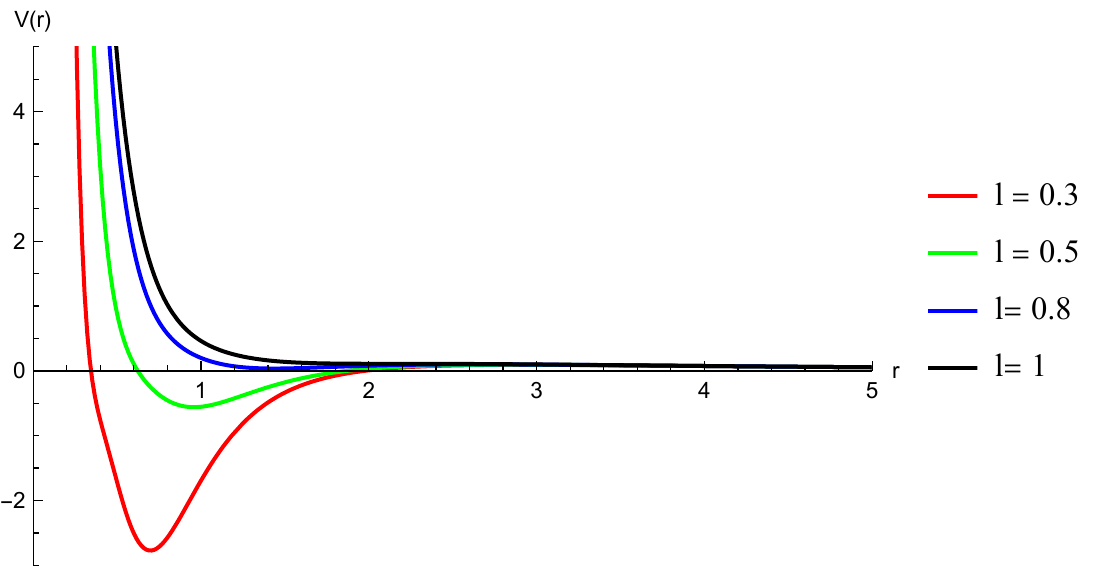}
   \caption{AMCNS black hole for $q=0.5$,  $\beta=0.001$, and $\ell=1$.} \label{pot2}
\end{subfigure}
  \caption{Potential plotted against radial distance.}
    \label{pot}
\end{figure}

The term $\ell(\ell+1)$ describes the discrete robustness of quantum mechanics in general relativity by providing a measure of the angular momentum carried by the particles that the black hole emits. Also called the angular momentum quantum number, the value that $\ell$ takes says quite a lot about the emitted particles and spectrum:
\begin{itemize}
    \item $~\ell=0$ insinuates that the emitted particles have zero angular momentum and are moving in a direction away from the black hole. The probability of a particle escaping the black hole in this case is maximum at low energies and it decreases as the energy of the particle increases. The spectrum of emitted radiation for the particles in this state is typically dominated by low-energy photons.\\
    \item $~\ell=1$ insinuates that the emitted particles have some angular momentum and are moving in a circular orbit around the black hole. The probability of a particle escaping the black hole in this case is maximum at intermediate energies and it decreases at both low and high energies. The spectrum of emitted radiation for the particles in this state typically peaks at intermediate energy. \\
%    \item $\ell=2$ insinuates that the emitted particles have substantial angular momentum and are moving in a rather elliptical orbit around the black hole. The probability of a particle escaping the black hole in this case is maximum at high energies and it decreases as the energy of the particle decreases. The spectrum of emitted radiation for the particles in this state typically peaks at high energy.

\end{itemize}
Therefore, the value of $\ell$ particularizes the detailed shape of the spectrum of the radiation that a black hole emits, granting information about the features of the black hole itself. The graphical representation of the results acquired for $V(r)$ and $T$ is presented in Figs.~\ref{pot}, \ref{pot1} and \ref{pot2}  respectively for an AMCNS black hole. 

Plotted to show the variation of the potential $V(r)$ in terms of the radial coordinate $r$ for varying $l$, Fig.~\ref{pot1} is perceived to show a negative intercept that rises to positive until a certain value before falling again when $\ell=0$. However, for $\ell=1$, Fig.~\ref{pot2} has a negative trough that briskly rises with $l$ to positive.

\begin{figure}
    \centering
   \begin{subfigure}{0.45\textwidth}
  \centering
  \includegraphics[width=1\linewidth]{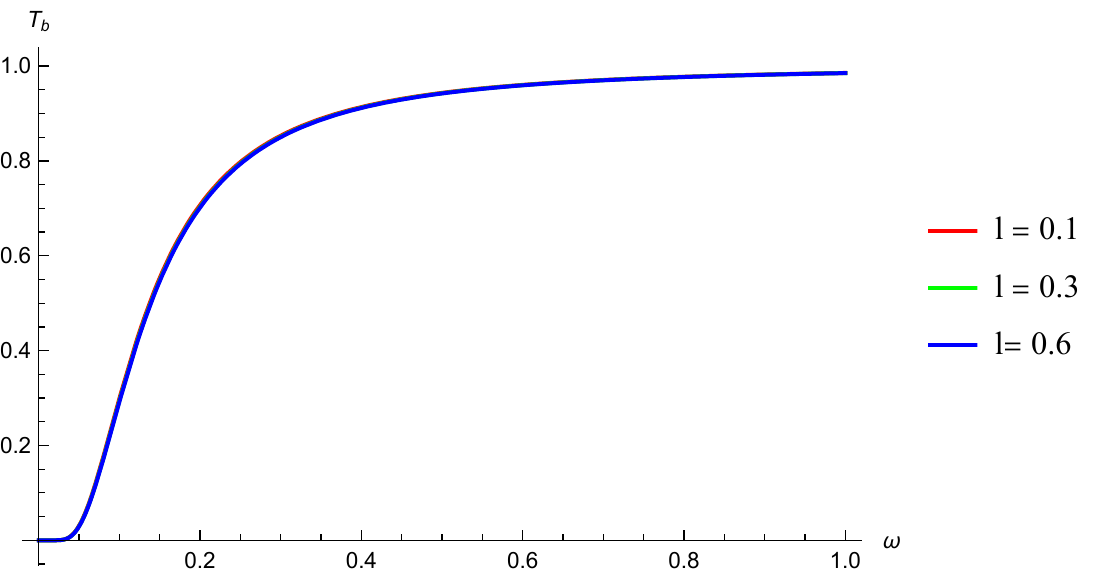}
    \caption{AMCNS black hole}
  \label{tc01}
\end{subfigure}
\begin{subfigure}{0.45\textwidth}
  \centering
  \includegraphics[width=1\linewidth]{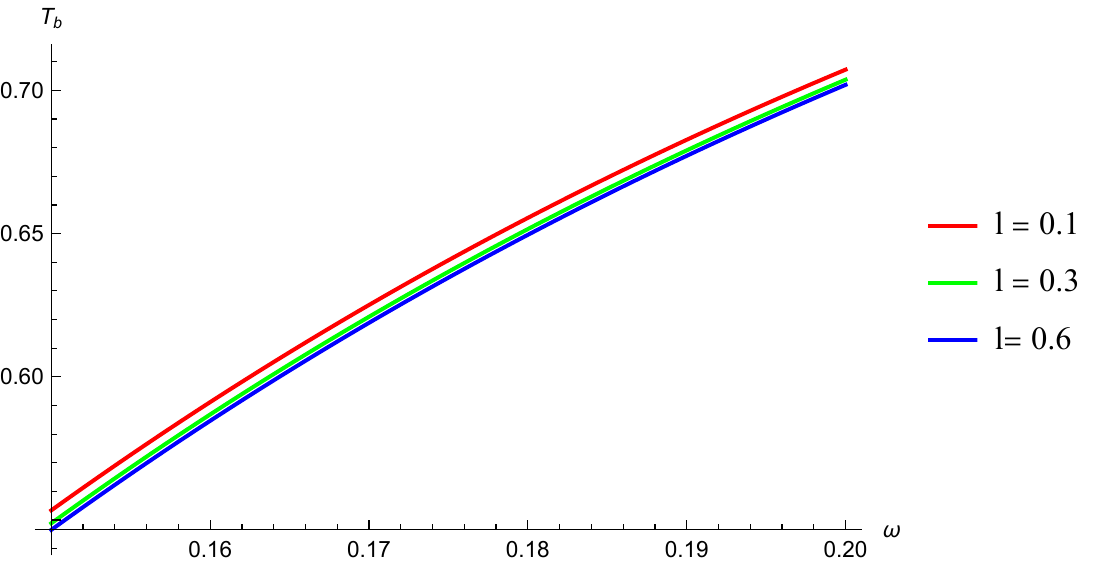}
    \caption{AMCNS black hole} \label{tc02}
\end{subfigure}
  \caption{Transmission probability plotted against frequency for $q=0.5$, and $\beta=0.001$ for $\ell=0$.}
  \label{tc0}
\end{figure}

Fig.~\ref{tc0} illustrates the variation of the Transmission probability $T$ in terms of the frequency $\omega$ for different values of $l$ for an AMCNS black hole when $\ell=0$. While the graph traits of all cases in Fig.~\ref{tc01} are analogous to each other with the same probability, a closer look at the ascension in the frequency range of $0.05 \leq \omega \leq 0.2$ infers that any and every variation (even the ones of the smallest magnitude) in the parameters of a black hole gives rise to an innate frequency.

\begin{figure}
    \centering
   \begin{subfigure}{0.45\textwidth}
  \centering
  \includegraphics[width=1\linewidth]{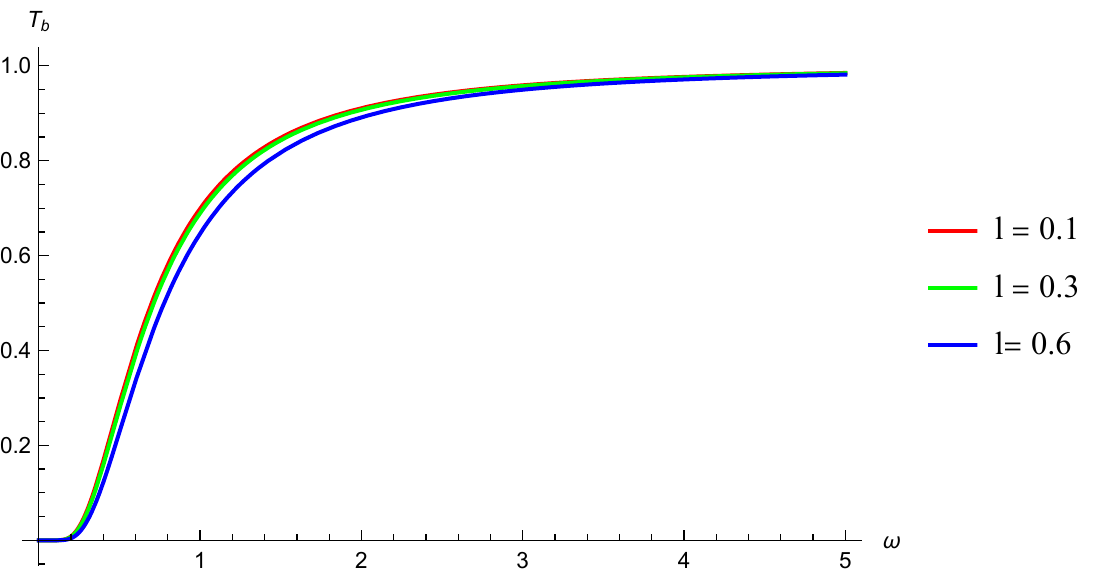}
    \caption{AMCNS black hole}
  \label{tc11}
\end{subfigure}
\begin{subfigure}{0.45\textwidth}
  \centering
  \includegraphics[width=1\linewidth]{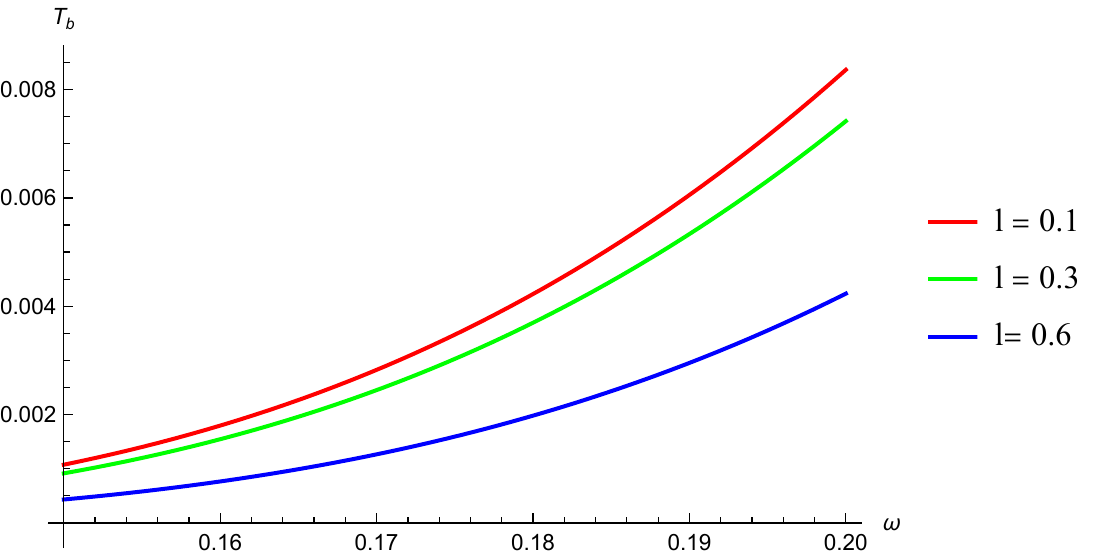}
    \caption{AMCNS black hole} \label{tc12}
\end{subfigure}
  \caption{Transmission probability plotted against frequency for $q=0.5$, and $\beta=0.001$ for $\ell=1$.}
  \label{tc1}
\end{figure}

Fig.~\ref{tc1} illustrates the variation of the Transmission probability $T$ in terms of the frequency $\omega$ for different values of $l$ for an AMCNS black hole when $\ell=1$. In Fig.~\ref{tc11}, the probability of transmission of each case is discrete from the others and is noticed to be higher for lower values of $l$, converging to unity as $\omega$ increases. The offsetting of every different $l$ in Fig.~\ref{tc12} is highly conspicuous attributable to the factor of angular momentum that prevails. Recalling that $l$ is a length-scale parameter, the decrease in $T_b$ with increasing $l$ can be explained by the inverse proportionality exhibited by frequency with respect to wavelength.

Nevertheless, the overall nature of the AMCNS black hole plots is adequately comparable to that of the Schwarzschild case and to Figs. 2 and 5 in %*ref* https://arxiv.org/pdf/2210.15620.pdf
in which the author handles a regular Hayward black hole.

\section{CONCLUSION}
\label{Sconc}

%Update conclusions
In this article, we have attempted to analyze the astrophysics of an asymptotic, magnetically-charged, non-singular black hole. o find its solution, we commenced with the Euler-Lagrange equation for nonlinear electrodynamics and applied it to the Hayward metric through the trace of the energy-momentum tensor. Establishing its connection to the energy density eased us in uncovering the mass function and, with additional computations, the metric function of the AMCNS black hole. 

This illation was then employed in the Gauss-Bonnet theorem and modified with the Gibbons and Werner method to obtain the weak deflection angle of the AMCNS black hole. This is done with the aim of examining the black hole's gravitational field, its intrinsic attributes such as mass, spin, etc., the nature of the light bending, and, ultimately, the validity of Einstein's theory deeper. These help us comprehend the dynamics of black holes better and prepares Observational Astrophysicists on what to or not to expect. The deflection angle that was obtained was observed to be vastly variant for the AMCNS black hole with an increase that is triggered by its governing parameters.

We then proceeded to calculate the effects of the dark matter medium on the deflection angle. Numerous theoretical and experimental endeavors are adopted to bridge the gap between physics and astrophysics; incorporating the consequences of dark matter majorly subsidizes that. Dark matter plays a key role in the intensified regime of black holes accreting and jets launching. The dark matter medium was perceived to increase the deflection more, and for the AMCNS balc hole, the slightest drop in the refractive index generated an immense upshot.

The greybody factor was then investigated using the Schr\"{o}dinger-like equation. The potential and the transmission probability were derived for the AMCNS black hole and were plotted against a Schwarzschild black hole. One can discern the sizable differences in each of the curves for every case of the azimuthal quantum number.

Finally, the black hole shadow was probed for the AMCNS case. Although many parameters affect the shadow size, the factors assumed here for this particular black hole display highly distinguished traits, making the black hole fairly distinctive from a Schwarzschild black hole. With the assistance of the Euler-Lagrange equation once again, the critical curve and the shadow's angular radius were determined. They were scrutinized thoroughly with stereographic projections; the singularity at the center of the Schwarzschild black hole curves and distorts its spacetime geometry more than that of a non-singular black hole, therefore, having a larger horizon and shadow size. The smoother spacetime geometry of the AMCNS black hole gives it a relatively smaller horizon and shadow size. 

Meticulous methodologies of maximal parametric inclusiveness facilitate us in refining our comprehension of gravitational lensing. The distortions that every black hole produces offer so much information about the source of the light rays along with the black hole, which in turn motivates us to come up with new hypotheses that could polish and perfect our perception: our future work is directed toward the same. we intend to gather as many contributing factors that affect our grasp of black hole theories as possible and enhance them, no matter how complex it is, so that both theoretical and observational astrophysics can use them as a resource for improvement. We hope to attain ways that could minimize our assumptions like zero spin, staticity, spherical symmetry, and so many other constraints, and deal with the black hole realistically instead to the greatest degree feasible.

\acknowledgments 
A. {\"O}. would like to acknowledge the contribution of the COST Action CA18108 - Quantum gravity phenomenology in the multi-messenger approach (QG-MM) and the COST Action CA21106 - COSMIC WISPers in the Dark Universe: Theory, astrophysics and experiments (CosmicWISPers). A. {\"O}. is funded by the Scientific and Technological Research Council of Turkey (TUBITAK).

\bibliography{ref.bib}

\begin{thebibliography}{162}%
\makeatletter
\providecommand \@ifxundefined [1]{%
 \@ifx{#1\undefined}
}%
\providecommand \@ifnum [1]{%
 \ifnum #1\expandafter \@firstoftwo
 \else \expandafter \@secondoftwo
 \fi
}%
\providecommand \@ifx [1]{%
 \ifx #1\expandafter \@firstoftwo
 \else \expandafter \@secondoftwo
 \fi
}%
\providecommand \natexlab [1]{#1}%
\providecommand \enquote  [1]{``#1''}%
\providecommand \bibnamefont  [1]{#1}%
\providecommand \bibfnamefont [1]{#1}%
\providecommand \citenamefont [1]{#1}%
\providecommand \href@noop [0]{\@secondoftwo}%
\providecommand \href [0]{\begingroup \@sanitize@url \@href}%
\providecommand \@href[1]{\@@startlink{#1}\@@href}%
\providecommand \@@href[1]{\endgroup#1\@@endlink}%
\providecommand \@sanitize@url [0]{\catcode `\\12\catcode `\$12\catcode
  `\&12\catcode `\#12\catcode `\^12\catcode `\_12\catcode `\%12\relax}%
\providecommand \@@startlink[1]{}%
\providecommand \@@endlink[0]{}%
\providecommand \url  [0]{\begingroup\@sanitize@url \@url }%
\providecommand \@url [1]{\endgroup\@href {#1}{\urlprefix }}%
\providecommand \urlprefix  [0]{URL }%
\providecommand \Eprint [0]{\href }%
\providecommand \doibase [0]{http://dx.doi.org/}%
\providecommand \selectlanguage [0]{\@gobble}%
\providecommand \bibinfo  [0]{\@secondoftwo}%
\providecommand \bibfield  [0]{\@secondoftwo}%
\providecommand \translation [1]{[#1]}%
\providecommand \BibitemOpen [0]{}%
\providecommand \bibitemStop [0]{}%
\providecommand \bibitemNoStop [0]{.\EOS\space}%
\providecommand \EOS [0]{\spacefactor3000\relax}%
\providecommand \BibitemShut  [1]{\csname bibitem#1\endcsname}%
\let\auto@bib@innerbib\@empty
%</preamble>
\bibitem [{\citenamefont {{Montgomery}}\ \emph {et~al.}(2009)\citenamefont
  {{Montgomery}}, \citenamefont {{Orchiston}},\ and\ \citenamefont
  {{Whittingham}}}]{michell}%
  \BibitemOpen
  \bibfield  {author} {\bibinfo {author} {\bibfnamefont {Colin}\ \bibnamefont
  {{Montgomery}}}, \bibinfo {author} {\bibfnamefont {Wayne}\ \bibnamefont
  {{Orchiston}}}, \ and\ \bibinfo {author} {\bibfnamefont {Ian}\ \bibnamefont
  {{Whittingham}}},\ }\bibfield  {title} {\enquote {\bibinfo {title} {{Michell,
  Laplace and the origin of the black hole concept}},}\ }\href@noop {}
  {\bibfield  {journal} {\bibinfo  {journal} {Journal of Astronomical History
  and Heritage}\ }\textbf {\bibinfo {volume} {12}},\ \bibinfo {pages} {90--96}
  (\bibinfo {year} {2009})}\BibitemShut {NoStop}%
\bibitem [{\citenamefont {de~Laplace}(1799)}]{de1799exposition}%
  \BibitemOpen
  \bibfield  {author} {\bibinfo {author} {\bibfnamefont {P.S.}\ \bibnamefont
  {de~Laplace}},\ }\href {https://books.google.com.cy/books?id=4GJUAAAAcAAJ}
  {\emph {\bibinfo {title} {Exposition du systeme du monde. Seconde edition
  revue et augmentee}}}\ (\bibinfo  {publisher} {Crapelet},\ \bibinfo {year}
  {1799})\BibitemShut {NoStop}%
\bibitem [{\citenamefont {Einstein}(1916)}]{Einstein:1916vd}%
  \BibitemOpen
  \bibfield  {author} {\bibinfo {author} {\bibfnamefont {Albert}\ \bibnamefont
  {Einstein}},\ }\bibfield  {title} {\enquote {\bibinfo {title} {{The
  Foundation of the General Theory of Relativity}},}\ }\href {\doibase
  10.1002/andp.19163540702} {\bibfield  {journal} {\bibinfo  {journal} {Annalen
  Phys.}\ }\textbf {\bibinfo {volume} {49}},\ \bibinfo {pages} {769--822}
  (\bibinfo {year} {1916})}\BibitemShut {NoStop}%
\bibitem [{\citenamefont {Schwarzschild}(1916)}]{Schwarzschild:1916uq}%
  \BibitemOpen
  \bibfield  {author} {\bibinfo {author} {\bibfnamefont {Karl}\ \bibnamefont
  {Schwarzschild}},\ }\bibfield  {title} {\enquote {\bibinfo {title} {{On the
  gravitational field of a mass point according to Einstein's theory}},}\
  }\href@noop {} {\bibfield  {journal} {\bibinfo  {journal} {Sitzungsber.
  Preuss. Akad. Wiss. Berlin (Math. Phys. )}\ }\textbf {\bibinfo {volume}
  {1916}},\ \bibinfo {pages} {189--196} (\bibinfo {year} {1916})},\ \Eprint
  {http://arxiv.org/abs/physics/9905030} {arXiv:physics/9905030} \BibitemShut
  {NoStop}%
\bibitem [{\citenamefont {{Nordstr{\"o}m}}(1918)}]{1918KNAB...20.1238N}%
  \BibitemOpen
  \bibfield  {author} {\bibinfo {author} {\bibfnamefont {G.}~\bibnamefont
  {{Nordstr{\"o}m}}},\ }\bibfield  {title} {\enquote {\bibinfo {title} {{On the
  Energy of the Gravitation field in Einstein's Theory}},}\ }\href@noop {}
  {\bibfield  {journal} {\bibinfo  {journal} {Koninklijke Nederlandse Akademie
  van Wetenschappen Proceedings Series B Physical Sciences}\ }\textbf {\bibinfo
  {volume} {20}},\ \bibinfo {pages} {1238--1245} (\bibinfo {year}
  {1918})}\BibitemShut {NoStop}%
\bibitem [{\citenamefont {Akiyama}\ \emph {et~al.}(2019)\citenamefont {Akiyama}
  \emph {et~al.}}]{EventHorizonTelescope:2019dse}%
  \BibitemOpen
  \bibfield  {author} {\bibinfo {author} {\bibfnamefont {Kazunori}\
  \bibnamefont {Akiyama}} \emph {et~al.} (\bibinfo {collaboration} {Event
  Horizon Telescope}),\ }\bibfield  {title} {\enquote {\bibinfo {title} {{First
  M87 Event Horizon Telescope Results. I. The Shadow of the Supermassive Black
  Hole}},}\ }\href {\doibase 10.3847/2041-8213/ab0ec7} {\bibfield  {journal}
  {\bibinfo  {journal} {Astrophys. J. Lett.}\ }\textbf {\bibinfo {volume}
  {875}},\ \bibinfo {pages} {L1} (\bibinfo {year} {2019})},\ \Eprint
  {http://arxiv.org/abs/1906.11238} {arXiv:1906.11238 [astro-ph.GA]}
  \BibitemShut {NoStop}%
\bibitem [{\citenamefont {Akiyama}\ \emph {et~al.}(2022)\citenamefont {Akiyama}
  \emph {et~al.}}]{EventHorizonTelescope:2022xnr}%
  \BibitemOpen
  \bibfield  {author} {\bibinfo {author} {\bibfnamefont {Kazunori}\
  \bibnamefont {Akiyama}} \emph {et~al.} (\bibinfo {collaboration} {Event
  Horizon Telescope}),\ }\bibfield  {title} {\enquote {\bibinfo {title} {{First
  Sagittarius A* Event Horizon Telescope Results. I. The Shadow of the
  Supermassive Black Hole in the Center of the Milky Way}},}\ }\href {\doibase
  10.3847/2041-8213/ac6674} {\bibfield  {journal} {\bibinfo  {journal}
  {Astrophys. J. Lett.}\ }\textbf {\bibinfo {volume} {930}},\ \bibinfo {pages}
  {L12} (\bibinfo {year} {2022})}\BibitemShut {NoStop}%
\bibitem [{\citenamefont {Kocherlakota}\ \emph
  {et~al.}(2021{\natexlab{a}})\citenamefont {Kocherlakota} \emph
  {et~al.}}]{EventHorizonTelescope:2021dqv}%
  \BibitemOpen
  \bibfield  {author} {\bibinfo {author} {\bibfnamefont {Prashant}\
  \bibnamefont {Kocherlakota}} \emph {et~al.} (\bibinfo {collaboration} {Event
  Horizon Telescope}),\ }\bibfield  {title} {\enquote {\bibinfo {title}
  {{Constraints on black-hole charges with the 2017 EHT observations of
  M87*}},}\ }\href {\doibase 10.1103/PhysRevD.103.104047} {\bibfield  {journal}
  {\bibinfo  {journal} {Phys. Rev. D}\ }\textbf {\bibinfo {volume} {103}},\
  \bibinfo {pages} {104047} (\bibinfo {year} {2021}{\natexlab{a}})},\ \Eprint
  {http://arxiv.org/abs/2105.09343} {arXiv:2105.09343 [gr-qc]} \BibitemShut
  {NoStop}%
\bibitem [{\citenamefont {Vagnozzi}\ \emph {et~al.}(2023)\citenamefont
  {Vagnozzi}, \citenamefont {Roy}, \citenamefont {Tsai}, \citenamefont
  {Visinelli}, \citenamefont {Afrin}, \citenamefont {Allahyari}, \citenamefont
  {Bambhaniya}, \citenamefont {Dey}, \citenamefont {Ghosh}, \citenamefont
  {Joshi}, \citenamefont {Jusufi}, \citenamefont {Khodadi}, \citenamefont
  {Walia}, \citenamefont {{\"O}vg{\"u}n},\ and\ \citenamefont
  {Bambi}}]{Vagnozzi:2022moj}%
  \BibitemOpen
  \bibfield  {author} {\bibinfo {author} {\bibfnamefont {Sunny}\ \bibnamefont
  {Vagnozzi}}, \bibinfo {author} {\bibfnamefont {Rittick}\ \bibnamefont {Roy}},
  \bibinfo {author} {\bibfnamefont {Yu-Dai}\ \bibnamefont {Tsai}}, \bibinfo
  {author} {\bibfnamefont {Luca}\ \bibnamefont {Visinelli}}, \bibinfo {author}
  {\bibfnamefont {Misba}\ \bibnamefont {Afrin}}, \bibinfo {author}
  {\bibfnamefont {Alireza}\ \bibnamefont {Allahyari}}, \bibinfo {author}
  {\bibfnamefont {Parth}\ \bibnamefont {Bambhaniya}}, \bibinfo {author}
  {\bibfnamefont {Dipanjan}\ \bibnamefont {Dey}}, \bibinfo {author}
  {\bibfnamefont {Sushant~G}\ \bibnamefont {Ghosh}}, \bibinfo {author}
  {\bibfnamefont {Pankaj~S.}\ \bibnamefont {Joshi}}, \bibinfo {author}
  {\bibfnamefont {Kimet}\ \bibnamefont {Jusufi}}, \bibinfo {author}
  {\bibfnamefont {Mohsen}\ \bibnamefont {Khodadi}}, \bibinfo {author}
  {\bibfnamefont {Rahul~Kumar}\ \bibnamefont {Walia}}, \bibinfo {author}
  {\bibfnamefont {Ali}\ \bibnamefont {{\"O}vg{\"u}n}}, \ and\ \bibinfo {author}
  {\bibfnamefont {Cosimo}\ \bibnamefont {Bambi}},\ }\bibfield  {title}
  {\enquote {\bibinfo {title} {Horizon-scale tests of gravity theories and
  fundamental physics from the event horizon telescope image of sagittarius
  a*},}\ }\href@noop {} {\bibfield  {journal} {\bibinfo  {journal} {Classical
  and Quantum Gravity}\ } (\bibinfo {year} {2023})},\ \Eprint
  {http://arxiv.org/abs/2205.07787} {arXiv:2205.07787 [gr-qc]} \BibitemShut
  {NoStop}%
\bibitem [{\citenamefont {Penrose}(1965)}]{Penrose:1964wq}%
  \BibitemOpen
  \bibfield  {author} {\bibinfo {author} {\bibfnamefont {Roger}\ \bibnamefont
  {Penrose}},\ }\bibfield  {title} {\enquote {\bibinfo {title} {{Gravitational
  collapse and space-time singularities}},}\ }\href {\doibase
  10.1103/PhysRevLett.14.57} {\bibfield  {journal} {\bibinfo  {journal} {Phys.
  Rev. Lett.}\ }\textbf {\bibinfo {volume} {14}},\ \bibinfo {pages} {57--59}
  (\bibinfo {year} {1965})}\BibitemShut {NoStop}%
\bibitem [{\citenamefont {Hawking}(1975)}]{Hawking:1975vcx}%
  \BibitemOpen
  \bibfield  {author} {\bibinfo {author} {\bibfnamefont {S.~W.}\ \bibnamefont
  {Hawking}},\ }\bibfield  {title} {\enquote {\bibinfo {title} {{Particle
  Creation by Black Holes}},}\ }\href {\doibase 10.1007/BF02345020} {\bibfield
  {journal} {\bibinfo  {journal} {Commun. Math. Phys.}\ }\textbf {\bibinfo
  {volume} {43}},\ \bibinfo {pages} {199--220} (\bibinfo {year} {1975})},\
  \bibinfo {note} {[Erratum: Commun.Math.Phys. 46, 206 (1976)]}\BibitemShut
  {NoStop}%
\bibitem [{\citenamefont {Bardeen}(1968)}]{bardeen1968proceedings}%
  \BibitemOpen
  \bibfield  {author} {\bibinfo {author} {\bibfnamefont {James~M}\ \bibnamefont
  {Bardeen}},\ }\bibfield  {title} {\enquote {\bibinfo {title} {Non-singular
  general-relativistic gravitational collapse},}\ }\href@noop {} {\bibfield
  {journal} {\bibinfo  {journal} {Proceedings of International Conference GR5
  (Tbilisi, USSR, 1968)}\ ,\ \bibinfo {pages} {174}} (\bibinfo {year}
  {1968})}\BibitemShut {NoStop}%
\bibitem [{\citenamefont {Carleo}\ \emph {et~al.}(2022)\citenamefont {Carleo},
  \citenamefont {Lambiase},\ and\ \citenamefont {\"Ovg\"un}}]{Carleo:2022ukm}%
  \BibitemOpen
  \bibfield  {author} {\bibinfo {author} {\bibfnamefont {Amodio}\ \bibnamefont
  {Carleo}}, \bibinfo {author} {\bibfnamefont {Gaetano}\ \bibnamefont
  {Lambiase}}, \ and\ \bibinfo {author} {\bibfnamefont {Ali}\ \bibnamefont
  {\"Ovg\"un}},\ }\bibfield  {title} {\enquote {\bibinfo {title} {{Non-linear
  Electrodynamics in Blandford-Znajeck Energy Extraction}},}\ }\href {\doibase
  10.1002/andp.202200635} {\  (\bibinfo {year} {2022}),\
  10.1002/andp.202200635},\ \Eprint {http://arxiv.org/abs/2210.11162}
  {arXiv:2210.11162 [gr-qc]} \BibitemShut {NoStop}%
\bibitem [{\citenamefont {Hayward}(2006)}]{Hayward:2005gi}%
  \BibitemOpen
  \bibfield  {author} {\bibinfo {author} {\bibfnamefont {Sean~A.}\ \bibnamefont
  {Hayward}},\ }\bibfield  {title} {\enquote {\bibinfo {title} {{Formation and
  evaporation of regular black holes}},}\ }\href {\doibase
  10.1103/PhysRevLett.96.031103} {\bibfield  {journal} {\bibinfo  {journal}
  {Phys. Rev. Lett.}\ }\textbf {\bibinfo {volume} {96}},\ \bibinfo {pages}
  {031103} (\bibinfo {year} {2006})},\ \Eprint
  {http://arxiv.org/abs/gr-qc/0506126} {arXiv:gr-qc/0506126} \BibitemShut
  {NoStop}%
\bibitem [{\citenamefont {Luminet}(1979)}]{Luminet:1979nyg}%
  \BibitemOpen
  \bibfield  {author} {\bibinfo {author} {\bibfnamefont {J.~P.}\ \bibnamefont
  {Luminet}},\ }\bibfield  {title} {\enquote {\bibinfo {title} {{Image of a
  spherical black hole with thin accretion disk}},}\ }\href@noop {} {\bibfield
  {journal} {\bibinfo  {journal} {Astron. Astrophys.}\ }\textbf {\bibinfo
  {volume} {75}},\ \bibinfo {pages} {228--235} (\bibinfo {year}
  {1979})}\BibitemShut {NoStop}%
\bibitem [{\citenamefont {Falcke}\ \emph {et~al.}(2000)\citenamefont {Falcke},
  \citenamefont {Melia},\ and\ \citenamefont {Agol}}]{Falcke:1999pj}%
  \BibitemOpen
  \bibfield  {author} {\bibinfo {author} {\bibfnamefont {Heino}\ \bibnamefont
  {Falcke}}, \bibinfo {author} {\bibfnamefont {Fulvio}\ \bibnamefont {Melia}},
  \ and\ \bibinfo {author} {\bibfnamefont {Eric}\ \bibnamefont {Agol}},\
  }\bibfield  {title} {\enquote {\bibinfo {title} {{Viewing the shadow of the
  black hole at the galactic center}},}\ }\href {\doibase 10.1086/312423}
  {\bibfield  {journal} {\bibinfo  {journal} {Astrophys. J. Lett.}\ }\textbf
  {\bibinfo {volume} {528}},\ \bibinfo {pages} {L13} (\bibinfo {year}
  {2000})},\ \Eprint {http://arxiv.org/abs/astro-ph/9912263}
  {arXiv:astro-ph/9912263} \BibitemShut {NoStop}%
\bibitem [{\citenamefont {Bronzwaer}\ and\ \citenamefont
  {Falcke}(2021)}]{Bronzwaer:2021lzo}%
  \BibitemOpen
  \bibfield  {author} {\bibinfo {author} {\bibfnamefont {Thomas}\ \bibnamefont
  {Bronzwaer}}\ and\ \bibinfo {author} {\bibfnamefont {Heino}\ \bibnamefont
  {Falcke}},\ }\bibfield  {title} {\enquote {\bibinfo {title} {{The Nature of
  Black Hole Shadows}},}\ }\href {\doibase 10.3847/1538-4357/ac1738} {\bibfield
   {journal} {\bibinfo  {journal} {Astrophys. J.}\ }\textbf {\bibinfo {volume}
  {920}},\ \bibinfo {pages} {155} (\bibinfo {year} {2021})},\ \Eprint
  {http://arxiv.org/abs/2108.03966} {arXiv:2108.03966 [astro-ph.HE]}
  \BibitemShut {NoStop}%
\bibitem [{\citenamefont {Narayan}\ \emph {et~al.}(2019)\citenamefont
  {Narayan}, \citenamefont {Johnson},\ and\ \citenamefont
  {Gammie}}]{Narayan:2019imo}%
  \BibitemOpen
  \bibfield  {author} {\bibinfo {author} {\bibfnamefont {Ramesh}\ \bibnamefont
  {Narayan}}, \bibinfo {author} {\bibfnamefont {Michael~D.}\ \bibnamefont
  {Johnson}}, \ and\ \bibinfo {author} {\bibfnamefont {Charles~F.}\
  \bibnamefont {Gammie}},\ }\bibfield  {title} {\enquote {\bibinfo {title}
  {{The Shadow of a Spherically Accreting Black Hole}},}\ }\href {\doibase
  10.3847/2041-8213/ab518c} {\bibfield  {journal} {\bibinfo  {journal}
  {Astrophys. J. Lett.}\ }\textbf {\bibinfo {volume} {885}},\ \bibinfo {pages}
  {L33} (\bibinfo {year} {2019})},\ \Eprint {http://arxiv.org/abs/1910.02957}
  {arXiv:1910.02957 [astro-ph.HE]} \BibitemShut {NoStop}%
\bibitem [{\citenamefont {Pantig}\ \emph {et~al.}(2023)\citenamefont {Pantig},
  \citenamefont {\"Ovg\"un},\ and\ \citenamefont {Demir}}]{Symmergent-bh3}%
  \BibitemOpen
  \bibfield  {author} {\bibinfo {author} {\bibfnamefont {Reggie~C.}\
  \bibnamefont {Pantig}}, \bibinfo {author} {\bibfnamefont {Ali}\ \bibnamefont
  {\"Ovg\"un}}, \ and\ \bibinfo {author} {\bibfnamefont {Durmu\c{s}}\
  \bibnamefont {Demir}},\ }\bibfield  {title} {\enquote {\bibinfo {title}
  {{Testing symmergent gravity through the shadow image and weak field photon
  deflection by a rotating black hole using the M87$^*$ and Sgr. $\hbox {A}^*$
  results}},}\ }\href {\doibase 10.1140/epjc/s10052-023-11400-6} {\bibfield
  {journal} {\bibinfo  {journal} {Eur. Phys. J. C}\ }\textbf {\bibinfo {volume}
  {83}},\ \bibinfo {pages} {250} (\bibinfo {year} {2023})},\ \Eprint
  {http://arxiv.org/abs/2208.02969} {arXiv:2208.02969 [gr-qc]} \BibitemShut
  {NoStop}%
\bibitem [{\citenamefont {\c{C}imdiker}\ \emph {et~al.}(2021)\citenamefont
  {\c{C}imdiker}, \citenamefont {Demir},\ and\ \citenamefont
  {\"Ovg\"un}}]{Symmergent-bh}%
  \BibitemOpen
  \bibfield  {author} {\bibinfo {author} {\bibfnamefont {\.Irfan}\ \bibnamefont
  {\c{C}imdiker}}, \bibinfo {author} {\bibfnamefont {Durmu\c{s}}\ \bibnamefont
  {Demir}}, \ and\ \bibinfo {author} {\bibfnamefont {Ali}\ \bibnamefont
  {\"Ovg\"un}},\ }\bibfield  {title} {\enquote {\bibinfo {title} {{Black hole
  shadow in symmergent gravity}},}\ }\href {\doibase
  10.1016/j.dark.2021.100900} {\bibfield  {journal} {\bibinfo  {journal} {Phys.
  Dark Univ.}\ }\textbf {\bibinfo {volume} {34}},\ \bibinfo {pages} {100900}
  (\bibinfo {year} {2021})},\ \Eprint {http://arxiv.org/abs/2110.11904}
  {arXiv:2110.11904 [gr-qc]} \BibitemShut {NoStop}%
\bibitem [{\citenamefont {Rayimbaev}\ \emph {et~al.}(2023)\citenamefont
  {Rayimbaev}, \citenamefont {Pantig}, \citenamefont {\"Ovg\"un}, \citenamefont
  {Abdujabbarov},\ and\ \citenamefont {Demir}}]{Symmergent-bh2}%
  \BibitemOpen
  \bibfield  {author} {\bibinfo {author} {\bibfnamefont {Javlon}\ \bibnamefont
  {Rayimbaev}}, \bibinfo {author} {\bibfnamefont {Reggie~C.}\ \bibnamefont
  {Pantig}}, \bibinfo {author} {\bibfnamefont {Ali}\ \bibnamefont {\"Ovg\"un}},
  \bibinfo {author} {\bibfnamefont {Ahmadjon}\ \bibnamefont {Abdujabbarov}}, \
  and\ \bibinfo {author} {\bibfnamefont {Durmu\c{s}}\ \bibnamefont {Demir}},\
  }\bibfield  {title} {\enquote {\bibinfo {title} {{Quasiperiodic oscillations,
  weak field lensing and shadow cast around black holes in Symmergent
  gravity}},}\ }\href {\doibase 10.1016/j.aop.2023.169335} {\bibfield
  {journal} {\bibinfo  {journal} {Annals of Physics}\ } (\bibinfo {year}
  {2023}),\ 10.1016/j.aop.2023.169335},\ \Eprint
  {http://arxiv.org/abs/2206.06599} {arXiv:2206.06599 [gr-qc]} \BibitemShut
  {NoStop}%
\bibitem [{\citenamefont {Ghosh}\ \emph {et~al.}(2021)\citenamefont {Ghosh},
  \citenamefont {Kumar},\ and\ \citenamefont {Islam}}]{Ghosh:2020spb}%
  \BibitemOpen
  \bibfield  {author} {\bibinfo {author} {\bibfnamefont {Sushant~G.}\
  \bibnamefont {Ghosh}}, \bibinfo {author} {\bibfnamefont {Rahul}\ \bibnamefont
  {Kumar}}, \ and\ \bibinfo {author} {\bibfnamefont {Shafqat~Ul}\ \bibnamefont
  {Islam}},\ }\bibfield  {title} {\enquote {\bibinfo {title} {{Parameters
  estimation and strong gravitational lensing of nonsingular Kerr-Sen black
  holes}},}\ }\href {\doibase 10.1088/1475-7516/2021/03/056} {\bibfield
  {journal} {\bibinfo  {journal} {JCAP}\ }\textbf {\bibinfo {volume} {03}},\
  \bibinfo {pages} {056} (\bibinfo {year} {2021})},\ \Eprint
  {http://arxiv.org/abs/2011.08023} {arXiv:2011.08023 [gr-qc]} \BibitemShut
  {NoStop}%
\bibitem [{\citenamefont {Allahyari}\ \emph {et~al.}(2020)\citenamefont
  {Allahyari}, \citenamefont {Khodadi}, \citenamefont {Vagnozzi},\ and\
  \citenamefont {Mota}}]{Allahyari:2019jqz}%
  \BibitemOpen
  \bibfield  {author} {\bibinfo {author} {\bibfnamefont {Alireza}\ \bibnamefont
  {Allahyari}}, \bibinfo {author} {\bibfnamefont {Mohsen}\ \bibnamefont
  {Khodadi}}, \bibinfo {author} {\bibfnamefont {Sunny}\ \bibnamefont
  {Vagnozzi}}, \ and\ \bibinfo {author} {\bibfnamefont {David~F.}\ \bibnamefont
  {Mota}},\ }\bibfield  {title} {\enquote {\bibinfo {title} {{Magnetically
  charged black holes from non-linear electrodynamics and the Event Horizon
  Telescope}},}\ }\href {\doibase 10.1088/1475-7516/2020/02/003} {\bibfield
  {journal} {\bibinfo  {journal} {JCAP}\ }\textbf {\bibinfo {volume} {02}},\
  \bibinfo {pages} {003} (\bibinfo {year} {2020})},\ \Eprint
  {http://arxiv.org/abs/1912.08231} {arXiv:1912.08231 [gr-qc]} \BibitemShut
  {NoStop}%
\bibitem [{\citenamefont {Bambi}\ \emph {et~al.}(2019)\citenamefont {Bambi},
  \citenamefont {Freese}, \citenamefont {Vagnozzi},\ and\ \citenamefont
  {Visinelli}}]{Bambi:2019tjh}%
  \BibitemOpen
  \bibfield  {author} {\bibinfo {author} {\bibfnamefont {Cosimo}\ \bibnamefont
  {Bambi}}, \bibinfo {author} {\bibfnamefont {Katherine}\ \bibnamefont
  {Freese}}, \bibinfo {author} {\bibfnamefont {Sunny}\ \bibnamefont
  {Vagnozzi}}, \ and\ \bibinfo {author} {\bibfnamefont {Luca}\ \bibnamefont
  {Visinelli}},\ }\bibfield  {title} {\enquote {\bibinfo {title} {{Testing the
  rotational nature of the supermassive object M87* from the circularity and
  size of its first image}},}\ }\href {\doibase 10.1103/PhysRevD.100.044057}
  {\bibfield  {journal} {\bibinfo  {journal} {Phys. Rev. D}\ }\textbf {\bibinfo
  {volume} {100}},\ \bibinfo {pages} {044057} (\bibinfo {year} {2019})},\
  \Eprint {http://arxiv.org/abs/1904.12983} {arXiv:1904.12983 [gr-qc]}
  \BibitemShut {NoStop}%
\bibitem [{\citenamefont {Kocherlakota}\ \emph
  {et~al.}(2021{\natexlab{b}})\citenamefont {Kocherlakota}, \citenamefont
  {Rezzolla}, \citenamefont {Falcke} \emph {et~al.}}]{Prashant2021}%
  \BibitemOpen
  \bibfield  {author} {\bibinfo {author} {\bibfnamefont {Prashant}\
  \bibnamefont {Kocherlakota}}, \bibinfo {author} {\bibfnamefont {Luciano}\
  \bibnamefont {Rezzolla}}, \bibinfo {author} {\bibfnamefont {Heino}\
  \bibnamefont {Falcke}},  \emph {et~al.} (\bibinfo {collaboration} {EHT
  Collaboration}),\ }\bibfield  {title} {\enquote {\bibinfo {title}
  {Constraints on black-hole charges with the 2017 eht observations of m87*},}\
  }\href {\doibase 10.1103/PhysRevD.103.104047} {\bibfield  {journal} {\bibinfo
   {journal} {Phys. Rev. D}\ }\textbf {\bibinfo {volume} {103}},\ \bibinfo
  {pages} {104047} (\bibinfo {year} {2021}{\natexlab{b}})}\BibitemShut
  {NoStop}%
\bibitem [{\citenamefont {\"Ovg\"un}\ \emph {et~al.}(2018)\citenamefont
  {\"Ovg\"un}, \citenamefont {Sakall\i{}},\ and\ \citenamefont
  {Saavedra}}]{Ovgun:2018tua}%
  \BibitemOpen
  \bibfield  {author} {\bibinfo {author} {\bibfnamefont {Ali}\ \bibnamefont
  {\"Ovg\"un}}, \bibinfo {author} {\bibfnamefont {\.Izzet}\ \bibnamefont
  {Sakall\i{}}}, \ and\ \bibinfo {author} {\bibfnamefont {Joel}\ \bibnamefont
  {Saavedra}},\ }\bibfield  {title} {\enquote {\bibinfo {title} {{Shadow cast
  and Deflection angle of Kerr-Newman-Kasuya spacetime}},}\ }\href {\doibase
  10.1088/1475-7516/2018/10/041} {\bibfield  {journal} {\bibinfo  {journal}
  {JCAP}\ }\textbf {\bibinfo {volume} {10}},\ \bibinfo {pages} {041} (\bibinfo
  {year} {2018})},\ \Eprint {http://arxiv.org/abs/1807.00388} {arXiv:1807.00388
  [gr-qc]} \BibitemShut {NoStop}%
\bibitem [{\citenamefont {\"Ovg\"un}\ and\ \citenamefont
  {Sakall\i{}}(2020)}]{Ovgun:2020gjz}%
  \BibitemOpen
  \bibfield  {author} {\bibinfo {author} {\bibfnamefont {Ali}\ \bibnamefont
  {\"Ovg\"un}}\ and\ \bibinfo {author} {\bibfnamefont {\.Izzet}\ \bibnamefont
  {Sakall\i{}}},\ }\bibfield  {title} {\enquote {\bibinfo {title} {{Testing
  generalized Einstein\textendash{}Cartan\textendash{}Kibble\textendash{}Sciama
  gravity using weak deflection angle and shadow cast}},}\ }\href {\doibase
  10.1088/1361-6382/abb579} {\bibfield  {journal} {\bibinfo  {journal} {Class.
  Quant. Grav.}\ }\textbf {\bibinfo {volume} {37}},\ \bibinfo {pages} {225003}
  (\bibinfo {year} {2020})},\ \Eprint {http://arxiv.org/abs/2005.00982}
  {arXiv:2005.00982 [gr-qc]} \BibitemShut {NoStop}%
\bibitem [{\citenamefont {\"Ovg\"un}\ \emph {et~al.}(2020)\citenamefont
  {\"Ovg\"un}, \citenamefont {Sakall\i{}}, \citenamefont {Saavedra},\ and\
  \citenamefont {Leiva}}]{Ovgun:2019jdo}%
  \BibitemOpen
  \bibfield  {author} {\bibinfo {author} {\bibfnamefont {Ali}\ \bibnamefont
  {\"Ovg\"un}}, \bibinfo {author} {\bibfnamefont {\.Izzet}\ \bibnamefont
  {Sakall\i{}}}, \bibinfo {author} {\bibfnamefont {Joel}\ \bibnamefont
  {Saavedra}}, \ and\ \bibinfo {author} {\bibfnamefont {Carlos}\ \bibnamefont
  {Leiva}},\ }\bibfield  {title} {\enquote {\bibinfo {title} {{Shadow cast of
  noncommutative black holes in Rastall gravity}},}\ }\href {\doibase
  10.1142/S0217732320501631} {\bibfield  {journal} {\bibinfo  {journal} {Mod.
  Phys. Lett. A}\ }\textbf {\bibinfo {volume} {35}},\ \bibinfo {pages}
  {2050163} (\bibinfo {year} {2020})},\ \Eprint
  {http://arxiv.org/abs/1906.05954} {arXiv:1906.05954 [hep-th]} \BibitemShut
  {NoStop}%
\bibitem [{\citenamefont {Kuang}\ and\ \citenamefont
  {\"Ovg\"un}(2022)}]{Kuang:2022xjp}%
  \BibitemOpen
  \bibfield  {author} {\bibinfo {author} {\bibfnamefont {Xiao-Mei}\
  \bibnamefont {Kuang}}\ and\ \bibinfo {author} {\bibfnamefont {Ali}\
  \bibnamefont {\"Ovg\"un}},\ }\bibfield  {title} {\enquote {\bibinfo {title}
  {{Strong gravitational lensing and shadow constraint from M87* of slowly
  rotating Kerr-like black hole}},}\ }\href@noop {} {\  (\bibinfo {year}
  {2022})},\ \Eprint {http://arxiv.org/abs/2205.11003} {arXiv:2205.11003
  [gr-qc]} \BibitemShut {NoStop}%
\bibitem [{\citenamefont {Kumaran}\ and\ \citenamefont
  {\"Ovg\"un}(2022)}]{Kumaran:2022soh}%
  \BibitemOpen
  \bibfield  {author} {\bibinfo {author} {\bibfnamefont {Yashmitha}\
  \bibnamefont {Kumaran}}\ and\ \bibinfo {author} {\bibfnamefont {Ali}\
  \bibnamefont {\"Ovg\"un}},\ }\bibfield  {title} {\enquote {\bibinfo {title}
  {{Deflection Angle and Shadow of the Reissner\textendash{}Nordstr\"om Black
  Hole with Higher-Order Magnetic Correction in Einstein-Nonlinear-Maxwell
  Fields}},}\ }\href {\doibase 10.3390/sym14102054} {\bibfield  {journal}
  {\bibinfo  {journal} {Symmetry}\ }\textbf {\bibinfo {volume} {14}},\ \bibinfo
  {pages} {2054} (\bibinfo {year} {2022})},\ \Eprint
  {http://arxiv.org/abs/2210.00468} {arXiv:2210.00468 [gr-qc]} \BibitemShut
  {NoStop}%
\bibitem [{\citenamefont {Mustafa}\ \emph {et~al.}(2022)\citenamefont
  {Mustafa}, \citenamefont {Atamurotov}, \citenamefont {Hussain}, \citenamefont
  {Shaymatov},\ and\ \citenamefont {\"Ovg\"un}}]{Mustafa:2022xod}%
  \BibitemOpen
  \bibfield  {author} {\bibinfo {author} {\bibfnamefont {Ghulam}\ \bibnamefont
  {Mustafa}}, \bibinfo {author} {\bibfnamefont {Farruh}\ \bibnamefont
  {Atamurotov}}, \bibinfo {author} {\bibfnamefont {Ibrar}\ \bibnamefont
  {Hussain}}, \bibinfo {author} {\bibfnamefont {Sanjar}\ \bibnamefont
  {Shaymatov}}, \ and\ \bibinfo {author} {\bibfnamefont {Ali}\ \bibnamefont
  {\"Ovg\"un}},\ }\bibfield  {title} {\enquote {\bibinfo {title} {{Shadows and
  gravitational weak lensing by the Schwarzschild black hole in the string
  cloud background with quintessential field*}},}\ }\href {\doibase
  10.1088/1674-1137/ac917f} {\bibfield  {journal} {\bibinfo  {journal} {Chin.
  Phys. C}\ }\textbf {\bibinfo {volume} {46}},\ \bibinfo {pages} {125107}
  (\bibinfo {year} {2022})},\ \Eprint {http://arxiv.org/abs/2207.07608}
  {arXiv:2207.07608 [gr-qc]} \BibitemShut {NoStop}%
\bibitem [{\citenamefont {Okyay}\ and\ \citenamefont
  {\"Ovg\"un}(2022)}]{Okyay:2021nnh}%
  \BibitemOpen
  \bibfield  {author} {\bibinfo {author} {\bibfnamefont {Mert}\ \bibnamefont
  {Okyay}}\ and\ \bibinfo {author} {\bibfnamefont {Ali}\ \bibnamefont
  {\"Ovg\"un}},\ }\bibfield  {title} {\enquote {\bibinfo {title} {{Nonlinear
  electrodynamics effects on the black hole shadow, deflection angle,
  quasinormal modes and greybody factors}},}\ }\href {\doibase
  10.1088/1475-7516/2022/01/009} {\bibfield  {journal} {\bibinfo  {journal}
  {JCAP}\ }\textbf {\bibinfo {volume} {01}},\ \bibinfo {pages} {009} (\bibinfo
  {year} {2022})},\ \Eprint {http://arxiv.org/abs/2108.07766} {arXiv:2108.07766
  [gr-qc]} \BibitemShut {NoStop}%
\bibitem [{\citenamefont {Atamurotov}\ \emph {et~al.}(2023)\citenamefont
  {Atamurotov}, \citenamefont {Hussain}, \citenamefont {Mustafa},\ and\
  \citenamefont {\"Ovg\"un}}]{Atamurotov:2022knb}%
  \BibitemOpen
  \bibfield  {author} {\bibinfo {author} {\bibfnamefont {Farruh}\ \bibnamefont
  {Atamurotov}}, \bibinfo {author} {\bibfnamefont {Ibrar}\ \bibnamefont
  {Hussain}}, \bibinfo {author} {\bibfnamefont {Ghulam}\ \bibnamefont
  {Mustafa}}, \ and\ \bibinfo {author} {\bibfnamefont {Ali}\ \bibnamefont
  {\"Ovg\"un}},\ }\bibfield  {title} {\enquote {\bibinfo {title} {{Weak
  deflection angle and shadow cast by the charged-Kiselev black hole with cloud
  of strings in plasma*}},}\ }\href {\doibase 10.1088/1674-1137/ac9fbb}
  {\bibfield  {journal} {\bibinfo  {journal} {Chin. Phys. C}\ }\textbf
  {\bibinfo {volume} {47}},\ \bibinfo {pages} {025102} (\bibinfo {year}
  {2023})}\BibitemShut {NoStop}%
\bibitem [{\citenamefont {Abdikamalov}\ \emph {et~al.}(2019)\citenamefont
  {Abdikamalov}, \citenamefont {Abdujabbarov}, \citenamefont {Ayzenberg},
  \citenamefont {Malafarina}, \citenamefont {Bambi},\ and\ \citenamefont
  {Ahmedov}}]{Abdikamalov:2019ztb}%
  \BibitemOpen
  \bibfield  {author} {\bibinfo {author} {\bibfnamefont {Askar~B.}\
  \bibnamefont {Abdikamalov}}, \bibinfo {author} {\bibfnamefont {Ahmadjon~A.}\
  \bibnamefont {Abdujabbarov}}, \bibinfo {author} {\bibfnamefont {Dimitry}\
  \bibnamefont {Ayzenberg}}, \bibinfo {author} {\bibfnamefont {Daniele}\
  \bibnamefont {Malafarina}}, \bibinfo {author} {\bibfnamefont {Cosimo}\
  \bibnamefont {Bambi}}, \ and\ \bibinfo {author} {\bibfnamefont {Bobomurat}\
  \bibnamefont {Ahmedov}},\ }\bibfield  {title} {\enquote {\bibinfo {title}
  {{Black hole mimicker hiding in the shadow: Optical properties of the
  $\gamma$ metric}},}\ }\href {\doibase 10.1103/PhysRevD.100.024014} {\bibfield
   {journal} {\bibinfo  {journal} {Phys. Rev. D}\ }\textbf {\bibinfo {volume}
  {100}},\ \bibinfo {pages} {024014} (\bibinfo {year} {2019})},\ \Eprint
  {http://arxiv.org/abs/1904.06207} {arXiv:1904.06207 [gr-qc]} \BibitemShut
  {NoStop}%
\bibitem [{\citenamefont {Abdujabbarov}\ \emph {et~al.}(2016)\citenamefont
  {Abdujabbarov}, \citenamefont {Juraev}, \citenamefont {Ahmedov},\ and\
  \citenamefont {Stuchl\'\i{}k}}]{Abdujabbarov:2016efm}%
  \BibitemOpen
  \bibfield  {author} {\bibinfo {author} {\bibfnamefont {Ahmadjon}\
  \bibnamefont {Abdujabbarov}}, \bibinfo {author} {\bibfnamefont {Bakhtinur}\
  \bibnamefont {Juraev}}, \bibinfo {author} {\bibfnamefont {Bobomurat}\
  \bibnamefont {Ahmedov}}, \ and\ \bibinfo {author} {\bibfnamefont
  {Zden\v{e}k}\ \bibnamefont {Stuchl\'\i{}k}},\ }\bibfield  {title} {\enquote
  {\bibinfo {title} {{Shadow of rotating wormhole in plasma environment}},}\
  }\href {\doibase 10.1007/s10509-016-2818-9} {\bibfield  {journal} {\bibinfo
  {journal} {Astrophys. Space Sci.}\ }\textbf {\bibinfo {volume} {361}},\
  \bibinfo {pages} {226} (\bibinfo {year} {2016})}\BibitemShut {NoStop}%
\bibitem [{\citenamefont {Atamurotov}\ and\ \citenamefont
  {Ahmedov}(2015)}]{Atamurotov:2015nra}%
  \BibitemOpen
  \bibfield  {author} {\bibinfo {author} {\bibfnamefont {Farruh}\ \bibnamefont
  {Atamurotov}}\ and\ \bibinfo {author} {\bibfnamefont {Bobomurat}\
  \bibnamefont {Ahmedov}},\ }\bibfield  {title} {\enquote {\bibinfo {title}
  {{Optical properties of black hole in the presence of plasma: shadow}},}\
  }\href {\doibase 10.1103/PhysRevD.92.084005} {\bibfield  {journal} {\bibinfo
  {journal} {Phys. Rev. D}\ }\textbf {\bibinfo {volume} {92}},\ \bibinfo
  {pages} {084005} (\bibinfo {year} {2015})},\ \Eprint
  {http://arxiv.org/abs/1507.08131} {arXiv:1507.08131 [gr-qc]} \BibitemShut
  {NoStop}%
\bibitem [{\citenamefont {Papnoi}\ \emph {et~al.}(2014)\citenamefont {Papnoi},
  \citenamefont {Atamurotov}, \citenamefont {Ghosh},\ and\ \citenamefont
  {Ahmedov}}]{Papnoi:2014aaa}%
  \BibitemOpen
  \bibfield  {author} {\bibinfo {author} {\bibfnamefont {Uma}\ \bibnamefont
  {Papnoi}}, \bibinfo {author} {\bibfnamefont {Farruh}\ \bibnamefont
  {Atamurotov}}, \bibinfo {author} {\bibfnamefont {Sushant~G.}\ \bibnamefont
  {Ghosh}}, \ and\ \bibinfo {author} {\bibfnamefont {Bobomurat}\ \bibnamefont
  {Ahmedov}},\ }\bibfield  {title} {\enquote {\bibinfo {title} {{Shadow of
  five-dimensional rotating Myers-Perry black hole}},}\ }\href {\doibase
  10.1103/PhysRevD.90.024073} {\bibfield  {journal} {\bibinfo  {journal} {Phys.
  Rev. D}\ }\textbf {\bibinfo {volume} {90}},\ \bibinfo {pages} {024073}
  (\bibinfo {year} {2014})},\ \Eprint {http://arxiv.org/abs/1407.0834}
  {arXiv:1407.0834 [gr-qc]} \BibitemShut {NoStop}%
\bibitem [{\citenamefont {Abdujabbarov}\ \emph {et~al.}(2013)\citenamefont
  {Abdujabbarov}, \citenamefont {Atamurotov}, \citenamefont {Kucukakca},
  \citenamefont {Ahmedov},\ and\ \citenamefont {Camci}}]{Abdujabbarov:2012bn}%
  \BibitemOpen
  \bibfield  {author} {\bibinfo {author} {\bibfnamefont {Ahmadjon}\
  \bibnamefont {Abdujabbarov}}, \bibinfo {author} {\bibfnamefont {Farruh}\
  \bibnamefont {Atamurotov}}, \bibinfo {author} {\bibfnamefont {Yusuf}\
  \bibnamefont {Kucukakca}}, \bibinfo {author} {\bibfnamefont {Bobomurat}\
  \bibnamefont {Ahmedov}}, \ and\ \bibinfo {author} {\bibfnamefont {Ugur}\
  \bibnamefont {Camci}},\ }\bibfield  {title} {\enquote {\bibinfo {title}
  {{Shadow of Kerr-Taub-NUT black hole}},}\ }\href {\doibase
  10.1007/s10509-012-1337-6} {\bibfield  {journal} {\bibinfo  {journal}
  {Astrophys. Space Sci.}\ }\textbf {\bibinfo {volume} {344}},\ \bibinfo
  {pages} {429--435} (\bibinfo {year} {2013})},\ \Eprint
  {http://arxiv.org/abs/1212.4949} {arXiv:1212.4949 [physics.gen-ph]}
  \BibitemShut {NoStop}%
\bibitem [{\citenamefont {Atamurotov}\ \emph {et~al.}(2013)\citenamefont
  {Atamurotov}, \citenamefont {Abdujabbarov},\ and\ \citenamefont
  {Ahmedov}}]{Atamurotov:2013sca}%
  \BibitemOpen
  \bibfield  {author} {\bibinfo {author} {\bibfnamefont {Farruh}\ \bibnamefont
  {Atamurotov}}, \bibinfo {author} {\bibfnamefont {Ahmadjon}\ \bibnamefont
  {Abdujabbarov}}, \ and\ \bibinfo {author} {\bibfnamefont {Bobomurat}\
  \bibnamefont {Ahmedov}},\ }\bibfield  {title} {\enquote {\bibinfo {title}
  {{Shadow of rotating non-Kerr black hole}},}\ }\href {\doibase
  10.1103/PhysRevD.88.064004} {\bibfield  {journal} {\bibinfo  {journal} {Phys.
  Rev. D}\ }\textbf {\bibinfo {volume} {88}},\ \bibinfo {pages} {064004}
  (\bibinfo {year} {2013})}\BibitemShut {NoStop}%
\bibitem [{\citenamefont {Cunha}\ and\ \citenamefont
  {Herdeiro}(2018)}]{Cunha:2018acu}%
  \BibitemOpen
  \bibfield  {author} {\bibinfo {author} {\bibfnamefont {Pedro V.~P.}\
  \bibnamefont {Cunha}}\ and\ \bibinfo {author} {\bibfnamefont {Carlos A.~R.}\
  \bibnamefont {Herdeiro}},\ }\bibfield  {title} {\enquote {\bibinfo {title}
  {{Shadows and strong gravitational lensing: a brief review}},}\ }\href
  {\doibase 10.1007/s10714-018-2361-9} {\bibfield  {journal} {\bibinfo
  {journal} {Gen. Rel. Grav.}\ }\textbf {\bibinfo {volume} {50}},\ \bibinfo
  {pages} {42} (\bibinfo {year} {2018})},\ \Eprint
  {http://arxiv.org/abs/1801.00860} {arXiv:1801.00860 [gr-qc]} \BibitemShut
  {NoStop}%
\bibitem [{\citenamefont {Gralla}\ \emph {et~al.}(2019)\citenamefont {Gralla},
  \citenamefont {Holz},\ and\ \citenamefont {Wald}}]{Gralla:2019xty}%
  \BibitemOpen
  \bibfield  {author} {\bibinfo {author} {\bibfnamefont {Samuel~E.}\
  \bibnamefont {Gralla}}, \bibinfo {author} {\bibfnamefont {Daniel~E.}\
  \bibnamefont {Holz}}, \ and\ \bibinfo {author} {\bibfnamefont {Robert~M.}\
  \bibnamefont {Wald}},\ }\bibfield  {title} {\enquote {\bibinfo {title}
  {{Black Hole Shadows, Photon Rings, and Lensing Rings}},}\ }\href {\doibase
  10.1103/PhysRevD.100.024018} {\bibfield  {journal} {\bibinfo  {journal}
  {Phys. Rev. D}\ }\textbf {\bibinfo {volume} {100}},\ \bibinfo {pages}
  {024018} (\bibinfo {year} {2019})},\ \Eprint
  {http://arxiv.org/abs/1906.00873} {arXiv:1906.00873 [astro-ph.HE]}
  \BibitemShut {NoStop}%
\bibitem [{\citenamefont {Belhaj}\ \emph {et~al.}(2021)\citenamefont {Belhaj},
  \citenamefont {Belmahi}, \citenamefont {Benali}, \citenamefont {El~Hadri},
  \citenamefont {El~Moumni},\ and\ \citenamefont
  {Torrente-Lujan}}]{Belhaj:2020okh}%
  \BibitemOpen
  \bibfield  {author} {\bibinfo {author} {\bibfnamefont {A.}~\bibnamefont
  {Belhaj}}, \bibinfo {author} {\bibfnamefont {H.}~\bibnamefont {Belmahi}},
  \bibinfo {author} {\bibfnamefont {M.}~\bibnamefont {Benali}}, \bibinfo
  {author} {\bibfnamefont {W.}~\bibnamefont {El~Hadri}}, \bibinfo {author}
  {\bibfnamefont {H.}~\bibnamefont {El~Moumni}}, \ and\ \bibinfo {author}
  {\bibfnamefont {E.}~\bibnamefont {Torrente-Lujan}},\ }\bibfield  {title}
  {\enquote {\bibinfo {title} {{Shadows of 5D black holes from string
  theory}},}\ }\href {\doibase 10.1016/j.physletb.2020.136025} {\bibfield
  {journal} {\bibinfo  {journal} {Phys. Lett. B}\ }\textbf {\bibinfo {volume}
  {812}},\ \bibinfo {pages} {136025} (\bibinfo {year} {2021})},\ \Eprint
  {http://arxiv.org/abs/2008.13478} {arXiv:2008.13478 [hep-th]} \BibitemShut
  {NoStop}%
\bibitem [{\citenamefont {Belhaj}\ \emph {et~al.}(2020)\citenamefont {Belhaj},
  \citenamefont {Benali}, \citenamefont {El~Balali}, \citenamefont
  {El~Moumni},\ and\ \citenamefont {Ennadifi}}]{Belhaj:2020rdb}%
  \BibitemOpen
  \bibfield  {author} {\bibinfo {author} {\bibfnamefont {A.}~\bibnamefont
  {Belhaj}}, \bibinfo {author} {\bibfnamefont {M.}~\bibnamefont {Benali}},
  \bibinfo {author} {\bibfnamefont {A.}~\bibnamefont {El~Balali}}, \bibinfo
  {author} {\bibfnamefont {H.}~\bibnamefont {El~Moumni}}, \ and\ \bibinfo
  {author} {\bibfnamefont {S.~E.}\ \bibnamefont {Ennadifi}},\ }\bibfield
  {title} {\enquote {\bibinfo {title} {{Deflection angle and shadow behaviors
  of quintessential black holes in arbitrary dimensions}},}\ }\href {\doibase
  10.1088/1361-6382/abbaa9} {\bibfield  {journal} {\bibinfo  {journal} {Class.
  Quant. Grav.}\ }\textbf {\bibinfo {volume} {37}},\ \bibinfo {pages} {215004}
  (\bibinfo {year} {2020})},\ \Eprint {http://arxiv.org/abs/2006.01078}
  {arXiv:2006.01078 [gr-qc]} \BibitemShut {NoStop}%
\bibitem [{\citenamefont {Konoplya}(2019)}]{Konoplya2019}%
  \BibitemOpen
  \bibfield  {author} {\bibinfo {author} {\bibfnamefont {R.~A.}\ \bibnamefont
  {Konoplya}},\ }\bibfield  {title} {\enquote {\bibinfo {title} {{Shadow of a
  black hole surrounded by dark matter}},}\ }\href {\doibase
  10.1016/j.physletb.2019.05.043} {\bibfield  {journal} {\bibinfo  {journal}
  {Phys. Lett. B}\ }\textbf {\bibinfo {volume} {795}},\ \bibinfo {pages} {1--6}
  (\bibinfo {year} {2019})},\ \Eprint {http://arxiv.org/abs/1905.00064}
  {arXiv:1905.00064 [gr-qc]} \BibitemShut {NoStop}%
\bibitem [{\citenamefont {Wei}\ \emph {et~al.}(2019)\citenamefont {Wei},
  \citenamefont {Zou}, \citenamefont {Liu},\ and\ \citenamefont
  {Mann}}]{Wei2019}%
  \BibitemOpen
  \bibfield  {author} {\bibinfo {author} {\bibfnamefont {Shao-Wen}\
  \bibnamefont {Wei}}, \bibinfo {author} {\bibfnamefont {Yuan-Chuan}\
  \bibnamefont {Zou}}, \bibinfo {author} {\bibfnamefont {Yu-Xiao}\ \bibnamefont
  {Liu}}, \ and\ \bibinfo {author} {\bibfnamefont {Robert~B.}\ \bibnamefont
  {Mann}},\ }\bibfield  {title} {\enquote {\bibinfo {title} {{Curvature radius
  and Kerr black hole shadow}},}\ }\href {\doibase
  10.1088/1475-7516/2019/08/030} {\bibfield  {journal} {\bibinfo  {journal}
  {JCAP}\ }\textbf {\bibinfo {volume} {08}},\ \bibinfo {pages} {030} (\bibinfo
  {year} {2019})},\ \Eprint {http://arxiv.org/abs/1904.07710} {arXiv:1904.07710
  [gr-qc]} \BibitemShut {NoStop}%
\bibitem [{\citenamefont {Ling}\ \emph {et~al.}(2021)\citenamefont {Ling},
  \citenamefont {Guo}, \citenamefont {Liu}, \citenamefont {Kuang},\ and\
  \citenamefont {Wang}}]{Ling:2021vgk}%
  \BibitemOpen
  \bibfield  {author} {\bibinfo {author} {\bibfnamefont {Ru}~\bibnamefont
  {Ling}}, \bibinfo {author} {\bibfnamefont {Hong}\ \bibnamefont {Guo}},
  \bibinfo {author} {\bibfnamefont {Hang}\ \bibnamefont {Liu}}, \bibinfo
  {author} {\bibfnamefont {Xiao-Mei}\ \bibnamefont {Kuang}}, \ and\ \bibinfo
  {author} {\bibfnamefont {Bin}\ \bibnamefont {Wang}},\ }\bibfield  {title}
  {\enquote {\bibinfo {title} {{Shadow and near-horizon characteristics of the
  acoustic charged black hole in curved spacetime}},}\ }\href {\doibase
  10.1103/PhysRevD.104.104003} {\bibfield  {journal} {\bibinfo  {journal}
  {Phys. Rev. D}\ }\textbf {\bibinfo {volume} {104}},\ \bibinfo {pages}
  {104003} (\bibinfo {year} {2021})},\ \Eprint
  {http://arxiv.org/abs/2107.05171} {arXiv:2107.05171 [gr-qc]} \BibitemShut
  {NoStop}%
\bibitem [{\citenamefont {Kumar}\ \emph {et~al.}(2020)\citenamefont {Kumar},
  \citenamefont {Ghosh},\ and\ \citenamefont {Wang}}]{Kumar:2020hgm}%
  \BibitemOpen
  \bibfield  {author} {\bibinfo {author} {\bibfnamefont {Rahul}\ \bibnamefont
  {Kumar}}, \bibinfo {author} {\bibfnamefont {Sushant~G.}\ \bibnamefont
  {Ghosh}}, \ and\ \bibinfo {author} {\bibfnamefont {Anzhong}\ \bibnamefont
  {Wang}},\ }\bibfield  {title} {\enquote {\bibinfo {title} {{Gravitational
  deflection of light and shadow cast by rotating Kalb-Ramond black holes}},}\
  }\href {\doibase 10.1103/PhysRevD.101.104001} {\bibfield  {journal} {\bibinfo
   {journal} {Phys. Rev. D}\ }\textbf {\bibinfo {volume} {101}},\ \bibinfo
  {pages} {104001} (\bibinfo {year} {2020})},\ \Eprint
  {http://arxiv.org/abs/2001.00460} {arXiv:2001.00460 [gr-qc]} \BibitemShut
  {NoStop}%
\bibitem [{\citenamefont {{Kumar}}\ and\ \citenamefont
  {{Ghosh}}(2017)}]{Kumar2017EPJC}%
  \BibitemOpen
  \bibfield  {author} {\bibinfo {author} {\bibfnamefont {Rahul}\ \bibnamefont
  {{Kumar}}}\ and\ \bibinfo {author} {\bibfnamefont {Sushant~G.}\ \bibnamefont
  {{Ghosh}}},\ }\bibfield  {title} {\enquote {\bibinfo {title} {{Accretion onto
  a noncommutative geometry inspired black hole}},}\ }\href {\doibase
  10.1140/epjc/s10052-017-5141-x} {\bibfield  {journal} {\bibinfo  {journal}
  {European Physical Journal C}\ }\textbf {\bibinfo {volume} {77}},\ \bibinfo
  {eid} {577} (\bibinfo {year} {2017})},\ \Eprint
  {http://arxiv.org/abs/1703.10479} {arXiv:1703.10479 [gr-qc]} \BibitemShut
  {NoStop}%
\bibitem [{\citenamefont {Cunha}\ \emph {et~al.}(2017)\citenamefont {Cunha},
  \citenamefont {Herdeiro}, \citenamefont {Kleihaus}, \citenamefont {Kunz},\
  and\ \citenamefont {Radu}}]{Cunha:2016wzk}%
  \BibitemOpen
  \bibfield  {author} {\bibinfo {author} {\bibfnamefont {Pedro V.~P.}\
  \bibnamefont {Cunha}}, \bibinfo {author} {\bibfnamefont {Carlos A.~R.}\
  \bibnamefont {Herdeiro}}, \bibinfo {author} {\bibfnamefont {Burkhard}\
  \bibnamefont {Kleihaus}}, \bibinfo {author} {\bibfnamefont {Jutta}\
  \bibnamefont {Kunz}}, \ and\ \bibinfo {author} {\bibfnamefont {Eugen}\
  \bibnamefont {Radu}},\ }\bibfield  {title} {\enquote {\bibinfo {title}
  {{Shadows of
  Einstein\textendash{}dilaton\textendash{}Gauss\textendash{}Bonnet black
  holes}},}\ }\href {\doibase 10.1016/j.physletb.2017.03.020} {\bibfield
  {journal} {\bibinfo  {journal} {Phys. Lett. B}\ }\textbf {\bibinfo {volume}
  {768}},\ \bibinfo {pages} {373--379} (\bibinfo {year} {2017})},\ \Eprint
  {http://arxiv.org/abs/1701.00079} {arXiv:1701.00079 [gr-qc]} \BibitemShut
  {NoStop}%
\bibitem [{\citenamefont {Cunha}\ \emph
  {et~al.}(2016{\natexlab{a}})\citenamefont {Cunha}, \citenamefont {Herdeiro},
  \citenamefont {Radu},\ and\ \citenamefont {Runarsson}}]{Cunha:2016bpi}%
  \BibitemOpen
  \bibfield  {author} {\bibinfo {author} {\bibfnamefont {Pedro V.~P.}\
  \bibnamefont {Cunha}}, \bibinfo {author} {\bibfnamefont {Carlos A.~R.}\
  \bibnamefont {Herdeiro}}, \bibinfo {author} {\bibfnamefont {Eugen}\
  \bibnamefont {Radu}}, \ and\ \bibinfo {author} {\bibfnamefont {Helgi~F.}\
  \bibnamefont {Runarsson}},\ }\bibfield  {title} {\enquote {\bibinfo {title}
  {{Shadows of Kerr black holes with and without scalar hair}},}\ }\href
  {\doibase 10.1142/S0218271816410212} {\bibfield  {journal} {\bibinfo
  {journal} {Int. J. Mod. Phys. D}\ }\textbf {\bibinfo {volume} {25}},\
  \bibinfo {pages} {1641021} (\bibinfo {year} {2016}{\natexlab{a}})},\ \Eprint
  {http://arxiv.org/abs/1605.08293} {arXiv:1605.08293 [gr-qc]} \BibitemShut
  {NoStop}%
\bibitem [{\citenamefont {Cunha}\ \emph
  {et~al.}(2016{\natexlab{b}})\citenamefont {Cunha}, \citenamefont {Grover},
  \citenamefont {Herdeiro}, \citenamefont {Radu}, \citenamefont {Runarsson},\
  and\ \citenamefont {Wittig}}]{Cunha:2016bjh}%
  \BibitemOpen
  \bibfield  {author} {\bibinfo {author} {\bibfnamefont {P.~V.~P.}\
  \bibnamefont {Cunha}}, \bibinfo {author} {\bibfnamefont {J.}~\bibnamefont
  {Grover}}, \bibinfo {author} {\bibfnamefont {C.}~\bibnamefont {Herdeiro}},
  \bibinfo {author} {\bibfnamefont {E.}~\bibnamefont {Radu}}, \bibinfo {author}
  {\bibfnamefont {H.}~\bibnamefont {Runarsson}}, \ and\ \bibinfo {author}
  {\bibfnamefont {A.}~\bibnamefont {Wittig}},\ }\bibfield  {title} {\enquote
  {\bibinfo {title} {{Chaotic lensing around boson stars and Kerr black holes
  with scalar hair}},}\ }\href {\doibase 10.1103/PhysRevD.94.104023} {\bibfield
   {journal} {\bibinfo  {journal} {Phys. Rev. D}\ }\textbf {\bibinfo {volume}
  {94}},\ \bibinfo {pages} {104023} (\bibinfo {year} {2016}{\natexlab{b}})},\
  \Eprint {http://arxiv.org/abs/1609.01340} {arXiv:1609.01340 [gr-qc]}
  \BibitemShut {NoStop}%
\bibitem [{\citenamefont {Zakharov}(2014)}]{Zakharov:2014lqa}%
  \BibitemOpen
  \bibfield  {author} {\bibinfo {author} {\bibfnamefont {Alexander~F.}\
  \bibnamefont {Zakharov}},\ }\bibfield  {title} {\enquote {\bibinfo {title}
  {{Constraints on a charge in the Reissner-Nordstr\"om metric for the black
  hole at the Galactic Center}},}\ }\href {\doibase 10.1103/PhysRevD.90.062007}
  {\bibfield  {journal} {\bibinfo  {journal} {Phys. Rev. D}\ }\textbf {\bibinfo
  {volume} {90}},\ \bibinfo {pages} {062007} (\bibinfo {year} {2014})},\
  \Eprint {http://arxiv.org/abs/1407.7457} {arXiv:1407.7457 [gr-qc]}
  \BibitemShut {NoStop}%
\bibitem [{\citenamefont {Tsukamoto}(2018)}]{Tsukamoto:2017fxq}%
  \BibitemOpen
  \bibfield  {author} {\bibinfo {author} {\bibfnamefont {Naoki}\ \bibnamefont
  {Tsukamoto}},\ }\bibfield  {title} {\enquote {\bibinfo {title} {{Black hole
  shadow in an asymptotically-flat, stationary, and axisymmetric spacetime: The
  Kerr-Newman and rotating regular black holes}},}\ }\href {\doibase
  10.1103/PhysRevD.97.064021} {\bibfield  {journal} {\bibinfo  {journal} {Phys.
  Rev. D}\ }\textbf {\bibinfo {volume} {97}},\ \bibinfo {pages} {064021}
  (\bibinfo {year} {2018})},\ \Eprint {http://arxiv.org/abs/1708.07427}
  {arXiv:1708.07427 [gr-qc]} \BibitemShut {NoStop}%
\bibitem [{\citenamefont {Chakhchi}\ \emph {et~al.}(2022)\citenamefont
  {Chakhchi}, \citenamefont {El~Moumni},\ and\ \citenamefont
  {Masmar}}]{Chakhchi:2022fls}%
  \BibitemOpen
  \bibfield  {author} {\bibinfo {author} {\bibfnamefont {L.}~\bibnamefont
  {Chakhchi}}, \bibinfo {author} {\bibfnamefont {H.}~\bibnamefont {El~Moumni}},
  \ and\ \bibinfo {author} {\bibfnamefont {K.}~\bibnamefont {Masmar}},\
  }\bibfield  {title} {\enquote {\bibinfo {title} {{Shadows and optical
  appearance of a power-Yang-Mills black hole surrounded by different accretion
  disk profiles}},}\ }\href {\doibase 10.1103/PhysRevD.105.064031} {\bibfield
  {journal} {\bibinfo  {journal} {Phys. Rev. D}\ }\textbf {\bibinfo {volume}
  {105}},\ \bibinfo {pages} {064031} (\bibinfo {year} {2022})}\BibitemShut
  {NoStop}%
\bibitem [{\citenamefont {Li}\ \emph {et~al.}(2020{\natexlab{a}})\citenamefont
  {Li}, \citenamefont {Guo},\ and\ \citenamefont {Chen}}]{Li2020}%
  \BibitemOpen
  \bibfield  {author} {\bibinfo {author} {\bibfnamefont {Peng-Cheng}\
  \bibnamefont {Li}}, \bibinfo {author} {\bibfnamefont {Minyong}\ \bibnamefont
  {Guo}}, \ and\ \bibinfo {author} {\bibfnamefont {Bin}\ \bibnamefont {Chen}},\
  }\bibfield  {title} {\enquote {\bibinfo {title} {{Shadow of a Spinning Black
  Hole in an Expanding Universe}},}\ }\href {\doibase
  10.1103/PhysRevD.101.084041} {\bibfield  {journal} {\bibinfo  {journal}
  {Phys. Rev. D}\ }\textbf {\bibinfo {volume} {101}},\ \bibinfo {pages}
  {084041} (\bibinfo {year} {2020}{\natexlab{a}})},\ \Eprint
  {http://arxiv.org/abs/2001.04231} {arXiv:2001.04231 [gr-qc]} \BibitemShut
  {NoStop}%
\bibitem [{\citenamefont {Pantig}\ and\ \citenamefont
  {\"Ovg\"un}(2023)}]{Pantig:2022ely}%
  \BibitemOpen
  \bibfield  {author} {\bibinfo {author} {\bibfnamefont {Reggie~C.}\
  \bibnamefont {Pantig}}\ and\ \bibinfo {author} {\bibfnamefont {Ali}\
  \bibnamefont {\"Ovg\"un}},\ }\bibfield  {title} {\enquote {\bibinfo {title}
  {{Testing dynamical torsion effects on the charged black
  hole\textquoteright{}s shadow, deflection angle and greybody with M87* and
  Sgr. A* from EHT}},}\ }\href {\doibase 10.1016/j.aop.2022.169197} {\bibfield
  {journal} {\bibinfo  {journal} {Annals Phys.}\ }\textbf {\bibinfo {volume}
  {448}},\ \bibinfo {pages} {169197} (\bibinfo {year} {2023})},\ \Eprint
  {http://arxiv.org/abs/2206.02161} {arXiv:2206.02161 [gr-qc]} \BibitemShut
  {NoStop}%
\bibitem [{\citenamefont {Pantig}\ \emph {et~al.}(2022)\citenamefont {Pantig},
  \citenamefont {Mastrototaro}, \citenamefont {Lambiase},\ and\ \citenamefont
  {\"Ovg\"un}}]{Pantig:2022gih}%
  \BibitemOpen
  \bibfield  {author} {\bibinfo {author} {\bibfnamefont {Reggie~C.}\
  \bibnamefont {Pantig}}, \bibinfo {author} {\bibfnamefont {Leonardo}\
  \bibnamefont {Mastrototaro}}, \bibinfo {author} {\bibfnamefont {Gaetano}\
  \bibnamefont {Lambiase}}, \ and\ \bibinfo {author} {\bibfnamefont {Ali}\
  \bibnamefont {\"Ovg\"un}},\ }\bibfield  {title} {\enquote {\bibinfo {title}
  {{Shadow, lensing, quasinormal modes, greybody bounds and neutrino
  propagation by dyonic ModMax black holes}},}\ }\href {\doibase
  10.1140/epjc/s10052-022-11125-y} {\bibfield  {journal} {\bibinfo  {journal}
  {Eur. Phys. J. C}\ }\textbf {\bibinfo {volume} {82}},\ \bibinfo {pages}
  {1155} (\bibinfo {year} {2022})},\ \Eprint {http://arxiv.org/abs/2208.06664}
  {arXiv:2208.06664 [gr-qc]} \BibitemShut {NoStop}%
\bibitem [{\citenamefont {Lobos}\ and\ \citenamefont
  {Pantig}(2022)}]{Lobos:2022jsz}%
  \BibitemOpen
  \bibfield  {author} {\bibinfo {author} {\bibfnamefont {Nikko John Leo~S.}\
  \bibnamefont {Lobos}}\ and\ \bibinfo {author} {\bibfnamefont {Reggie~C.}\
  \bibnamefont {Pantig}},\ }\bibfield  {title} {\enquote {\bibinfo {title}
  {Generalized extended uncertainty principle black holes: Shadow and lensing
  in the macro- and microscopic realms},}\ }\href {\doibase
  10.3390/physics4040084} {\bibfield  {journal} {\bibinfo  {journal} {Physics}\
  }\textbf {\bibinfo {volume} {4}},\ \bibinfo {pages} {1318--1330} (\bibinfo
  {year} {2022})}\BibitemShut {NoStop}%
\bibitem [{\citenamefont {Uniyal}\ \emph
  {et~al.}(2023{\natexlab{a}})\citenamefont {Uniyal}, \citenamefont {Pantig},\
  and\ \citenamefont {\"Ovg\"un}}]{Uniyal:2022vdu}%
  \BibitemOpen
  \bibfield  {author} {\bibinfo {author} {\bibfnamefont {Akhil}\ \bibnamefont
  {Uniyal}}, \bibinfo {author} {\bibfnamefont {Reggie~C.}\ \bibnamefont
  {Pantig}}, \ and\ \bibinfo {author} {\bibfnamefont {Ali}\ \bibnamefont
  {\"Ovg\"un}},\ }\bibfield  {title} {\enquote {\bibinfo {title} {{Probing a
  non-linear electrodynamics black hole with thin accretion disk, shadow, and
  deflection angle with M87* and Sgr A* from EHT}},}\ }\href {\doibase
  10.1016/j.dark.2023.101178} {\bibfield  {journal} {\bibinfo  {journal} {Phys.
  Dark Univ.}\ }\textbf {\bibinfo {volume} {40}},\ \bibinfo {pages} {101178}
  (\bibinfo {year} {2023}{\natexlab{a}})},\ \Eprint
  {http://arxiv.org/abs/2205.11072} {arXiv:2205.11072 [gr-qc]} \BibitemShut
  {NoStop}%
\bibitem [{\citenamefont {\"Ovg\"un}\ \emph {et~al.}(2023)\citenamefont
  {\"Ovg\"un}, \citenamefont {Pantig},\ and\ \citenamefont
  {Rinc\'on}}]{Ovgun:2023ego}%
  \BibitemOpen
  \bibfield  {author} {\bibinfo {author} {\bibfnamefont {Ali}\ \bibnamefont
  {\"Ovg\"un}}, \bibinfo {author} {\bibfnamefont {Reggie~C.}\ \bibnamefont
  {Pantig}}, \ and\ \bibinfo {author} {\bibfnamefont {\'Angel}\ \bibnamefont
  {Rinc\'on}},\ }\bibfield  {title} {\enquote {\bibinfo {title} {{4D
  scale-dependent Schwarzschild-AdS/dS black holes: study of shadow and weak
  deflection angle and greybody bounding}},}\ }\href {\doibase
  10.1140/epjp/s13360-023-03793-w} {\bibfield  {journal} {\bibinfo  {journal}
  {Eur. Phys. J. Plus}\ }\textbf {\bibinfo {volume} {138}},\ \bibinfo {pages}
  {192} (\bibinfo {year} {2023})},\ \Eprint {http://arxiv.org/abs/2303.01696}
  {arXiv:2303.01696 [gr-qc]} \BibitemShut {NoStop}%
\bibitem [{\citenamefont {Uniyal}\ \emph
  {et~al.}(2023{\natexlab{b}})\citenamefont {Uniyal}, \citenamefont
  {Chakrabarti}, \citenamefont {Pantig},\ and\ \citenamefont
  {\"Ovg\"un}}]{Uniyal:2023inx}%
  \BibitemOpen
  \bibfield  {author} {\bibinfo {author} {\bibfnamefont {Akhil}\ \bibnamefont
  {Uniyal}}, \bibinfo {author} {\bibfnamefont {Sayan}\ \bibnamefont
  {Chakrabarti}}, \bibinfo {author} {\bibfnamefont {Reggie~C.}\ \bibnamefont
  {Pantig}}, \ and\ \bibinfo {author} {\bibfnamefont {Ali}\ \bibnamefont
  {\"Ovg\"un}},\ }\bibfield  {title} {\enquote {\bibinfo {title} {{Nonlinearly
  charged black holes: Shadow and Thin-accretion disk}},}\ }\href@noop {} {\
  (\bibinfo {year} {2023}{\natexlab{b}})},\ \Eprint
  {http://arxiv.org/abs/2303.07174} {arXiv:2303.07174 [gr-qc]} \BibitemShut
  {NoStop}%
\bibitem [{\citenamefont {Panotopoulos}\ \emph {et~al.}(2021)\citenamefont
  {Panotopoulos}, \citenamefont {Rinc\'on},\ and\ \citenamefont
  {Lopes}}]{Panotopoulos:2021tkk}%
  \BibitemOpen
  \bibfield  {author} {\bibinfo {author} {\bibfnamefont {Grigoris}\
  \bibnamefont {Panotopoulos}}, \bibinfo {author} {\bibfnamefont {\'Angel}\
  \bibnamefont {Rinc\'on}}, \ and\ \bibinfo {author} {\bibfnamefont {Ilidio}\
  \bibnamefont {Lopes}},\ }\bibfield  {title} {\enquote {\bibinfo {title}
  {{Orbits of light rays in scale-dependent gravity: Exact analytical solutions
  to the null geodesic equations}},}\ }\href {\doibase
  10.1103/PhysRevD.103.104040} {\bibfield  {journal} {\bibinfo  {journal}
  {Phys. Rev. D}\ }\textbf {\bibinfo {volume} {103}},\ \bibinfo {pages}
  {104040} (\bibinfo {year} {2021})},\ \Eprint
  {http://arxiv.org/abs/2104.13611} {arXiv:2104.13611 [gr-qc]} \BibitemShut
  {NoStop}%
\bibitem [{\citenamefont {Panotopoulos}\ and\ \citenamefont
  {Rincon}(2022)}]{Panotopoulos:2022bky}%
  \BibitemOpen
  \bibfield  {author} {\bibinfo {author} {\bibfnamefont {Grigoris}\
  \bibnamefont {Panotopoulos}}\ and\ \bibinfo {author} {\bibfnamefont {Angel}\
  \bibnamefont {Rincon}},\ }\bibfield  {title} {\enquote {\bibinfo {title}
  {{Orbits of light rays in (1＋2)-dimensional
  Einstein\textendash{}power\textendash{}Maxwell gravity: Exact analytical
  solution to the null geodesic equations}},}\ }\href {\doibase
  10.1016/j.aop.2022.168947} {\bibfield  {journal} {\bibinfo  {journal} {Annals
  Phys.}\ }\textbf {\bibinfo {volume} {443}},\ \bibinfo {pages} {168947}
  (\bibinfo {year} {2022})},\ \Eprint {http://arxiv.org/abs/2206.03437}
  {arXiv:2206.03437 [gr-qc]} \BibitemShut {NoStop}%
\bibitem [{\citenamefont {Khodadi}\ and\ \citenamefont
  {Lambiase}(2022)}]{Khodadi:2022pqh}%
  \BibitemOpen
  \bibfield  {author} {\bibinfo {author} {\bibfnamefont {Mohsen}\ \bibnamefont
  {Khodadi}}\ and\ \bibinfo {author} {\bibfnamefont {Gaetano}\ \bibnamefont
  {Lambiase}},\ }\bibfield  {title} {\enquote {\bibinfo {title} {{Probing
  Lorentz symmetry violation using the first image of Sagittarius A:
  Constraints on standard model extension coefficients}},}\ }\href {\doibase
  10.1103/PhysRevD.106.104050} {\bibfield  {journal} {\bibinfo  {journal}
  {Phys. Rev. D}\ }\textbf {\bibinfo {volume} {106}},\ \bibinfo {pages}
  {104050} (\bibinfo {year} {2022})},\ \Eprint
  {http://arxiv.org/abs/2206.08601} {arXiv:2206.08601 [gr-qc]} \BibitemShut
  {NoStop}%
\bibitem [{\citenamefont {Khodadi}\ \emph {et~al.}(2021)\citenamefont
  {Khodadi}, \citenamefont {Lambiase},\ and\ \citenamefont
  {Mota}}]{Khodadi:2021gbc}%
  \BibitemOpen
  \bibfield  {author} {\bibinfo {author} {\bibfnamefont {Mohsen}\ \bibnamefont
  {Khodadi}}, \bibinfo {author} {\bibfnamefont {Gaetano}\ \bibnamefont
  {Lambiase}}, \ and\ \bibinfo {author} {\bibfnamefont {David~F.}\ \bibnamefont
  {Mota}},\ }\bibfield  {title} {\enquote {\bibinfo {title} {{No-hair theorem
  in the wake of Event Horizon Telescope}},}\ }\href {\doibase
  10.1088/1475-7516/2021/09/028} {\bibfield  {journal} {\bibinfo  {journal}
  {JCAP}\ }\textbf {\bibinfo {volume} {09}},\ \bibinfo {pages} {028} (\bibinfo
  {year} {2021})},\ \Eprint {http://arxiv.org/abs/2107.00834} {arXiv:2107.00834
  [gr-qc]} \BibitemShut {NoStop}%
\bibitem [{\citenamefont {Meng}\ \emph {et~al.}(2023)\citenamefont {Meng},
  \citenamefont {Kuang}, \citenamefont {Wang},\ and\ \citenamefont
  {Wu}}]{Meng:2023unt}%
  \BibitemOpen
  \bibfield  {author} {\bibinfo {author} {\bibfnamefont {Yuan}\ \bibnamefont
  {Meng}}, \bibinfo {author} {\bibfnamefont {Xiao-Mei}\ \bibnamefont {Kuang}},
  \bibinfo {author} {\bibfnamefont {Xi-Jing}\ \bibnamefont {Wang}}, \ and\
  \bibinfo {author} {\bibfnamefont {Jian-Pin}\ \bibnamefont {Wu}},\ }\bibfield
  {title} {\enquote {\bibinfo {title} {{Shadow revisiting and weak
  gravitational lensing with Chern-Simons modification}},}\ }\href {\doibase
  10.1016/j.physletb.2023.137940} {\  (\bibinfo {year} {2023}),\
  10.1016/j.physletb.2023.137940},\ \Eprint {http://arxiv.org/abs/2305.04210}
  {arXiv:2305.04210 [gr-qc]} \BibitemShut {NoStop}%
\bibitem [{\citenamefont {Pantig}\ and\ \citenamefont
  {\"Ovg\"un}(2022{\natexlab{a}})}]{Pantig:2022whj}%
  \BibitemOpen
  \bibfield  {author} {\bibinfo {author} {\bibfnamefont {Reggie~C.}\
  \bibnamefont {Pantig}}\ and\ \bibinfo {author} {\bibfnamefont {Ali}\
  \bibnamefont {\"Ovg\"un}},\ }\bibfield  {title} {\enquote {\bibinfo {title}
  {{Dehnen halo effect on a black hole in an ultra-faint dwarf galaxy}},}\
  }\href {\doibase 10.1088/1475-7516/2022/08/056} {\bibfield  {journal}
  {\bibinfo  {journal} {JCAP}\ }\textbf {\bibinfo {volume} {08}},\ \bibinfo
  {pages} {056} (\bibinfo {year} {2022}{\natexlab{a}})},\ \Eprint
  {http://arxiv.org/abs/2202.07404} {arXiv:2202.07404 [astro-ph.GA]}
  \BibitemShut {NoStop}%
\bibitem [{\citenamefont {Pantig}\ and\ \citenamefont
  {\"Ovg\"un}(2022{\natexlab{b}})}]{Pantig:2022sjb}%
  \BibitemOpen
  \bibfield  {author} {\bibinfo {author} {\bibfnamefont {Reggie~C.}\
  \bibnamefont {Pantig}}\ and\ \bibinfo {author} {\bibfnamefont {Ali}\
  \bibnamefont {\"Ovg\"un}},\ }\bibfield  {title} {\enquote {\bibinfo {title}
  {{Black hole in quantum wave dark matter}},}\ }\href {\doibase
  10.1002/prop.202200164} {\bibfield  {journal} {\bibinfo  {journal} {Fortsch.
  Phys.}\ }\textbf {\bibinfo {volume} {2022}},\ \bibinfo {pages} {2200164}
  (\bibinfo {year} {2022}{\natexlab{b}})},\ \Eprint
  {http://arxiv.org/abs/2210.00523} {arXiv:2210.00523 [gr-qc]} \BibitemShut
  {NoStop}%
\bibitem [{\citenamefont {Pantig}(2023)}]{Pantig:2023yer}%
  \BibitemOpen
  \bibfield  {author} {\bibinfo {author} {\bibfnamefont {Reggie~C.}\
  \bibnamefont {Pantig}},\ }\bibfield  {title} {\enquote {\bibinfo {title}
  {{Constraining a one-dimensional wave-type gravitational wave parameter
  through the shadow of M87* via Event Horizon Telescope}},}\ }\href@noop {} {\
   (\bibinfo {year} {2023})},\ \Eprint {http://arxiv.org/abs/2303.01698}
  {arXiv:2303.01698 [gr-qc]} \BibitemShut {NoStop}%
\bibitem [{\citenamefont {Wang}\ \emph {et~al.}(2021)\citenamefont {Wang},
  \citenamefont {Chen},\ and\ \citenamefont {Jing}}]{Wang:2019skw}%
  \BibitemOpen
  \bibfield  {author} {\bibinfo {author} {\bibfnamefont {Mingzhi}\ \bibnamefont
  {Wang}}, \bibinfo {author} {\bibfnamefont {Songbai}\ \bibnamefont {Chen}}, \
  and\ \bibinfo {author} {\bibfnamefont {Jiliang}\ \bibnamefont {Jing}},\
  }\bibfield  {title} {\enquote {\bibinfo {title} {{Effect of gravitational
  wave on shadow of a Schwarzschild black hole}},}\ }\href {\doibase
  10.1140/epjc/s10052-021-09287-2} {\bibfield  {journal} {\bibinfo  {journal}
  {Eur. Phys. J. C}\ }\textbf {\bibinfo {volume} {81}},\ \bibinfo {pages} {509}
  (\bibinfo {year} {2021})},\ \Eprint {http://arxiv.org/abs/1908.04527}
  {arXiv:1908.04527 [gr-qc]} \BibitemShut {NoStop}%
\bibitem [{\citenamefont {Roy}\ and\ \citenamefont
  {Chakrabarti}(2020)}]{Roy:2020dyy}%
  \BibitemOpen
  \bibfield  {author} {\bibinfo {author} {\bibfnamefont {Rittick}\ \bibnamefont
  {Roy}}\ and\ \bibinfo {author} {\bibfnamefont {Sayan}\ \bibnamefont
  {Chakrabarti}},\ }\bibfield  {title} {\enquote {\bibinfo {title} {{Study on
  black hole shadows in asymptotically de Sitter spacetimes}},}\ }\href
  {\doibase 10.1103/PhysRevD.102.024059} {\bibfield  {journal} {\bibinfo
  {journal} {Phys. Rev. D}\ }\textbf {\bibinfo {volume} {102}},\ \bibinfo
  {pages} {024059} (\bibinfo {year} {2020})},\ \Eprint
  {http://arxiv.org/abs/2003.14107} {arXiv:2003.14107 [gr-qc]} \BibitemShut
  {NoStop}%
\bibitem [{\citenamefont {Konoplya}(2021)}]{Konoplya:2021ube}%
  \BibitemOpen
  \bibfield  {author} {\bibinfo {author} {\bibfnamefont {R.~A.}\ \bibnamefont
  {Konoplya}},\ }\bibfield  {title} {\enquote {\bibinfo {title} {{Black holes
  in galactic centers: Quasinormal ringing, grey-body factors and Unruh
  temperature}},}\ }\href {\doibase 10.1016/j.physletb.2021.136734} {\bibfield
  {journal} {\bibinfo  {journal} {Phys. Lett. B}\ }\textbf {\bibinfo {volume}
  {823}},\ \bibinfo {pages} {136734} (\bibinfo {year} {2021})},\ \Eprint
  {http://arxiv.org/abs/2109.01640} {arXiv:2109.01640 [gr-qc]} \BibitemShut
  {NoStop}%
\bibitem [{\citenamefont {Anjum}\ \emph {et~al.}(2023)\citenamefont {Anjum},
  \citenamefont {Afrin},\ and\ \citenamefont {Ghosh}}]{Anjum:2023axh}%
  \BibitemOpen
  \bibfield  {author} {\bibinfo {author} {\bibfnamefont {Arshia}\ \bibnamefont
  {Anjum}}, \bibinfo {author} {\bibfnamefont {Misba}\ \bibnamefont {Afrin}}, \
  and\ \bibinfo {author} {\bibfnamefont {Sushant~G.}\ \bibnamefont {Ghosh}},\
  }\bibfield  {title} {\enquote {\bibinfo {title} {{Astrophysical consequences
  of dark matter for photon orbits and shadows of supermassive black holes}},}\
  }\href@noop {} {\  (\bibinfo {year} {2023})},\ \Eprint
  {http://arxiv.org/abs/2301.06373} {arXiv:2301.06373 [gr-qc]} \BibitemShut
  {NoStop}%
\bibitem [{\citenamefont {Hou}\ \emph {et~al.}(2018)\citenamefont {Hou},
  \citenamefont {Xu}, \citenamefont {Zhou},\ and\ \citenamefont
  {Wang}}]{Hou:2018bar}%
  \BibitemOpen
  \bibfield  {author} {\bibinfo {author} {\bibfnamefont {Xian}\ \bibnamefont
  {Hou}}, \bibinfo {author} {\bibfnamefont {Zhaoyi}\ \bibnamefont {Xu}},
  \bibinfo {author} {\bibfnamefont {Ming}\ \bibnamefont {Zhou}}, \ and\
  \bibinfo {author} {\bibfnamefont {Jiancheng}\ \bibnamefont {Wang}},\
  }\bibfield  {title} {\enquote {\bibinfo {title} {{Black hole shadow of Sgr
  A$^{*}$ in dark matter halo}},}\ }\href {\doibase
  10.1088/1475-7516/2018/07/015} {\bibfield  {journal} {\bibinfo  {journal}
  {JCAP}\ }\textbf {\bibinfo {volume} {07}},\ \bibinfo {pages} {015} (\bibinfo
  {year} {2018})},\ \Eprint {http://arxiv.org/abs/1804.08110} {arXiv:1804.08110
  [gr-qc]} \BibitemShut {NoStop}%
\bibitem [{\citenamefont {Lambiase}\ \emph {et~al.}(2023)\citenamefont
  {Lambiase}, \citenamefont {Pantig}, \citenamefont {Gogoi},\ and\
  \citenamefont {\"Ovg\"un}}]{Lambiase:2023fbd}%
  \BibitemOpen
  \bibfield  {author} {\bibinfo {author} {\bibfnamefont {Gaetano}\ \bibnamefont
  {Lambiase}}, \bibinfo {author} {\bibfnamefont {Reggie~C.}\ \bibnamefont
  {Pantig}}, \bibinfo {author} {\bibfnamefont {Dhruba~Jyoti}\ \bibnamefont
  {Gogoi}}, \ and\ \bibinfo {author} {\bibfnamefont {Ali}\ \bibnamefont
  {\"Ovg\"un}},\ }\bibfield  {title} {\enquote {\bibinfo {title}
  {{Investigating the Connection between Generalized Uncertainty Principle and
  Asymptotically Safe Gravity in Black Hole Signatures through Shadow and
  Quasinormal Modes}},}\ }\href@noop {} {\  (\bibinfo {year} {2023})},\ \Eprint
  {http://arxiv.org/abs/2304.00183} {arXiv:2304.00183 [gr-qc]} \BibitemShut
  {NoStop}%
\bibitem [{\citenamefont {Shaikh}(2022)}]{Shaikh:2022ivr}%
  \BibitemOpen
  \bibfield  {author} {\bibinfo {author} {\bibfnamefont {Rajibul}\ \bibnamefont
  {Shaikh}},\ }\bibfield  {title} {\enquote {\bibinfo {title} {{Testing black
  hole mimickers with the Event Horizon Telescope image of Sagittarius
  A$^*$}},}\ }\href {\doibase 10.1093/mnras/stad1383} {\  (\bibinfo {year}
  {2022}),\ 10.1093/mnras/stad1383},\ \Eprint {http://arxiv.org/abs/2208.01995}
  {arXiv:2208.01995 [gr-qc]} \BibitemShut {NoStop}%
\bibitem [{\citenamefont {Shaikh}\ \emph {et~al.}(2022)\citenamefont {Shaikh},
  \citenamefont {Paul}, \citenamefont {Banerjee},\ and\ \citenamefont
  {Sarkar}}]{Shaikh:2021cvl}%
  \BibitemOpen
  \bibfield  {author} {\bibinfo {author} {\bibfnamefont {Rajibul}\ \bibnamefont
  {Shaikh}}, \bibinfo {author} {\bibfnamefont {Suvankar}\ \bibnamefont {Paul}},
  \bibinfo {author} {\bibfnamefont {Pritam}\ \bibnamefont {Banerjee}}, \ and\
  \bibinfo {author} {\bibfnamefont {Tapobrata}\ \bibnamefont {Sarkar}},\
  }\bibfield  {title} {\enquote {\bibinfo {title} {{Shadows and thin accretion
  disk images of the $\gamma $-metric}},}\ }\href {\doibase
  10.1140/epjc/s10052-022-10664-8} {\bibfield  {journal} {\bibinfo  {journal}
  {Eur. Phys. J. C}\ }\textbf {\bibinfo {volume} {82}},\ \bibinfo {pages} {696}
  (\bibinfo {year} {2022})},\ \Eprint {http://arxiv.org/abs/2105.12057}
  {arXiv:2105.12057 [gr-qc]} \BibitemShut {NoStop}%
\bibitem [{\citenamefont {Shaikh}\ \emph {et~al.}(2019)\citenamefont {Shaikh},
  \citenamefont {Kocherlakota}, \citenamefont {Narayan},\ and\ \citenamefont
  {Joshi}}]{Shaikh:2018lcc}%
  \BibitemOpen
  \bibfield  {author} {\bibinfo {author} {\bibfnamefont {Rajibul}\ \bibnamefont
  {Shaikh}}, \bibinfo {author} {\bibfnamefont {Prashant}\ \bibnamefont
  {Kocherlakota}}, \bibinfo {author} {\bibfnamefont {Ramesh}\ \bibnamefont
  {Narayan}}, \ and\ \bibinfo {author} {\bibfnamefont {Pankaj~S.}\ \bibnamefont
  {Joshi}},\ }\bibfield  {title} {\enquote {\bibinfo {title} {{Shadows of
  spherically symmetric black holes and naked singularities}},}\ }\href
  {\doibase 10.1093/mnras/sty2624} {\bibfield  {journal} {\bibinfo  {journal}
  {Mon. Not. Roy. Astron. Soc.}\ }\textbf {\bibinfo {volume} {482}},\ \bibinfo
  {pages} {52--64} (\bibinfo {year} {2019})},\ \Eprint
  {http://arxiv.org/abs/1802.08060} {arXiv:1802.08060 [astro-ph.HE]}
  \BibitemShut {NoStop}%
\bibitem [{\citenamefont {Shaikh}(2019)}]{Shaikh:2019fpu}%
  \BibitemOpen
  \bibfield  {author} {\bibinfo {author} {\bibfnamefont {Rajibul}\ \bibnamefont
  {Shaikh}},\ }\bibfield  {title} {\enquote {\bibinfo {title} {{Black hole
  shadow in a general rotating spacetime obtained through Newman-Janis
  algorithm}},}\ }\href {\doibase 10.1103/PhysRevD.100.024028} {\bibfield
  {journal} {\bibinfo  {journal} {Phys. Rev. D}\ }\textbf {\bibinfo {volume}
  {100}},\ \bibinfo {pages} {024028} (\bibinfo {year} {2019})},\ \Eprint
  {http://arxiv.org/abs/1904.08322} {arXiv:1904.08322 [gr-qc]} \BibitemShut
  {NoStop}%
\bibitem [{\citenamefont {Shaikh}\ and\ \citenamefont
  {Joshi}(2019)}]{Shaikh:2019hbm}%
  \BibitemOpen
  \bibfield  {author} {\bibinfo {author} {\bibfnamefont {Rajibul}\ \bibnamefont
  {Shaikh}}\ and\ \bibinfo {author} {\bibfnamefont {Pankaj~S.}\ \bibnamefont
  {Joshi}},\ }\bibfield  {title} {\enquote {\bibinfo {title} {{Can we
  distinguish black holes from naked singularities by the images of their
  accretion disks?}}}\ }\href {\doibase 10.1088/1475-7516/2019/10/064}
  {\bibfield  {journal} {\bibinfo  {journal} {JCAP}\ }\textbf {\bibinfo
  {volume} {10}},\ \bibinfo {pages} {064} (\bibinfo {year} {2019})},\ \Eprint
  {http://arxiv.org/abs/1909.10322} {arXiv:1909.10322 [gr-qc]} \BibitemShut
  {NoStop}%
\bibitem [{\citenamefont {Shaikh}\ \emph {et~al.}(2021)\citenamefont {Shaikh},
  \citenamefont {Pal}, \citenamefont {Pal},\ and\ \citenamefont
  {Sarkar}}]{Shaikh:2021yux}%
  \BibitemOpen
  \bibfield  {author} {\bibinfo {author} {\bibfnamefont {Rajibul}\ \bibnamefont
  {Shaikh}}, \bibinfo {author} {\bibfnamefont {Kunal}\ \bibnamefont {Pal}},
  \bibinfo {author} {\bibfnamefont {Kuntal}\ \bibnamefont {Pal}}, \ and\
  \bibinfo {author} {\bibfnamefont {Tapobrata}\ \bibnamefont {Sarkar}},\
  }\bibfield  {title} {\enquote {\bibinfo {title} {{Constraining alternatives
  to the Kerr black hole}},}\ }\href {\doibase 10.1093/mnras/stab1779}
  {\bibfield  {journal} {\bibinfo  {journal} {Mon. Not. Roy. Astron. Soc.}\
  }\textbf {\bibinfo {volume} {506}},\ \bibinfo {pages} {1229--1236} (\bibinfo
  {year} {2021})},\ \Eprint {http://arxiv.org/abs/2102.04299} {arXiv:2102.04299
  [gr-qc]} \BibitemShut {NoStop}%
\bibitem [{\citenamefont {Rahaman}\ \emph {et~al.}(2021)\citenamefont
  {Rahaman}, \citenamefont {Manna}, \citenamefont {Shaikh}, \citenamefont
  {Aktar}, \citenamefont {Mondal},\ and\ \citenamefont
  {Samanta}}]{Rahaman:2021kge}%
  \BibitemOpen
  \bibfield  {author} {\bibinfo {author} {\bibfnamefont {Farook}\ \bibnamefont
  {Rahaman}}, \bibinfo {author} {\bibfnamefont {Tuhina}\ \bibnamefont {Manna}},
  \bibinfo {author} {\bibfnamefont {Rajibul}\ \bibnamefont {Shaikh}}, \bibinfo
  {author} {\bibfnamefont {Somi}\ \bibnamefont {Aktar}}, \bibinfo {author}
  {\bibfnamefont {Monimala}\ \bibnamefont {Mondal}}, \ and\ \bibinfo {author}
  {\bibfnamefont {Bidisha}\ \bibnamefont {Samanta}},\ }\bibfield  {title}
  {\enquote {\bibinfo {title} {{Thin accretion disks around traversable
  wormholes}},}\ }\href {\doibase 10.1016/j.nuclphysb.2021.115548} {\bibfield
  {journal} {\bibinfo  {journal} {Nucl. Phys. B}\ }\textbf {\bibinfo {volume}
  {972}},\ \bibinfo {pages} {115548} (\bibinfo {year} {2021})},\ \Eprint
  {http://arxiv.org/abs/2110.09820} {arXiv:2110.09820 [gr-qc]} \BibitemShut
  {NoStop}%
\bibitem [{\citenamefont {Virbhadra}\ and\ \citenamefont
  {Ellis}(2000)}]{Virbhadra:1999nm}%
  \BibitemOpen
  \bibfield  {author} {\bibinfo {author} {\bibfnamefont {K.~S.}\ \bibnamefont
  {Virbhadra}}\ and\ \bibinfo {author} {\bibfnamefont {George F.~R.}\
  \bibnamefont {Ellis}},\ }\bibfield  {title} {\enquote {\bibinfo {title}
  {{Schwarzschild black hole lensing}},}\ }\href {\doibase
  10.1103/PhysRevD.62.084003} {\bibfield  {journal} {\bibinfo  {journal} {Phys.
  Rev. D}\ }\textbf {\bibinfo {volume} {62}},\ \bibinfo {pages} {084003}
  (\bibinfo {year} {2000})},\ \Eprint {http://arxiv.org/abs/astro-ph/9904193}
  {arXiv:astro-ph/9904193} \BibitemShut {NoStop}%
\bibitem [{\citenamefont {Virbhadra}\ and\ \citenamefont
  {Ellis}(2002)}]{Virbhadra:2002ju}%
  \BibitemOpen
  \bibfield  {author} {\bibinfo {author} {\bibfnamefont {K.~S.}\ \bibnamefont
  {Virbhadra}}\ and\ \bibinfo {author} {\bibfnamefont {G.~F.~R.}\ \bibnamefont
  {Ellis}},\ }\bibfield  {title} {\enquote {\bibinfo {title} {{Gravitational
  lensing by naked singularities}},}\ }\href {\doibase
  10.1103/PhysRevD.65.103004} {\bibfield  {journal} {\bibinfo  {journal} {Phys.
  Rev. D}\ }\textbf {\bibinfo {volume} {65}},\ \bibinfo {pages} {103004}
  (\bibinfo {year} {2002})}\BibitemShut {NoStop}%
\bibitem [{\citenamefont {Adler}\ and\ \citenamefont
  {Virbhadra}(2022)}]{Adler:2022qtb}%
  \BibitemOpen
  \bibfield  {author} {\bibinfo {author} {\bibfnamefont {Stephen~L.}\
  \bibnamefont {Adler}}\ and\ \bibinfo {author} {\bibfnamefont {K.~S.}\
  \bibnamefont {Virbhadra}},\ }\bibfield  {title} {\enquote {\bibinfo {title}
  {{Cosmological constant corrections to the photon sphere and black hole
  shadow radii}},}\ }\href@noop {} {\  (\bibinfo {year} {2022})},\ \Eprint
  {http://arxiv.org/abs/2205.04628} {arXiv:2205.04628 [gr-qc]} \BibitemShut
  {NoStop}%
\bibitem [{\citenamefont {Bozza}\ \emph {et~al.}(2001)\citenamefont {Bozza},
  \citenamefont {Capozziello}, \citenamefont {Iovane},\ and\ \citenamefont
  {Scarpetta}}]{Bozza:2001xd}%
  \BibitemOpen
  \bibfield  {author} {\bibinfo {author} {\bibfnamefont {V.}~\bibnamefont
  {Bozza}}, \bibinfo {author} {\bibfnamefont {S.}~\bibnamefont {Capozziello}},
  \bibinfo {author} {\bibfnamefont {G.}~\bibnamefont {Iovane}}, \ and\ \bibinfo
  {author} {\bibfnamefont {G.}~\bibnamefont {Scarpetta}},\ }\bibfield  {title}
  {\enquote {\bibinfo {title} {{Strong field limit of black hole gravitational
  lensing}},}\ }\href {\doibase 10.1023/A:1012292927358} {\bibfield  {journal}
  {\bibinfo  {journal} {Gen. Rel. Grav.}\ }\textbf {\bibinfo {volume} {33}},\
  \bibinfo {pages} {1535--1548} (\bibinfo {year} {2001})},\ \Eprint
  {http://arxiv.org/abs/gr-qc/0102068} {arXiv:gr-qc/0102068} \BibitemShut
  {NoStop}%
\bibitem [{\citenamefont {Bozza}(2002)}]{Bozza:2002zj}%
  \BibitemOpen
  \bibfield  {author} {\bibinfo {author} {\bibfnamefont {V.}~\bibnamefont
  {Bozza}},\ }\bibfield  {title} {\enquote {\bibinfo {title} {{Gravitational
  lensing in the strong field limit}},}\ }\href {\doibase
  10.1103/PhysRevD.66.103001} {\bibfield  {journal} {\bibinfo  {journal} {Phys.
  Rev. D}\ }\textbf {\bibinfo {volume} {66}},\ \bibinfo {pages} {103001}
  (\bibinfo {year} {2002})},\ \Eprint {http://arxiv.org/abs/gr-qc/0208075}
  {arXiv:gr-qc/0208075} \BibitemShut {NoStop}%
\bibitem [{\citenamefont {Perlick}(2004)}]{Perlick:2003vg}%
  \BibitemOpen
  \bibfield  {author} {\bibinfo {author} {\bibfnamefont {Volker}\ \bibnamefont
  {Perlick}},\ }\bibfield  {title} {\enquote {\bibinfo {title} {{On the Exact
  gravitational lens equation in spherically symmetric and static
  space-times}},}\ }\href {\doibase 10.1103/PhysRevD.69.064017} {\bibfield
  {journal} {\bibinfo  {journal} {Phys. Rev. D}\ }\textbf {\bibinfo {volume}
  {69}},\ \bibinfo {pages} {064017} (\bibinfo {year} {2004})},\ \Eprint
  {http://arxiv.org/abs/gr-qc/0307072} {arXiv:gr-qc/0307072} \BibitemShut
  {NoStop}%
\bibitem [{\citenamefont {He}\ \emph {et~al.}(2020)\citenamefont {He},
  \citenamefont {Zhou}, \citenamefont {Feng}, \citenamefont {Mu}, \citenamefont
  {Wang}, \citenamefont {Li}, \citenamefont {Pan},\ and\ \citenamefont
  {Lin}}]{He:2020eah}%
  \BibitemOpen
  \bibfield  {author} {\bibinfo {author} {\bibfnamefont {Guansheng}\
  \bibnamefont {He}}, \bibinfo {author} {\bibfnamefont {Xia}\ \bibnamefont
  {Zhou}}, \bibinfo {author} {\bibfnamefont {Zhongwen}\ \bibnamefont {Feng}},
  \bibinfo {author} {\bibfnamefont {Xueling}\ \bibnamefont {Mu}}, \bibinfo
  {author} {\bibfnamefont {Hui}\ \bibnamefont {Wang}}, \bibinfo {author}
  {\bibfnamefont {Weijun}\ \bibnamefont {Li}}, \bibinfo {author} {\bibfnamefont
  {Chaohong}\ \bibnamefont {Pan}}, \ and\ \bibinfo {author} {\bibfnamefont
  {Wenbin}\ \bibnamefont {Lin}},\ }\bibfield  {title} {\enquote {\bibinfo
  {title} {{Gravitational deflection of massive particles in Schwarzschild-de
  Sitter spacetime}},}\ }\href {\doibase 10.1140/epjc/s10052-020-8382-z}
  {\bibfield  {journal} {\bibinfo  {journal} {Eur. Phys. J. C}\ }\textbf
  {\bibinfo {volume} {80}},\ \bibinfo {pages} {835} (\bibinfo {year}
  {2020})}\BibitemShut {NoStop}%
\bibitem [{\citenamefont {Virbhadra}(2022{\natexlab{a}})}]{Virbhadra:2022ybp}%
  \BibitemOpen
  \bibfield  {author} {\bibinfo {author} {\bibfnamefont {K.~S.}\ \bibnamefont
  {Virbhadra}},\ }\bibfield  {title} {\enquote {\bibinfo {title} {{Compactness
  of supermassive dark objects at galactic centers}},}\ }\href@noop {} {\
  (\bibinfo {year} {2022}{\natexlab{a}})},\ \Eprint
  {http://arxiv.org/abs/2204.01792} {arXiv:2204.01792 [gr-qc]} \BibitemShut
  {NoStop}%
\bibitem [{\citenamefont {Virbhadra}(2022{\natexlab{b}})}]{Virbhadra:2022iiy}%
  \BibitemOpen
  \bibfield  {author} {\bibinfo {author} {\bibfnamefont {K.~S.}\ \bibnamefont
  {Virbhadra}},\ }\bibfield  {title} {\enquote {\bibinfo {title} {{Distortions
  of images of Schwarzschild lensing}},}\ }\href@noop {} {\  (\bibinfo {year}
  {2022}{\natexlab{b}})},\ \Eprint {http://arxiv.org/abs/2204.01879}
  {arXiv:2204.01879 [gr-qc]} \BibitemShut {NoStop}%
\bibitem [{\citenamefont {Gibbons}\ and\ \citenamefont
  {Werner}(2008)}]{Gibbons:2008rj}%
  \BibitemOpen
  \bibfield  {author} {\bibinfo {author} {\bibfnamefont {G.~W.}\ \bibnamefont
  {Gibbons}}\ and\ \bibinfo {author} {\bibfnamefont {M.~C.}\ \bibnamefont
  {Werner}},\ }\bibfield  {title} {\enquote {\bibinfo {title} {{Applications of
  the Gauss-Bonnet theorem to gravitational lensing}},}\ }\href {\doibase
  10.1088/0264-9381/25/23/235009} {\bibfield  {journal} {\bibinfo  {journal}
  {Class. Quant. Grav.}\ }\textbf {\bibinfo {volume} {25}},\ \bibinfo {pages}
  {235009} (\bibinfo {year} {2008})},\ \Eprint {http://arxiv.org/abs/0807.0854}
  {arXiv:0807.0854 [gr-qc]} \BibitemShut {NoStop}%
\bibitem [{\citenamefont {\"Ovg\"un}\ \emph {et~al.}(2022)\citenamefont
  {\"Ovg\"un}, \citenamefont {Kumaran}, \citenamefont {Javed},\ and\
  \citenamefont {Abbas}}]{Ovgun:2022opq}%
  \BibitemOpen
  \bibfield  {author} {\bibinfo {author} {\bibfnamefont {Ali}\ \bibnamefont
  {\"Ovg\"un}}, \bibinfo {author} {\bibfnamefont {Yashmitha}\ \bibnamefont
  {Kumaran}}, \bibinfo {author} {\bibfnamefont {Wajiha}\ \bibnamefont {Javed}},
  \ and\ \bibinfo {author} {\bibfnamefont {Jameela}\ \bibnamefont {Abbas}},\
  }\bibfield  {title} {\enquote {\bibinfo {title} {{Effect of Horndeski theory
  on weak deflection angle using the Gauss\textendash{}Bonnet theorem}},}\
  }\href {\doibase 10.1142/S0219887822501924} {\bibfield  {journal} {\bibinfo
  {journal} {Int. J. Geom. Meth. Mod. Phys.}\ }\textbf {\bibinfo {volume}
  {19}},\ \bibinfo {pages} {2250192} (\bibinfo {year} {2022})}\BibitemShut
  {NoStop}%
\bibitem [{\citenamefont {Kumaran}\ and\ \citenamefont
  {\"Ovg\"un}(2021)}]{Kumaran:2021rgj}%
  \BibitemOpen
  \bibfield  {author} {\bibinfo {author} {\bibfnamefont {Yashmitha}\
  \bibnamefont {Kumaran}}\ and\ \bibinfo {author} {\bibfnamefont {Ali}\
  \bibnamefont {\"Ovg\"un}},\ }\bibfield  {title} {\enquote {\bibinfo {title}
  {{Deriving weak deflection angle by black holes or wormholes using
  Gauss-Bonnet theorem}},}\ }\href {\doibase 10.3906/fiz-2110-16} {\bibfield
  {journal} {\bibinfo  {journal} {Turk. J. Phys.}\ }\textbf {\bibinfo {volume}
  {45}},\ \bibinfo {pages} {247--267} (\bibinfo {year} {2021})},\ \Eprint
  {http://arxiv.org/abs/2111.02805} {arXiv:2111.02805 [gr-qc]} \BibitemShut
  {NoStop}%
\bibitem [{\citenamefont {Javed}\ \emph {et~al.}(2021)\citenamefont {Javed},
  \citenamefont {Abbas}, \citenamefont {Kumaran},\ and\ \citenamefont
  {\"Ovg\"un}}]{Javed:2020pyz}%
  \BibitemOpen
  \bibfield  {author} {\bibinfo {author} {\bibfnamefont {Wajiha}\ \bibnamefont
  {Javed}}, \bibinfo {author} {\bibfnamefont {Jameela}\ \bibnamefont {Abbas}},
  \bibinfo {author} {\bibfnamefont {Yashmitha}\ \bibnamefont {Kumaran}}, \ and\
  \bibinfo {author} {\bibfnamefont {Ali}\ \bibnamefont {\"Ovg\"un}},\
  }\bibfield  {title} {\enquote {\bibinfo {title} {{Weak deflection angle by
  asymptotically flat black holes in Horndeski theory using Gauss-Bonnet
  theorem}},}\ }\href {\doibase 10.1142/S0219887821500031} {\bibfield
  {journal} {\bibinfo  {journal} {Int. J. Geom. Meth. Mod. Phys.}\ }\textbf
  {\bibinfo {volume} {18}},\ \bibinfo {pages} {2150003} (\bibinfo {year}
  {2021})},\ \Eprint {http://arxiv.org/abs/2102.02812} {arXiv:2102.02812
  [gr-qc]} \BibitemShut {NoStop}%
\bibitem [{\citenamefont {Kumaran}\ and\ \citenamefont
  {\"Ovg\"un}(2020)}]{Kumaran:2019qqp}%
  \BibitemOpen
  \bibfield  {author} {\bibinfo {author} {\bibfnamefont {Yashmitha}\
  \bibnamefont {Kumaran}}\ and\ \bibinfo {author} {\bibfnamefont {Ali}\
  \bibnamefont {\"Ovg\"un}},\ }\bibfield  {title} {\enquote {\bibinfo {title}
  {{Weak Deflection Angle of Extended Uncertainty Principle Black Holes}},}\
  }\href {\doibase 10.1088/1674-1137/44/2/025101} {\bibfield  {journal}
  {\bibinfo  {journal} {Chin. Phys. C}\ }\textbf {\bibinfo {volume} {44}},\
  \bibinfo {pages} {025101} (\bibinfo {year} {2020})},\ \Eprint
  {http://arxiv.org/abs/1905.11710} {arXiv:1905.11710 [gr-qc]} \BibitemShut
  {NoStop}%
\bibitem [{\citenamefont {\"Ovg\"un}(2018)}]{Ovgun:2018fnk}%
  \BibitemOpen
  \bibfield  {author} {\bibinfo {author} {\bibfnamefont {Ali}\ \bibnamefont
  {\"Ovg\"un}},\ }\bibfield  {title} {\enquote {\bibinfo {title} {{Light
  deflection by Damour-Solodukhin wormholes and Gauss-Bonnet theorem}},}\
  }\href {\doibase 10.1103/PhysRevD.98.044033} {\bibfield  {journal} {\bibinfo
  {journal} {Phys. Rev. D}\ }\textbf {\bibinfo {volume} {98}},\ \bibinfo
  {pages} {044033} (\bibinfo {year} {2018})},\ \Eprint
  {http://arxiv.org/abs/1805.06296} {arXiv:1805.06296 [gr-qc]} \BibitemShut
  {NoStop}%
\bibitem [{\citenamefont {\"Ovg\"un}(2019{\natexlab{a}})}]{Ovgun:2019wej}%
  \BibitemOpen
  \bibfield  {author} {\bibinfo {author} {\bibfnamefont {A.}~\bibnamefont
  {\"Ovg\"un}},\ }\bibfield  {title} {\enquote {\bibinfo {title} {{Weak field
  deflection angle by regular black holes with cosmic strings using the
  Gauss-Bonnet theorem}},}\ }\href {\doibase 10.1103/PhysRevD.99.104075}
  {\bibfield  {journal} {\bibinfo  {journal} {Phys. Rev. D}\ }\textbf {\bibinfo
  {volume} {99}},\ \bibinfo {pages} {104075} (\bibinfo {year}
  {2019}{\natexlab{a}})},\ \Eprint {http://arxiv.org/abs/1902.04411}
  {arXiv:1902.04411 [gr-qc]} \BibitemShut {NoStop}%
\bibitem [{\citenamefont {\"Ovg\"un}(2019{\natexlab{b}})}]{Ovgun:2018oxk}%
  \BibitemOpen
  \bibfield  {author} {\bibinfo {author} {\bibfnamefont {Ali}\ \bibnamefont
  {\"Ovg\"un}},\ }\bibfield  {title} {\enquote {\bibinfo {title} {{Deflection
  Angle of Photons through Dark Matter by Black Holes and Wormholes Using
  Gauss\textendash{}Bonnet Theorem}},}\ }\href {\doibase
  10.3390/universe5050115} {\bibfield  {journal} {\bibinfo  {journal}
  {Universe}\ }\textbf {\bibinfo {volume} {5}},\ \bibinfo {pages} {115}
  (\bibinfo {year} {2019}{\natexlab{b}})},\ \Eprint
  {http://arxiv.org/abs/1806.05549} {arXiv:1806.05549 [physics.gen-ph]}
  \BibitemShut {NoStop}%
\bibitem [{\citenamefont {Javed}\ \emph {et~al.}(2019)\citenamefont {Javed},
  \citenamefont {Babar},\ and\ \citenamefont {\"Ovg\"un}}]{Javed:2019ynm}%
  \BibitemOpen
  \bibfield  {author} {\bibinfo {author} {\bibfnamefont {Wajiha}\ \bibnamefont
  {Javed}}, \bibinfo {author} {\bibfnamefont {Rimsha}\ \bibnamefont {Babar}}, \
  and\ \bibinfo {author} {\bibfnamefont {Al\"\i{}}\ \bibnamefont {\"Ovg\"un}},\
  }\bibfield  {title} {\enquote {\bibinfo {title} {{Effect of the dilaton field
  and plasma medium on deflection angle by black holes in
  Einstein-Maxwell-dilaton-axion theory}},}\ }\href {\doibase
  10.1103/PhysRevD.100.104032} {\bibfield  {journal} {\bibinfo  {journal}
  {Phys. Rev. D}\ }\textbf {\bibinfo {volume} {100}},\ \bibinfo {pages}
  {104032} (\bibinfo {year} {2019})},\ \Eprint
  {http://arxiv.org/abs/1910.11697} {arXiv:1910.11697 [gr-qc]} \BibitemShut
  {NoStop}%
\bibitem [{\citenamefont {Werner}(2012)}]{Werner2012}%
  \BibitemOpen
  \bibfield  {author} {\bibinfo {author} {\bibfnamefont {M.~C.}\ \bibnamefont
  {Werner}},\ }\bibfield  {title} {\enquote {\bibinfo {title} {{Gravitational
  lensing in the Kerr-Randers optical geometry}},}\ }\href {\doibase
  10.1007/s10714-012-1458-9} {\bibfield  {journal} {\bibinfo  {journal} {Gen.
  Rel. Grav.}\ }\textbf {\bibinfo {volume} {44}},\ \bibinfo {pages}
  {3047--3057} (\bibinfo {year} {2012})},\ \Eprint
  {http://arxiv.org/abs/1205.3876} {arXiv:1205.3876 [gr-qc]} \BibitemShut
  {NoStop}%
\bibitem [{\citenamefont {Ishihara}\ \emph {et~al.}(2016)\citenamefont
  {Ishihara}, \citenamefont {Suzuki}, \citenamefont {Ono}, \citenamefont
  {Kitamura},\ and\ \citenamefont {Asada}}]{Ishihara:2016vdc}%
  \BibitemOpen
  \bibfield  {author} {\bibinfo {author} {\bibfnamefont {Asahi}\ \bibnamefont
  {Ishihara}}, \bibinfo {author} {\bibfnamefont {Yusuke}\ \bibnamefont
  {Suzuki}}, \bibinfo {author} {\bibfnamefont {Toshiaki}\ \bibnamefont {Ono}},
  \bibinfo {author} {\bibfnamefont {Takao}\ \bibnamefont {Kitamura}}, \ and\
  \bibinfo {author} {\bibfnamefont {Hideki}\ \bibnamefont {Asada}},\ }\bibfield
   {title} {\enquote {\bibinfo {title} {{Gravitational bending angle of light
  for finite distance and the Gauss-Bonnet theorem}},}\ }\href {\doibase
  10.1103/PhysRevD.94.084015} {\bibfield  {journal} {\bibinfo  {journal} {Phys.
  Rev. D}\ }\textbf {\bibinfo {volume} {94}},\ \bibinfo {pages} {084015}
  (\bibinfo {year} {2016})},\ \Eprint {http://arxiv.org/abs/1604.08308}
  {arXiv:1604.08308 [gr-qc]} \BibitemShut {NoStop}%
\bibitem [{\citenamefont {Ono}\ \emph {et~al.}(2017)\citenamefont {Ono},
  \citenamefont {Ishihara},\ and\ \citenamefont {Asada}}]{Ono:2017pie}%
  \BibitemOpen
  \bibfield  {author} {\bibinfo {author} {\bibfnamefont {Toshiaki}\
  \bibnamefont {Ono}}, \bibinfo {author} {\bibfnamefont {Asahi}\ \bibnamefont
  {Ishihara}}, \ and\ \bibinfo {author} {\bibfnamefont {Hideki}\ \bibnamefont
  {Asada}},\ }\bibfield  {title} {\enquote {\bibinfo {title} {{Gravitomagnetic
  bending angle of light with finite-distance corrections in stationary
  axisymmetric spacetimes}},}\ }\href {\doibase 10.1103/PhysRevD.96.104037}
  {\bibfield  {journal} {\bibinfo  {journal} {Phys. Rev. D}\ }\textbf {\bibinfo
  {volume} {96}},\ \bibinfo {pages} {104037} (\bibinfo {year} {2017})},\
  \Eprint {http://arxiv.org/abs/1704.05615} {arXiv:1704.05615 [gr-qc]}
  \BibitemShut {NoStop}%
\bibitem [{\citenamefont {Li}\ and\ \citenamefont
  {\"Ovg\"un}(2020)}]{Li:2020dln}%
  \BibitemOpen
  \bibfield  {author} {\bibinfo {author} {\bibfnamefont {Zonghai}\ \bibnamefont
  {Li}}\ and\ \bibinfo {author} {\bibfnamefont {Ali}\ \bibnamefont
  {\"Ovg\"un}},\ }\bibfield  {title} {\enquote {\bibinfo {title}
  {{Finite-distance gravitational deflection of massive particles by a
  Kerr-like black hole in the bumblebee gravity model}},}\ }\href {\doibase
  10.1103/PhysRevD.101.024040} {\bibfield  {journal} {\bibinfo  {journal}
  {Phys. Rev. D}\ }\textbf {\bibinfo {volume} {101}},\ \bibinfo {pages}
  {024040} (\bibinfo {year} {2020})},\ \Eprint
  {http://arxiv.org/abs/2001.02074} {arXiv:2001.02074 [gr-qc]} \BibitemShut
  {NoStop}%
\bibitem [{\citenamefont {Li}\ \emph {et~al.}(2020{\natexlab{b}})\citenamefont
  {Li}, \citenamefont {Zhang},\ and\ \citenamefont {\"Ovg\"un}}]{Li:2020wvn}%
  \BibitemOpen
  \bibfield  {author} {\bibinfo {author} {\bibfnamefont {Zonghai}\ \bibnamefont
  {Li}}, \bibinfo {author} {\bibfnamefont {Guodong}\ \bibnamefont {Zhang}}, \
  and\ \bibinfo {author} {\bibfnamefont {Ali}\ \bibnamefont {\"Ovg\"un}},\
  }\bibfield  {title} {\enquote {\bibinfo {title} {{Circular Orbit of a
  Particle and Weak Gravitational Lensing}},}\ }\href {\doibase
  10.1103/PhysRevD.101.124058} {\bibfield  {journal} {\bibinfo  {journal}
  {Phys. Rev. D}\ }\textbf {\bibinfo {volume} {101}},\ \bibinfo {pages}
  {124058} (\bibinfo {year} {2020}{\natexlab{b}})},\ \Eprint
  {http://arxiv.org/abs/2006.13047} {arXiv:2006.13047 [gr-qc]} \BibitemShut
  {NoStop}%
\bibitem [{\citenamefont {Belhaj}\ \emph {et~al.}(2022)\citenamefont {Belhaj},
  \citenamefont {Belmahi}, \citenamefont {Benali},\ and\ \citenamefont
  {Moumni~El}}]{Belhaj:2022vte}%
  \BibitemOpen
  \bibfield  {author} {\bibinfo {author} {\bibfnamefont {A.}~\bibnamefont
  {Belhaj}}, \bibinfo {author} {\bibfnamefont {H.}~\bibnamefont {Belmahi}},
  \bibinfo {author} {\bibfnamefont {M.}~\bibnamefont {Benali}}, \ and\ \bibinfo
  {author} {\bibfnamefont {H.}~\bibnamefont {Moumni~El}},\ }\bibfield  {title}
  {\enquote {\bibinfo {title} {{Light Deflection by Rotating Regular Black
  Holes with a Cosmological Constant}},}\ }\href@noop {} {\  (\bibinfo {year}
  {2022})},\ \Eprint {http://arxiv.org/abs/2204.10150} {arXiv:2204.10150
  [gr-qc]} \BibitemShut {NoStop}%
\bibitem [{\citenamefont {Pantig}\ and\ \citenamefont
  {\"Ovg\"un}(2022{\natexlab{c}})}]{Pantig:2022toh}%
  \BibitemOpen
  \bibfield  {author} {\bibinfo {author} {\bibfnamefont {Reggie~C.}\
  \bibnamefont {Pantig}}\ and\ \bibinfo {author} {\bibfnamefont {Ali}\
  \bibnamefont {\"Ovg\"un}},\ }\bibfield  {title} {\enquote {\bibinfo {title}
  {{Dark matter effect on the weak deflection angle by black holes at the
  center of Milky Way and M87 galaxies}},}\ }\href {\doibase
  10.1140/epjc/s10052-022-10319-8} {\bibfield  {journal} {\bibinfo  {journal}
  {Eur. Phys. J. C}\ }\textbf {\bibinfo {volume} {82}},\ \bibinfo {pages} {391}
  (\bibinfo {year} {2022}{\natexlab{c}})},\ \Eprint
  {http://arxiv.org/abs/2201.03365} {arXiv:2201.03365 [gr-qc]} \BibitemShut
  {NoStop}%
\bibitem [{\citenamefont {Javed}\ \emph {et~al.}(2023)\citenamefont {Javed},
  \citenamefont {Atique}, \citenamefont {Pantig},\ and\ \citenamefont
  {\"Ovg\"un}}]{Javed:2023iih}%
  \BibitemOpen
  \bibfield  {author} {\bibinfo {author} {\bibfnamefont {Wajiha}\ \bibnamefont
  {Javed}}, \bibinfo {author} {\bibfnamefont {Mehak}\ \bibnamefont {Atique}},
  \bibinfo {author} {\bibfnamefont {Reggie~C.}\ \bibnamefont {Pantig}}, \ and\
  \bibinfo {author} {\bibfnamefont {Ali}\ \bibnamefont {\"Ovg\"un}},\
  }\bibfield  {title} {\enquote {\bibinfo {title} {{Weak Deflection Angle,
  Hawking Radiation and Greybody Bound of Reissner-Nordstr\"om Black Hole
  Corrected by Bounce Parameter}},}\ }\href {\doibase 10.3390/sym15010148}
  {\bibfield  {journal} {\bibinfo  {journal} {Symmetry}\ }\textbf {\bibinfo
  {volume} {15}},\ \bibinfo {pages} {148} (\bibinfo {year} {2023})},\ \Eprint
  {http://arxiv.org/abs/2301.01855} {arXiv:2301.01855 [gr-qc]} \BibitemShut
  {NoStop}%
\bibitem [{\citenamefont {Javed}\ \emph
  {et~al.}(2022{\natexlab{a}})\citenamefont {Javed}, \citenamefont {Atique},
  \citenamefont {Pantig},\ and\ \citenamefont {\"Ovg\"un}}]{Javed:2023IJGMMP}%
  \BibitemOpen
  \bibfield  {author} {\bibinfo {author} {\bibfnamefont {Wajiha}\ \bibnamefont
  {Javed}}, \bibinfo {author} {\bibfnamefont {Mehak}\ \bibnamefont {Atique}},
  \bibinfo {author} {\bibfnamefont {Reggie~C.}\ \bibnamefont {Pantig}}, \ and\
  \bibinfo {author} {\bibfnamefont {Ali}\ \bibnamefont {\"Ovg\"un}},\
  }\bibfield  {title} {\enquote {\bibinfo {title} {{Weak lensing, Hawking
  radiation and greybody factor bound by a charged black holes with non-linear
  electrodynamics corrections}},}\ }\href {\doibase 10.1142/s0219887823500408}
  {\bibfield  {journal} {\bibinfo  {journal} {International Journal of
  Geometric Methods in Modern Physics}\ ,\ \bibinfo {pages} {2350040}}
  (\bibinfo {year} {2022}{\natexlab{a}})}\BibitemShut {NoStop}%
\bibitem [{\citenamefont {Javed}\ \emph
  {et~al.}(2022{\natexlab{b}})\citenamefont {Javed}, \citenamefont {Riaz},
  \citenamefont {Pantig},\ and\ \citenamefont {\"Ovg\"un}}]{Javed:2022fsn}%
  \BibitemOpen
  \bibfield  {author} {\bibinfo {author} {\bibfnamefont {Wajiha}\ \bibnamefont
  {Javed}}, \bibinfo {author} {\bibfnamefont {Sibgha}\ \bibnamefont {Riaz}},
  \bibinfo {author} {\bibfnamefont {Reggie~C.}\ \bibnamefont {Pantig}}, \ and\
  \bibinfo {author} {\bibfnamefont {Ali}\ \bibnamefont {\"Ovg\"un}},\
  }\bibfield  {title} {\enquote {\bibinfo {title} {{Weak gravitational lensing
  in dark matter and plasma mediums for wormhole-like static aether
  solution}},}\ }\href {\doibase 10.1140/epjc/s10052-022-11030-4} {\bibfield
  {journal} {\bibinfo  {journal} {Eur. Phys. J. C}\ }\textbf {\bibinfo {volume}
  {82}},\ \bibinfo {pages} {1057} (\bibinfo {year} {2022}{\natexlab{b}})},\
  \Eprint {http://arxiv.org/abs/2212.00804} {arXiv:2212.00804 [gr-qc]}
  \BibitemShut {NoStop}%
\bibitem [{\citenamefont {Javed}\ \emph
  {et~al.}(2022{\natexlab{c}})\citenamefont {Javed}, \citenamefont {Irshad},
  \citenamefont {Pantig},\ and\ \citenamefont {\"Ovg\"un}}]{Javed:2022gtz}%
  \BibitemOpen
  \bibfield  {author} {\bibinfo {author} {\bibfnamefont {Wajiha}\ \bibnamefont
  {Javed}}, \bibinfo {author} {\bibfnamefont {Hafsa}\ \bibnamefont {Irshad}},
  \bibinfo {author} {\bibfnamefont {Reggie~C.}\ \bibnamefont {Pantig}}, \ and\
  \bibinfo {author} {\bibfnamefont {Ali}\ \bibnamefont {\"Ovg\"un}},\
  }\bibfield  {title} {\enquote {\bibinfo {title} {{Weak Deflection Angle by
  Kalb\textendash{}Ramond Traversable Wormhole in Plasma and Dark Matter
  Mediums}},}\ }\href {\doibase 10.3390/universe8110599} {\bibfield  {journal}
  {\bibinfo  {journal} {Universe}\ }\textbf {\bibinfo {volume} {8}},\ \bibinfo
  {pages} {599} (\bibinfo {year} {2022}{\natexlab{c}})},\ \Eprint
  {http://arxiv.org/abs/2211.07009} {arXiv:2211.07009 [gr-qc]} \BibitemShut
  {NoStop}%
\bibitem [{\citenamefont {Breton}(2003)}]{Breton:2003tk}%
  \BibitemOpen
  \bibfield  {author} {\bibinfo {author} {\bibfnamefont {Nora}\ \bibnamefont
  {Breton}},\ }\bibfield  {title} {\enquote {\bibinfo {title} {{Born-Infeld
  black hole in the isolated horizon framework}},}\ }\href {\doibase
  10.1103/PhysRevD.67.124004} {\bibfield  {journal} {\bibinfo  {journal} {Phys.
  Rev. D}\ }\textbf {\bibinfo {volume} {67}},\ \bibinfo {pages} {124004}
  (\bibinfo {year} {2003})},\ \Eprint {http://arxiv.org/abs/hep-th/0301254}
  {arXiv:hep-th/0301254} \BibitemShut {NoStop}%
\bibitem [{\citenamefont {Hendi}(2013)}]{Hendi:2013dwa}%
  \BibitemOpen
  \bibfield  {author} {\bibinfo {author} {\bibfnamefont {S.~H.}\ \bibnamefont
  {Hendi}},\ }\bibfield  {title} {\enquote {\bibinfo {title} {{Asymptotic
  Reissner-Nordstroem black holes}},}\ }\href {\doibase
  10.1016/j.aop.2013.03.008} {\bibfield  {journal} {\bibinfo  {journal} {Annals
  Phys.}\ }\textbf {\bibinfo {volume} {333}},\ \bibinfo {pages} {282--289}
  (\bibinfo {year} {2013})},\ \Eprint {http://arxiv.org/abs/1405.5359}
  {arXiv:1405.5359 [gr-qc]} \BibitemShut {NoStop}%
\bibitem [{\citenamefont {Kruglov}(2015)}]{Kruglov:2015yua}%
  \BibitemOpen
  \bibfield  {author} {\bibinfo {author} {\bibfnamefont {S.~I.}\ \bibnamefont
  {Kruglov}},\ }\bibfield  {title} {\enquote {\bibinfo {title} {{Nonlinear
  electrodynamics and black holes}},}\ }\href {\doibase
  10.1142/S0219887815500735} {\bibfield  {journal} {\bibinfo  {journal} {Int.
  J. Geom. Meth. Mod. Phys.}\ }\textbf {\bibinfo {volume} {12}},\ \bibinfo
  {pages} {1550073} (\bibinfo {year} {2015})},\ \Eprint
  {http://arxiv.org/abs/1504.03941} {arXiv:1504.03941 [physics.gen-ph]}
  \BibitemShut {NoStop}%
\bibitem [{\citenamefont {Kruglov}(2016{\natexlab{a}})}]{Kruglov:2016ymq}%
  \BibitemOpen
  \bibfield  {author} {\bibinfo {author} {\bibfnamefont {S.~I.}\ \bibnamefont
  {Kruglov}},\ }\bibfield  {title} {\enquote {\bibinfo {title} {{Asymptotic
  Reissner-Nordstr\"om solution within nonlinear electrodynamics}},}\ }\href
  {\doibase 10.1103/PhysRevD.94.044026} {\bibfield  {journal} {\bibinfo
  {journal} {Phys. Rev. D}\ }\textbf {\bibinfo {volume} {94}},\ \bibinfo
  {pages} {044026} (\bibinfo {year} {2016}{\natexlab{a}})},\ \Eprint
  {http://arxiv.org/abs/1608.04275} {arXiv:1608.04275 [gr-qc]} \BibitemShut
  {NoStop}%
\bibitem [{\citenamefont {Kruglov}(2016{\natexlab{b}})}]{Kruglov:2016ezw}%
  \BibitemOpen
  \bibfield  {author} {\bibinfo {author} {\bibfnamefont {S.~I.}\ \bibnamefont
  {Kruglov}},\ }\bibfield  {title} {\enquote {\bibinfo {title} {{Nonlinear
  arcsin-electrodynamics and asymptotic Reissner-Nordstr\"om black holes}},}\
  }\href {\doibase 10.1002/andp.201600027} {\bibfield  {journal} {\bibinfo
  {journal} {Annalen Phys.}\ }\textbf {\bibinfo {volume} {528}},\ \bibinfo
  {pages} {588--596} (\bibinfo {year} {2016}{\natexlab{b}})},\ \Eprint
  {http://arxiv.org/abs/1607.07726} {arXiv:1607.07726 [gr-qc]} \BibitemShut
  {NoStop}%
\bibitem [{\citenamefont {Kruglov}(2017)}]{Kruglov:2017fck}%
  \BibitemOpen
  \bibfield  {author} {\bibinfo {author} {\bibfnamefont {S.~I.}\ \bibnamefont
  {Kruglov}},\ }\bibfield  {title} {\enquote {\bibinfo {title} {{Black hole as
  a magnetic monopole within exponential nonlinear electrodynamics}},}\ }\href
  {\doibase 10.1016/j.aop.2016.12.036} {\bibfield  {journal} {\bibinfo
  {journal} {Annals Phys.}\ }\textbf {\bibinfo {volume} {378}},\ \bibinfo
  {pages} {59--70} (\bibinfo {year} {2017})},\ \Eprint
  {http://arxiv.org/abs/1703.02029} {arXiv:1703.02029 [gr-qc]} \BibitemShut
  {NoStop}%
\bibitem [{\citenamefont {\"Ovg\"un}(2021)}]{Ovgun:2021ttv}%
  \BibitemOpen
  \bibfield  {author} {\bibinfo {author} {\bibfnamefont {A.}~\bibnamefont
  {\"Ovg\"un}},\ }\bibfield  {title} {\enquote {\bibinfo {title} {{Black hole
  with confining electric potential in scalar-tensor description of regularized
  4-dimensional Einstein-Gauss-Bonnet gravity}},}\ }\href {\doibase
  10.1016/j.physletb.2021.136517} {\bibfield  {journal} {\bibinfo  {journal}
  {Phys. Lett. B}\ }\textbf {\bibinfo {volume} {820}},\ \bibinfo {pages}
  {136517} (\bibinfo {year} {2021})},\ \Eprint
  {http://arxiv.org/abs/2105.05035} {arXiv:2105.05035 [gr-qc]} \BibitemShut
  {NoStop}%
\bibitem [{\citenamefont {Ali}\ and\ \citenamefont
  {Saifullah}(2019)}]{Ali:2019bcn}%
  \BibitemOpen
  \bibfield  {author} {\bibinfo {author} {\bibfnamefont {Askar}\ \bibnamefont
  {Ali}}\ and\ \bibinfo {author} {\bibfnamefont {Khalid}\ \bibnamefont
  {Saifullah}},\ }\bibfield  {title} {\enquote {\bibinfo {title} {{Asymptotic
  magnetically charged non-singular black hole and its thermodynamics}},}\
  }\href {\doibase 10.1016/j.physletb.2019.03.041} {\bibfield  {journal}
  {\bibinfo  {journal} {Phys. Lett. B}\ }\textbf {\bibinfo {volume} {792}},\
  \bibinfo {pages} {276--283} (\bibinfo {year} {2019})},\ \Eprint
  {http://arxiv.org/abs/1904.05727} {arXiv:1904.05727 [gr-qc]} \BibitemShut
  {NoStop}%
\bibitem [{\citenamefont {Bera}\ \emph {et~al.}(2020)\citenamefont {Bera},
  \citenamefont {Ghosh},\ and\ \citenamefont {Majhi}}]{Bera:2019oxg}%
  \BibitemOpen
  \bibfield  {author} {\bibinfo {author} {\bibfnamefont {Avijit}\ \bibnamefont
  {Bera}}, \bibinfo {author} {\bibfnamefont {Subir}\ \bibnamefont {Ghosh}}, \
  and\ \bibinfo {author} {\bibfnamefont {Bibhas~Ranjan}\ \bibnamefont
  {Majhi}},\ }\bibfield  {title} {\enquote {\bibinfo {title} {{Hawking
  radiation in a non-covariant frame: the Jacobi metric approach}},}\ }\href
  {\doibase 10.1140/epjp/s13360-020-00693-1} {\bibfield  {journal} {\bibinfo
  {journal} {Eur. Phys. J. Plus}\ }\textbf {\bibinfo {volume} {135}},\ \bibinfo
  {pages} {670} (\bibinfo {year} {2020})},\ \Eprint
  {http://arxiv.org/abs/1909.12607} {arXiv:1909.12607 [gr-qc]} \BibitemShut
  {NoStop}%
\bibitem [{\citenamefont {Gibbons}(2016)}]{Gibbons:2015qja}%
  \BibitemOpen
  \bibfield  {author} {\bibinfo {author} {\bibfnamefont {G.~W.}\ \bibnamefont
  {Gibbons}},\ }\bibfield  {title} {\enquote {\bibinfo {title} {{The
  Jacobi-metric for timelike geodesics in static spacetimes}},}\ }\href
  {\doibase 10.1088/0264-9381/33/2/025004} {\bibfield  {journal} {\bibinfo
  {journal} {Class. Quant. Grav.}\ }\textbf {\bibinfo {volume} {33}},\ \bibinfo
  {pages} {025004} (\bibinfo {year} {2016})},\ \Eprint
  {http://arxiv.org/abs/1508.06755} {arXiv:1508.06755 [gr-qc]} \BibitemShut
  {NoStop}%
\bibitem [{\citenamefont {Chanda}\ \emph {et~al.}(2017)\citenamefont {Chanda},
  \citenamefont {Gibbons},\ and\ \citenamefont {Guha}}]{Chanda:2016aph}%
  \BibitemOpen
  \bibfield  {author} {\bibinfo {author} {\bibfnamefont {Sumanto}\ \bibnamefont
  {Chanda}}, \bibinfo {author} {\bibfnamefont {G.~W.}\ \bibnamefont {Gibbons}},
  \ and\ \bibinfo {author} {\bibfnamefont {Partha}\ \bibnamefont {Guha}},\
  }\bibfield  {title} {\enquote {\bibinfo {title} {{Jacobi-Maupertuis-Eisenhart
  metric and geodesic flows}},}\ }\href {\doibase 10.1063/1.4978333} {\bibfield
   {journal} {\bibinfo  {journal} {J. Math. Phys.}\ }\textbf {\bibinfo {volume}
  {58}},\ \bibinfo {pages} {032503} (\bibinfo {year} {2017})},\ \Eprint
  {http://arxiv.org/abs/1612.00375} {arXiv:1612.00375 [math-ph]} \BibitemShut
  {NoStop}%
\bibitem [{\citenamefont {Das}\ \emph {et~al.}(2017)\citenamefont {Das},
  \citenamefont {Sk},\ and\ \citenamefont {Ghosh}}]{Das:2016opi}%
  \BibitemOpen
  \bibfield  {author} {\bibinfo {author} {\bibfnamefont {Praloy}\ \bibnamefont
  {Das}}, \bibinfo {author} {\bibfnamefont {Ripon}\ \bibnamefont {Sk}}, \ and\
  \bibinfo {author} {\bibfnamefont {Subir}\ \bibnamefont {Ghosh}},\ }\bibfield
  {title} {\enquote {\bibinfo {title} {{Motion of charged particle in
  Reissner\textendash{}Nordstr\"om spacetime: a Jacobi-metric approach}},}\
  }\href {\doibase 10.1140/epjc/s10052-017-5295-6} {\bibfield  {journal}
  {\bibinfo  {journal} {Eur. Phys. J. C}\ }\textbf {\bibinfo {volume} {77}},\
  \bibinfo {pages} {735} (\bibinfo {year} {2017})},\ \Eprint
  {http://arxiv.org/abs/1609.04577} {arXiv:1609.04577 [gr-qc]} \BibitemShut
  {NoStop}%
\bibitem [{\citenamefont {Srinivasan}\ and\ \citenamefont
  {Padmanabhan}(1999)}]{Srinivasan:1998ty}%
  \BibitemOpen
  \bibfield  {author} {\bibinfo {author} {\bibfnamefont {K.}~\bibnamefont
  {Srinivasan}}\ and\ \bibinfo {author} {\bibfnamefont {T.}~\bibnamefont
  {Padmanabhan}},\ }\bibfield  {title} {\enquote {\bibinfo {title} {{Particle
  production and complex path analysis}},}\ }\href {\doibase
  10.1103/PhysRevD.60.024007} {\bibfield  {journal} {\bibinfo  {journal} {Phys.
  Rev. D}\ }\textbf {\bibinfo {volume} {60}},\ \bibinfo {pages} {024007}
  (\bibinfo {year} {1999})},\ \Eprint {http://arxiv.org/abs/gr-qc/9812028}
  {arXiv:gr-qc/9812028} \BibitemShut {NoStop}%
\bibitem [{\citenamefont {Robinson}\ and\ \citenamefont
  {Wilczek}(2005)}]{Robinson:2005pd}%
  \BibitemOpen
  \bibfield  {author} {\bibinfo {author} {\bibfnamefont {Sean~P.}\ \bibnamefont
  {Robinson}}\ and\ \bibinfo {author} {\bibfnamefont {Frank}\ \bibnamefont
  {Wilczek}},\ }\bibfield  {title} {\enquote {\bibinfo {title} {{A Relationship
  between Hawking radiation and gravitational anomalies}},}\ }\href {\doibase
  10.1103/PhysRevLett.95.011303} {\bibfield  {journal} {\bibinfo  {journal}
  {Phys. Rev. Lett.}\ }\textbf {\bibinfo {volume} {95}},\ \bibinfo {pages}
  {011303} (\bibinfo {year} {2005})},\ \Eprint
  {http://arxiv.org/abs/gr-qc/0502074} {arXiv:gr-qc/0502074} \BibitemShut
  {NoStop}%
\bibitem [{\citenamefont {Iso}\ \emph {et~al.}(2006)\citenamefont {Iso},
  \citenamefont {Umetsu},\ and\ \citenamefont {Wilczek}}]{Iso:2006wa}%
  \BibitemOpen
  \bibfield  {author} {\bibinfo {author} {\bibfnamefont {Satoshi}\ \bibnamefont
  {Iso}}, \bibinfo {author} {\bibfnamefont {Hiroshi}\ \bibnamefont {Umetsu}}, \
  and\ \bibinfo {author} {\bibfnamefont {Frank}\ \bibnamefont {Wilczek}},\
  }\bibfield  {title} {\enquote {\bibinfo {title} {{Hawking radiation from
  charged black holes via gauge and gravitational anomalies}},}\ }\href
  {\doibase 10.1103/PhysRevLett.96.151302} {\bibfield  {journal} {\bibinfo
  {journal} {Phys. Rev. Lett.}\ }\textbf {\bibinfo {volume} {96}},\ \bibinfo
  {pages} {151302} (\bibinfo {year} {2006})},\ \Eprint
  {http://arxiv.org/abs/hep-th/0602146} {arXiv:hep-th/0602146} \BibitemShut
  {NoStop}%
\bibitem [{\citenamefont {Majhi}(2010)}]{Majhi:2010onr}%
  \BibitemOpen
  \bibfield  {author} {\bibinfo {author} {\bibfnamefont {Bibhas~Ranjan}\
  \bibnamefont {Majhi}},\ }\emph {\bibinfo {title} {{Quantum Tunneling in Black
  Holes}}},\ \href@noop {} {Ph.D. thesis},\ \bibinfo  {school} {Calcutta U.}
  (\bibinfo {year} {2010}),\ \Eprint {http://arxiv.org/abs/1110.6008}
  {arXiv:1110.6008 [gr-qc]} \BibitemShut {NoStop}%
\bibitem [{\citenamefont {Virbhadra}\ \emph {et~al.}(1998)\citenamefont
  {Virbhadra}, \citenamefont {Narasimha},\ and\ \citenamefont
  {Chitre}}]{Virbhadra:1998dy}%
  \BibitemOpen
  \bibfield  {author} {\bibinfo {author} {\bibfnamefont {K.~S.}\ \bibnamefont
  {Virbhadra}}, \bibinfo {author} {\bibfnamefont {D.}~\bibnamefont
  {Narasimha}}, \ and\ \bibinfo {author} {\bibfnamefont {S.~M.}\ \bibnamefont
  {Chitre}},\ }\bibfield  {title} {\enquote {\bibinfo {title} {{Role of the
  scalar field in gravitational lensing}},}\ }\href@noop {} {\bibfield
  {journal} {\bibinfo  {journal} {Astron. Astrophys.}\ }\textbf {\bibinfo
  {volume} {337}},\ \bibinfo {pages} {1--8} (\bibinfo {year} {1998})},\ \Eprint
  {http://arxiv.org/abs/astro-ph/9801174} {arXiv:astro-ph/9801174} \BibitemShut
  {NoStop}%
\bibitem [{\citenamefont {Weinberg}(1972)}]{Weinberg:1972kfs}%
  \BibitemOpen
  \bibfield  {author} {\bibinfo {author} {\bibfnamefont {Steven}\ \bibnamefont
  {Weinberg}},\ }\href@noop {} {\emph {\bibinfo {title} {{Gravitation and
  Cosmology}: {Principles and Applications of the General Theory of
  Relativity}}}}\ (\bibinfo  {publisher} {John Wiley and Sons},\ \bibinfo
  {address} {New York},\ \bibinfo {year} {1972})\BibitemShut {NoStop}%
\bibitem [{\citenamefont {Lu}\ and\ \citenamefont {Xie}(2019)}]{Lu:2019ush}%
  \BibitemOpen
  \bibfield  {author} {\bibinfo {author} {\bibfnamefont {Xu}~\bibnamefont
  {Lu}}\ and\ \bibinfo {author} {\bibfnamefont {Yi}~\bibnamefont {Xie}},\
  }\bibfield  {title} {\enquote {\bibinfo {title} {{Weak and strong deflection
  gravitational lensing by a renormalization group improved Schwarzschild black
  hole}},}\ }\href {\doibase 10.1140/epjc/s10052-019-7537-2} {\bibfield
  {journal} {\bibinfo  {journal} {Eur. Phys. J. C}\ }\textbf {\bibinfo {volume}
  {79}},\ \bibinfo {pages} {1016} (\bibinfo {year} {2019})}\BibitemShut
  {NoStop}%
\bibitem [{\citenamefont {Keeton}\ and\ \citenamefont
  {Petters}(2005)}]{keeton}%
  \BibitemOpen
  \bibfield  {author} {\bibinfo {author} {\bibfnamefont {Charles~R.}\
  \bibnamefont {Keeton}}\ and\ \bibinfo {author} {\bibfnamefont {A.~O.}\
  \bibnamefont {Petters}},\ }\bibfield  {title} {\enquote {\bibinfo {title}
  {{Formalism for testing theories of gravity using lensing by compact objects.
  I. Static, spherically symmetric case}},}\ }\href {\doibase
  10.1103/PhysRevD.72.104006} {\bibfield  {journal} {\bibinfo  {journal} {Phys.
  Rev. D}\ }\textbf {\bibinfo {volume} {72}},\ \bibinfo {pages} {104006}
  (\bibinfo {year} {2005})},\ \Eprint {http://arxiv.org/abs/gr-qc/0511019}
  {arXiv:gr-qc/0511019} \BibitemShut {NoStop}%
\bibitem [{\citenamefont {Latimer}(2013)}]{Latimer:2013rja}%
  \BibitemOpen
  \bibfield  {author} {\bibinfo {author} {\bibfnamefont {David~C.}\
  \bibnamefont {Latimer}},\ }\bibfield  {title} {\enquote {\bibinfo {title}
  {{Dispersive Light Propagation at Cosmological Distances: Matter Effects}},}\
  }\href {\doibase 10.1103/PhysRevD.88.063517} {\bibfield  {journal} {\bibinfo
  {journal} {Phys. Rev. D}\ }\textbf {\bibinfo {volume} {88}},\ \bibinfo
  {pages} {063517} (\bibinfo {year} {2013})},\ \Eprint
  {http://arxiv.org/abs/1308.1112} {arXiv:1308.1112 [hep-ph]} \BibitemShut
  {NoStop}%
\bibitem [{\citenamefont {\"Ovg\"un}(2020)}]{Ovgun:2020yuv}%
  \BibitemOpen
  \bibfield  {author} {\bibinfo {author} {\bibfnamefont {Ali}\ \bibnamefont
  {\"Ovg\"un}},\ }\bibfield  {title} {\enquote {\bibinfo {title} {{Weak
  Deflection Angle of Black-bounce Traversable Wormholes Using Gauss-Bonnet
  Theorem in the Dark Matter Medium}},}\ }\href {\doibase
  10.20944/preprints202008.0512.v1} {\bibfield  {journal} {\bibinfo  {journal}
  {Turk. J. Phys.}\ }\textbf {\bibinfo {volume} {44}},\ \bibinfo {pages}
  {465--471} (\bibinfo {year} {2020})},\ \Eprint
  {http://arxiv.org/abs/2011.04423} {arXiv:2011.04423 [gr-qc]} \BibitemShut
  {NoStop}%
\bibitem [{\citenamefont {Crisnejo}\ and\ \citenamefont
  {Gallo}(2018)}]{Crisnejo:2018uyn}%
  \BibitemOpen
  \bibfield  {author} {\bibinfo {author} {\bibfnamefont {Gabriel}\ \bibnamefont
  {Crisnejo}}\ and\ \bibinfo {author} {\bibfnamefont {Emanuel}\ \bibnamefont
  {Gallo}},\ }\bibfield  {title} {\enquote {\bibinfo {title} {{Weak lensing in
  a plasma medium and gravitational deflection of massive particles using the
  Gauss-Bonnet theorem. A unified treatment}},}\ }\href {\doibase
  10.1103/PhysRevD.97.124016} {\bibfield  {journal} {\bibinfo  {journal} {Phys.
  Rev. D}\ }\textbf {\bibinfo {volume} {97}},\ \bibinfo {pages} {124016}
  (\bibinfo {year} {2018})},\ \Eprint {http://arxiv.org/abs/1804.05473}
  {arXiv:1804.05473 [gr-qc]} \BibitemShut {NoStop}%
\bibitem [{\citenamefont {Bambi}(2012)}]{Bambi:2012tg}%
  \BibitemOpen
  \bibfield  {author} {\bibinfo {author} {\bibfnamefont {Cosimo}\ \bibnamefont
  {Bambi}},\ }\bibfield  {title} {\enquote {\bibinfo {title} {{A code to
  compute the emission of thin accretion disks in non-Kerr space-times and test
  the nature of black hole candidates}},}\ }\href {\doibase
  10.1088/0004-637X/761/2/174} {\bibfield  {journal} {\bibinfo  {journal}
  {Astrophys. J.}\ }\textbf {\bibinfo {volume} {761}},\ \bibinfo {pages} {174}
  (\bibinfo {year} {2012})},\ \Eprint {http://arxiv.org/abs/1210.5679}
  {arXiv:1210.5679 [gr-qc]} \BibitemShut {NoStop}%
\bibitem [{\citenamefont {Bambi}(2013)}]{Bambi:2013nla}%
  \BibitemOpen
  \bibfield  {author} {\bibinfo {author} {\bibfnamefont {Cosimo}\ \bibnamefont
  {Bambi}},\ }\bibfield  {title} {\enquote {\bibinfo {title} {{Can the
  supermassive objects at the centers of galaxies be traversable wormholes? The
  first test of strong gravity for mm/sub-mm very long baseline interferometry
  facilities}},}\ }\href {\doibase 10.1103/PhysRevD.87.107501} {\bibfield
  {journal} {\bibinfo  {journal} {Phys. Rev. D}\ }\textbf {\bibinfo {volume}
  {87}},\ \bibinfo {pages} {107501} (\bibinfo {year} {2013})},\ \Eprint
  {http://arxiv.org/abs/1304.5691} {arXiv:1304.5691 [gr-qc]} \BibitemShut
  {NoStop}%
\bibitem [{\citenamefont {Jaroszynski}\ and\ \citenamefont
  {Kurpiewski}(1997)}]{Jaroszynski:1997bw}%
  \BibitemOpen
  \bibfield  {author} {\bibinfo {author} {\bibfnamefont {M.}~\bibnamefont
  {Jaroszynski}}\ and\ \bibinfo {author} {\bibfnamefont {A.}~\bibnamefont
  {Kurpiewski}},\ }\bibfield  {title} {\enquote {\bibinfo {title} {{Optics near
  kerr black holes: spectra of advection dominated accretion flows}},}\
  }\href@noop {} {\bibfield  {journal} {\bibinfo  {journal} {Astron.
  Astrophys.}\ }\textbf {\bibinfo {volume} {326}},\ \bibinfo {pages} {419}
  (\bibinfo {year} {1997})},\ \Eprint {http://arxiv.org/abs/astro-ph/9705044}
  {arXiv:astro-ph/9705044} \BibitemShut {NoStop}%
\bibitem [{\citenamefont {Perlick}\ and\ \citenamefont
  {Tsupko}(2022)}]{Perlick:2021aok}%
  \BibitemOpen
  \bibfield  {author} {\bibinfo {author} {\bibfnamefont {Volker}\ \bibnamefont
  {Perlick}}\ and\ \bibinfo {author} {\bibfnamefont {Oleg~Yu.}\ \bibnamefont
  {Tsupko}},\ }\bibfield  {title} {\enquote {\bibinfo {title} {{Calculating
  black hole shadows: Review of analytical studies}},}\ }\href {\doibase
  10.1016/j.physrep.2021.10.004} {\bibfield  {journal} {\bibinfo  {journal}
  {Phys. Rept.}\ }\textbf {\bibinfo {volume} {947}},\ \bibinfo {pages} {1--39}
  (\bibinfo {year} {2022})},\ \Eprint {http://arxiv.org/abs/2105.07101}
  {arXiv:2105.07101 [gr-qc]} \BibitemShut {NoStop}%
\bibitem [{\citenamefont {Bapat}\ \emph {et~al.}(2021)\citenamefont {Bapat}
  \emph {et~al.}}]{shreyas_bapat_2021_4739508}%
  \BibitemOpen
  \bibfield  {author} {\bibinfo {author} {\bibfnamefont {Shreyas}\ \bibnamefont
  {Bapat}} \emph {et~al.},\ }\href {\doibase 10.5281/zenodo.4739508} {\enquote
  {\bibinfo {title} {einsteinpy/einsteinpy: Einsteinpy 0.4.0},}\ } (\bibinfo
  {year} {2021})\BibitemShut {NoStop}%
\bibitem [{\citenamefont {Akhmedov}\ \emph {et~al.}(2006)\citenamefont
  {Akhmedov}, \citenamefont {Akhmedova},\ and\ \citenamefont
  {Singleton}}]{Akhmedov:2006pg}%
  \BibitemOpen
  \bibfield  {author} {\bibinfo {author} {\bibfnamefont {Emil~T.}\ \bibnamefont
  {Akhmedov}}, \bibinfo {author} {\bibfnamefont {Valeria}\ \bibnamefont
  {Akhmedova}}, \ and\ \bibinfo {author} {\bibfnamefont {Douglas}\ \bibnamefont
  {Singleton}},\ }\bibfield  {title} {\enquote {\bibinfo {title} {{Hawking
  temperature in the tunneling picture}},}\ }\href {\doibase
  10.1016/j.physletb.2006.09.028} {\bibfield  {journal} {\bibinfo  {journal}
  {Phys. Lett. B}\ }\textbf {\bibinfo {volume} {642}},\ \bibinfo {pages}
  {124--128} (\bibinfo {year} {2006})},\ \Eprint
  {http://arxiv.org/abs/hep-th/0608098} {arXiv:hep-th/0608098} \BibitemShut
  {NoStop}%
\bibitem [{\citenamefont {Maldacena}\ and\ \citenamefont
  {Strominger}(1997)}]{Maldacena:1996ix}%
  \BibitemOpen
  \bibfield  {author} {\bibinfo {author} {\bibfnamefont {Juan~Martin}\
  \bibnamefont {Maldacena}}\ and\ \bibinfo {author} {\bibfnamefont {Andrew}\
  \bibnamefont {Strominger}},\ }\bibfield  {title} {\enquote {\bibinfo {title}
  {{Black hole grey body factors and d-brane spectroscopy}},}\ }\href {\doibase
  10.1103/PhysRevD.55.861} {\bibfield  {journal} {\bibinfo  {journal} {Phys.
  Rev. D}\ }\textbf {\bibinfo {volume} {55}},\ \bibinfo {pages} {861--870}
  (\bibinfo {year} {1997})},\ \Eprint {http://arxiv.org/abs/hep-th/9609026}
  {arXiv:hep-th/9609026} \BibitemShut {NoStop}%
\bibitem [{\citenamefont {Harmark}\ \emph {et~al.}(2010)\citenamefont
  {Harmark}, \citenamefont {Natario},\ and\ \citenamefont
  {Schiappa}}]{Harmark:2007jy}%
  \BibitemOpen
  \bibfield  {author} {\bibinfo {author} {\bibfnamefont {Troels}\ \bibnamefont
  {Harmark}}, \bibinfo {author} {\bibfnamefont {Jose}\ \bibnamefont {Natario}},
  \ and\ \bibinfo {author} {\bibfnamefont {Ricardo}\ \bibnamefont {Schiappa}},\
  }\bibfield  {title} {\enquote {\bibinfo {title} {{Greybody Factors for
  d-Dimensional Black Holes}},}\ }\href {\doibase 10.4310/ATMP.2010.v14.n3.a1}
  {\bibfield  {journal} {\bibinfo  {journal} {Adv. Theor. Math. Phys.}\
  }\textbf {\bibinfo {volume} {14}},\ \bibinfo {pages} {727--794} (\bibinfo
  {year} {2010})},\ \Eprint {http://arxiv.org/abs/0708.0017} {arXiv:0708.0017
  [hep-th]} \BibitemShut {NoStop}%
\bibitem [{\citenamefont {Rinc\'on}\ and\ \citenamefont
  {Panotopoulos}(2018)}]{Rincon:2018ktz}%
  \BibitemOpen
  \bibfield  {author} {\bibinfo {author} {\bibfnamefont {\'Angel}\ \bibnamefont
  {Rinc\'on}}\ and\ \bibinfo {author} {\bibfnamefont {Grigoris}\ \bibnamefont
  {Panotopoulos}},\ }\bibfield  {title} {\enquote {\bibinfo {title} {{Greybody
  factors and quasinormal modes for a nonminimally coupled scalar field in a
  cloud of strings in (2+1)-dimensional background}},}\ }\href {\doibase
  10.1140/epjc/s10052-018-6352-5} {\bibfield  {journal} {\bibinfo  {journal}
  {Eur. Phys. J. C}\ }\textbf {\bibinfo {volume} {78}},\ \bibinfo {pages} {858}
  (\bibinfo {year} {2018})},\ \Eprint {http://arxiv.org/abs/1810.08822}
  {arXiv:1810.08822 [gr-qc]} \BibitemShut {NoStop}%
\bibitem [{\citenamefont {Rinc\'on}\ and\ \citenamefont
  {Santos}(2020)}]{Rincon:2020cos}%
  \BibitemOpen
  \bibfield  {author} {\bibinfo {author} {\bibfnamefont {\'Angel}\ \bibnamefont
  {Rinc\'on}}\ and\ \bibinfo {author} {\bibfnamefont {Victor}\ \bibnamefont
  {Santos}},\ }\bibfield  {title} {\enquote {\bibinfo {title} {{Greybody factor
  and quasinormal modes of Regular Black Holes}},}\ }\href {\doibase
  10.1140/epjc/s10052-020-08445-2} {\bibfield  {journal} {\bibinfo  {journal}
  {Eur. Phys. J. C}\ }\textbf {\bibinfo {volume} {80}},\ \bibinfo {pages} {910}
  (\bibinfo {year} {2020})},\ \Eprint {http://arxiv.org/abs/2009.04386}
  {arXiv:2009.04386 [gr-qc]} \BibitemShut {NoStop}%
\bibitem [{\citenamefont {Panotopoulos}\ and\ \citenamefont
  {Rinc\'on}(2018)}]{Panotopoulos:2018pvu}%
  \BibitemOpen
  \bibfield  {author} {\bibinfo {author} {\bibfnamefont {Grigoris}\
  \bibnamefont {Panotopoulos}}\ and\ \bibinfo {author} {\bibfnamefont {Angel}\
  \bibnamefont {Rinc\'on}},\ }\bibfield  {title} {\enquote {\bibinfo {title}
  {{Greybody factors for a minimally coupled scalar field in three-dimensional
  Einstein-power-Maxwell black hole background}},}\ }\href {\doibase
  10.1103/PhysRevD.97.085014} {\bibfield  {journal} {\bibinfo  {journal} {Phys.
  Rev. D}\ }\textbf {\bibinfo {volume} {97}},\ \bibinfo {pages} {085014}
  (\bibinfo {year} {2018})},\ \Eprint {http://arxiv.org/abs/1804.04684}
  {arXiv:1804.04684 [hep-th]} \BibitemShut {NoStop}%
\bibitem [{\citenamefont {Panotopoulos}\ and\ \citenamefont
  {Rinc\'on}(2017)}]{Panotopoulos:2017yoe}%
  \BibitemOpen
  \bibfield  {author} {\bibinfo {author} {\bibfnamefont {Grigoris}\
  \bibnamefont {Panotopoulos}}\ and\ \bibinfo {author} {\bibfnamefont
  {\'Angel}\ \bibnamefont {Rinc\'on}},\ }\bibfield  {title} {\enquote {\bibinfo
  {title} {{Greybody factors for a minimally coupled massless scalar field in
  Einstein-Born-Infeld dilaton spacetime}},}\ }\href {\doibase
  10.1103/PhysRevD.96.025009} {\bibfield  {journal} {\bibinfo  {journal} {Phys.
  Rev. D}\ }\textbf {\bibinfo {volume} {96}},\ \bibinfo {pages} {025009}
  (\bibinfo {year} {2017})},\ \Eprint {http://arxiv.org/abs/1706.07455}
  {arXiv:1706.07455 [hep-th]} \BibitemShut {NoStop}%
\bibitem [{\citenamefont {Klebanov}\ and\ \citenamefont
  {Mathur}(1997)}]{Klebanov:1997cx}%
  \BibitemOpen
  \bibfield  {author} {\bibinfo {author} {\bibfnamefont {Igor~R.}\ \bibnamefont
  {Klebanov}}\ and\ \bibinfo {author} {\bibfnamefont {Samir~D.}\ \bibnamefont
  {Mathur}},\ }\bibfield  {title} {\enquote {\bibinfo {title} {{Black hole grey
  body factors and absorption of scalars by effective strings}},}\ }\href
  {\doibase 10.1016/S0550-3213(97)00287-3} {\bibfield  {journal} {\bibinfo
  {journal} {Nucl. Phys. B}\ }\textbf {\bibinfo {volume} {500}},\ \bibinfo
  {pages} {115--132} (\bibinfo {year} {1997})},\ \Eprint
  {http://arxiv.org/abs/hep-th/9701187} {arXiv:hep-th/9701187} \BibitemShut
  {NoStop}%
\bibitem [{\citenamefont {Liu}\ \emph {et~al.}(2022)\citenamefont {Liu},
  \citenamefont {Yang}, \citenamefont {\"Ovg\"un}, \citenamefont {Long},\ and\
  \citenamefont {Xu}}]{Liu:2022ygf}%
  \BibitemOpen
  \bibfield  {author} {\bibinfo {author} {\bibfnamefont {Dong}\ \bibnamefont
  {Liu}}, \bibinfo {author} {\bibfnamefont {Yi}~\bibnamefont {Yang}}, \bibinfo
  {author} {\bibfnamefont {Ali}\ \bibnamefont {\"Ovg\"un}}, \bibinfo {author}
  {\bibfnamefont {Zheng-Wen}\ \bibnamefont {Long}}, \ and\ \bibinfo {author}
  {\bibfnamefont {Zhaoyi}\ \bibnamefont {Xu}},\ }\bibfield  {title} {\enquote
  {\bibinfo {title} {{Quasinormal Modes and Greybody Bounds of Rotating Black
  Holes in a Dark Matter Halo}},}\ }\href@noop {} {\  (\bibinfo {year}
  {2022})},\ \Eprint {http://arxiv.org/abs/2204.11563} {arXiv:2204.11563
  [gr-qc]} \BibitemShut {NoStop}%
\bibitem [{\citenamefont {Fernando}(2005)}]{Fernando:2004ay}%
  \BibitemOpen
  \bibfield  {author} {\bibinfo {author} {\bibfnamefont {Sharmanthie}\
  \bibnamefont {Fernando}},\ }\bibfield  {title} {\enquote {\bibinfo {title}
  {{Greybody factors of charged dilaton black holes in 2 + 1 dimensions}},}\
  }\href {\doibase 10.1007/s10714-005-0035-x} {\bibfield  {journal} {\bibinfo
  {journal} {Gen. Rel. Grav.}\ }\textbf {\bibinfo {volume} {37}},\ \bibinfo
  {pages} {461--481} (\bibinfo {year} {2005})},\ \Eprint
  {http://arxiv.org/abs/hep-th/0407163} {arXiv:hep-th/0407163} \BibitemShut
  {NoStop}%
\bibitem [{\citenamefont {Konoplya}\ \emph {et~al.}(2019)\citenamefont
  {Konoplya}, \citenamefont {Zhidenko},\ and\ \citenamefont
  {Zinhailo}}]{Konoplya:2019hlu}%
  \BibitemOpen
  \bibfield  {author} {\bibinfo {author} {\bibfnamefont {R.~A.}\ \bibnamefont
  {Konoplya}}, \bibinfo {author} {\bibfnamefont {A.}~\bibnamefont {Zhidenko}},
  \ and\ \bibinfo {author} {\bibfnamefont {A.~F.}\ \bibnamefont {Zinhailo}},\
  }\bibfield  {title} {\enquote {\bibinfo {title} {{Higher order WKB formula
  for quasinormal modes and grey-body factors: recipes for quick and accurate
  calculations}},}\ }\href {\doibase 10.1088/1361-6382/ab2e25} {\bibfield
  {journal} {\bibinfo  {journal} {Class. Quant. Grav.}\ }\textbf {\bibinfo
  {volume} {36}},\ \bibinfo {pages} {155002} (\bibinfo {year} {2019})},\
  \Eprint {http://arxiv.org/abs/1904.10333} {arXiv:1904.10333 [gr-qc]}
  \BibitemShut {NoStop}%
\bibitem [{\citenamefont {Javed}\ \emph
  {et~al.}(2022{\natexlab{d}})\citenamefont {Javed}, \citenamefont {Riaz},\
  and\ \citenamefont {\"Ovg\"un}}]{Javed:2022rrs}%
  \BibitemOpen
  \bibfield  {author} {\bibinfo {author} {\bibfnamefont {Wajiha}\ \bibnamefont
  {Javed}}, \bibinfo {author} {\bibfnamefont {Sibgha}\ \bibnamefont {Riaz}}, \
  and\ \bibinfo {author} {\bibfnamefont {Ali}\ \bibnamefont {\"Ovg\"un}},\
  }\bibfield  {title} {\enquote {\bibinfo {title} {{Weak Deflection Angle and
  Greybody Bound of Magnetized Regular Black Hole}},}\ }\href {\doibase
  10.3390/universe8050262} {\bibfield  {journal} {\bibinfo  {journal}
  {Universe}\ }\textbf {\bibinfo {volume} {8}},\ \bibinfo {pages} {262}
  (\bibinfo {year} {2022}{\natexlab{d}})},\ \Eprint
  {http://arxiv.org/abs/2205.02229} {arXiv:2205.02229 [gr-qc]} \BibitemShut
  {NoStop}%
\bibitem [{\citenamefont {Javed}\ \emph
  {et~al.}(2022{\natexlab{e}})\citenamefont {Javed}, \citenamefont {Aqib},\
  and\ \citenamefont {\"Ovg\"un}}]{Javed:2022kzf}%
  \BibitemOpen
  \bibfield  {author} {\bibinfo {author} {\bibfnamefont {Wajiha}\ \bibnamefont
  {Javed}}, \bibinfo {author} {\bibfnamefont {Muhammad}\ \bibnamefont {Aqib}},
  \ and\ \bibinfo {author} {\bibfnamefont {Ali}\ \bibnamefont {\"Ovg\"un}},\
  }\bibfield  {title} {\enquote {\bibinfo {title} {{Effect of the magnetic
  charge on weak deflection angle and greybody bound of the black hole in
  Einstein-Gauss-Bonnet gravity}},}\ }\href {\doibase
  10.1016/j.physletb.2022.137114} {\bibfield  {journal} {\bibinfo  {journal}
  {Phys. Lett. B}\ }\textbf {\bibinfo {volume} {829}},\ \bibinfo {pages}
  {137114} (\bibinfo {year} {2022}{\natexlab{e}})},\ \Eprint
  {http://arxiv.org/abs/2204.07864} {arXiv:2204.07864 [gr-qc]} \BibitemShut
  {NoStop}%
\bibitem [{\citenamefont {Yang}\ \emph {et~al.}(2023)\citenamefont {Yang},
  \citenamefont {Liu}, \citenamefont {\"Ovg\"un}, \citenamefont {Long},\ and\
  \citenamefont {Xu}}]{Yang:2022ifo}%
  \BibitemOpen
  \bibfield  {author} {\bibinfo {author} {\bibfnamefont {Yi}~\bibnamefont
  {Yang}}, \bibinfo {author} {\bibfnamefont {Dong}\ \bibnamefont {Liu}},
  \bibinfo {author} {\bibfnamefont {Ali}\ \bibnamefont {\"Ovg\"un}}, \bibinfo
  {author} {\bibfnamefont {Zheng-Wen}\ \bibnamefont {Long}}, \ and\ \bibinfo
  {author} {\bibfnamefont {Zhaoyi}\ \bibnamefont {Xu}},\ }\bibfield  {title}
  {\enquote {\bibinfo {title} {{Probing hairy black holes caused by
  gravitational decoupling using quasinormal modes and greybody bounds}},}\
  }\href {\doibase 10.1103/PhysRevD.107.064042} {\bibfield  {journal} {\bibinfo
   {journal} {Phys. Rev. D}\ }\textbf {\bibinfo {volume} {107}},\ \bibinfo
  {pages} {064042} (\bibinfo {year} {2023})},\ \Eprint
  {http://arxiv.org/abs/2203.11551} {arXiv:2203.11551 [gr-qc]} \BibitemShut
  {NoStop}%
\bibitem [{\citenamefont {Javed}\ \emph
  {et~al.}(2022{\natexlab{f}})\citenamefont {Javed}, \citenamefont {Hussain},\
  and\ \citenamefont {\"Ovg\"un}}]{Javed:2021ymu}%
  \BibitemOpen
  \bibfield  {author} {\bibinfo {author} {\bibfnamefont {Wajiha}\ \bibnamefont
  {Javed}}, \bibinfo {author} {\bibfnamefont {Iqra}\ \bibnamefont {Hussain}}, \
  and\ \bibinfo {author} {\bibfnamefont {Ali}\ \bibnamefont {\"Ovg\"un}},\
  }\bibfield  {title} {\enquote {\bibinfo {title} {{Weak deflection angle of
  Kazakov\textendash{}Solodukhin black hole in plasma medium using
  Gauss\textendash{}Bonnet theorem and its greybody bonding}},}\ }\href
  {\doibase 10.1140/epjp/s13360-022-02374-7} {\bibfield  {journal} {\bibinfo
  {journal} {Eur. Phys. J. Plus}\ }\textbf {\bibinfo {volume} {137}},\ \bibinfo
  {pages} {148} (\bibinfo {year} {2022}{\natexlab{f}})},\ \Eprint
  {http://arxiv.org/abs/2201.09879} {arXiv:2201.09879 [gr-qc]} \BibitemShut
  {NoStop}%
\bibitem [{\citenamefont {Mangut}\ \emph {et~al.}(2023)\citenamefont {Mangut},
  \citenamefont {G\"ursel}, \citenamefont {Kanzi},\ and\ \citenamefont
  {Sakall\i{}}}]{Mangut:2023oou}%
  \BibitemOpen
  \bibfield  {author} {\bibinfo {author} {\bibfnamefont {M.}~\bibnamefont
  {Mangut}}, \bibinfo {author} {\bibfnamefont {H.}~\bibnamefont {G\"ursel}},
  \bibinfo {author} {\bibfnamefont {S.}~\bibnamefont {Kanzi}}, \ and\ \bibinfo
  {author} {\bibfnamefont {\.I.}\ \bibnamefont {Sakall\i{}}},\ }\bibfield
  {title} {\enquote {\bibinfo {title} {{Probing the Lorentz Invariance
  Violation via Gravitational Lensing and Analytical Eigenmodes of Perturbed
  Slowly Rotating Bumblebee Black Holes}},}\ }\href {\doibase
  10.3390/universe9050225} {\  (\bibinfo {year} {2023}),\
  10.3390/universe9050225},\ \Eprint {http://arxiv.org/abs/2305.10815}
  {arXiv:2305.10815 [gr-qc]} \BibitemShut {NoStop}%
\bibitem [{\citenamefont {Kanzi}\ \emph {et~al.}(2023)\citenamefont {Kanzi},
  \citenamefont {Sakall\i{}},\ and\ \citenamefont
  {Pourhassan}}]{Kanzi:2023itu}%
  \BibitemOpen
  \bibfield  {author} {\bibinfo {author} {\bibfnamefont {Sara}\ \bibnamefont
  {Kanzi}}, \bibinfo {author} {\bibfnamefont {\.Izzet}\ \bibnamefont
  {Sakall\i{}}}, \ and\ \bibinfo {author} {\bibfnamefont {Behnam}\ \bibnamefont
  {Pourhassan}},\ }\bibfield  {title} {\enquote {\bibinfo {title}
  {{Superradiant (In)stability, Greybody Radiation, and Quasinormal Modes of
  Rotating Black Holes in non-linear Maxwell f (R) Gravity}},}\ }\href
  {\doibase 10.3390/sym15040873} {\  (\bibinfo {year} {2023}),\
  10.3390/sym15040873},\ \Eprint {http://arxiv.org/abs/2301.03866}
  {arXiv:2301.03866 [hep-th]} \BibitemShut {NoStop}%
\bibitem [{\citenamefont {Al-Badawi}\ \emph {et~al.}(2022)\citenamefont
  {Al-Badawi}, \citenamefont {Kanzi},\ and\ \citenamefont
  {Sakall\i{}}}]{Al-Badawi:2021wdm}%
  \BibitemOpen
  \bibfield  {author} {\bibinfo {author} {\bibfnamefont {Ahmad}\ \bibnamefont
  {Al-Badawi}}, \bibinfo {author} {\bibfnamefont {Sara}\ \bibnamefont {Kanzi}},
  \ and\ \bibinfo {author} {\bibfnamefont {\.Izzet}\ \bibnamefont
  {Sakall\i{}}},\ }\bibfield  {title} {\enquote {\bibinfo {title} {{Greybody
  radiation of scalar and Dirac perturbations of NUT black holes}},}\ }\href
  {\doibase 10.1140/epjp/s13360-021-02227-9} {\bibfield  {journal} {\bibinfo
  {journal} {Eur. Phys. J. Plus}\ }\textbf {\bibinfo {volume} {137}},\ \bibinfo
  {pages} {94} (\bibinfo {year} {2022})},\ \Eprint
  {http://arxiv.org/abs/2111.15005} {arXiv:2111.15005 [gr-qc]} \BibitemShut
  {NoStop}%
\bibitem [{\citenamefont {Al-Badawi}(2022)}]{Al-Badawi:2022uwh}%
  \BibitemOpen
  \bibfield  {author} {\bibinfo {author} {\bibfnamefont {Ahmad}\ \bibnamefont
  {Al-Badawi}},\ }\bibfield  {title} {\enquote {\bibinfo {title} {{Greybody
  factor and perturbation of a Schwarzschild black hole with string clouds and
  quintessence}},}\ }\href {\doibase 10.1007/s10714-022-02900-z} {\bibfield
  {journal} {\bibinfo  {journal} {Gen. Rel. Grav.}\ }\textbf {\bibinfo {volume}
  {54}},\ \bibinfo {pages} {11} (\bibinfo {year} {2022})},\ \Eprint
  {http://arxiv.org/abs/2201.09106} {arXiv:2201.09106 [gr-qc]} \BibitemShut
  {NoStop}%
\bibitem [{\citenamefont {Al-Badawi}\ \emph {et~al.}(2023)\citenamefont
  {Al-Badawi}, \citenamefont {Kanzi},\ and\ \citenamefont
  {Sakall\i{}}}]{Al-Badawi:2022aby}%
  \BibitemOpen
  \bibfield  {author} {\bibinfo {author} {\bibfnamefont {Ahmad}\ \bibnamefont
  {Al-Badawi}}, \bibinfo {author} {\bibfnamefont {Sara}\ \bibnamefont {Kanzi}},
  \ and\ \bibinfo {author} {\bibfnamefont {\.Izzet}\ \bibnamefont
  {Sakall\i{}}},\ }\bibfield  {title} {\enquote {\bibinfo {title} {{Fermionic
  and bosonic greybody factors as well as quasinormal modes for charged Taub
  NUT black holes}},}\ }\href {\doibase 10.1016/j.aop.2023.169294} {\bibfield
  {journal} {\bibinfo  {journal} {Annals Phys.}\ }\textbf {\bibinfo {volume}
  {452}},\ \bibinfo {pages} {169294} (\bibinfo {year} {2023})},\ \Eprint
  {http://arxiv.org/abs/2203.04140} {arXiv:2203.04140 [hep-th]} \BibitemShut
  {NoStop}%
\bibitem [{\citenamefont {Visser}(1999)}]{Visser:1998ke}%
  \BibitemOpen
  \bibfield  {author} {\bibinfo {author} {\bibfnamefont {Matt}\ \bibnamefont
  {Visser}},\ }\bibfield  {title} {\enquote {\bibinfo {title} {{Some general
  bounds for 1-D scattering}},}\ }\href {\doibase 10.1103/PhysRevA.59.427}
  {\bibfield  {journal} {\bibinfo  {journal} {Phys. Rev. A}\ }\textbf {\bibinfo
  {volume} {59}},\ \bibinfo {pages} {427--438} (\bibinfo {year} {1999})},\
  \Eprint {http://arxiv.org/abs/quant-ph/9901030} {arXiv:quant-ph/9901030}
  \BibitemShut {NoStop}%
\bibitem [{\citenamefont {Boonserm}\ and\ \citenamefont
  {Visser}(2008)}]{Boonserm:2008zg}%
  \BibitemOpen
  \bibfield  {author} {\bibinfo {author} {\bibfnamefont {Petarpa}\ \bibnamefont
  {Boonserm}}\ and\ \bibinfo {author} {\bibfnamefont {Matt}\ \bibnamefont
  {Visser}},\ }\bibfield  {title} {\enquote {\bibinfo {title} {{Bounding the
  greybody factors for Schwarzschild black holes}},}\ }\href {\doibase
  10.1103/PhysRevD.78.101502} {\bibfield  {journal} {\bibinfo  {journal} {Phys.
  Rev. D}\ }\textbf {\bibinfo {volume} {78}},\ \bibinfo {pages} {101502}
  (\bibinfo {year} {2008})},\ \Eprint {http://arxiv.org/abs/0806.2209}
  {arXiv:0806.2209 [gr-qc]} \BibitemShut {NoStop}%
\bibitem [{\citenamefont {Boonserm}\ \emph {et~al.}(2018)\citenamefont
  {Boonserm}, \citenamefont {Ngampitipan},\ and\ \citenamefont
  {Wongjun}}]{Boonserm:2017qcq}%
  \BibitemOpen
  \bibfield  {author} {\bibinfo {author} {\bibfnamefont {Petarpa}\ \bibnamefont
  {Boonserm}}, \bibinfo {author} {\bibfnamefont {Tritos}\ \bibnamefont
  {Ngampitipan}}, \ and\ \bibinfo {author} {\bibfnamefont {Pitayuth}\
  \bibnamefont {Wongjun}},\ }\bibfield  {title} {\enquote {\bibinfo {title}
  {{Greybody factor for black holes in dRGT massive gravity}},}\ }\href
  {\doibase 10.1140/epjc/s10052-018-5975-x} {\bibfield  {journal} {\bibinfo
  {journal} {Eur. Phys. J. C}\ }\textbf {\bibinfo {volume} {78}},\ \bibinfo
  {pages} {492} (\bibinfo {year} {2018})},\ \Eprint
  {http://arxiv.org/abs/1705.03278} {arXiv:1705.03278 [gr-qc]} \BibitemShut
  {NoStop}%
\end{thebibliography}%

\end{document}